    \documentclass[prl,amsmath,amssymb,letterpaper,onecolumn]{revtex4-2} 
    
    \usepackage{amsmath}
    \usepackage{mathrsfs}
    \usepackage{amsfonts}
    \usepackage{tabularx}
    \usepackage{times}
    \usepackage{graphicx}
    \usepackage{color}
    \usepackage{txfonts}
    
    \newcommand{\eq}{\begin{equation}}
    \newcommand{\xeq}{\end{equation}}

\begin{document}
                
    \author{X. Y. Jin}
    \affiliation{Associate of the National Institute of Standards and Technology, Boulder, Colorado 80305, USA}
    \affiliation{Department of Physics, University of Colorado, Boulder, Colorado 80309, USA}

    \author{K. Cicak}
    \affiliation{National Institute of Standards and Technology, Boulder, Colorado 80305, USA}

    \author{Z. Parrott}
    \affiliation{Associate of the National Institute of Standards and Technology, Boulder, Colorado 80305, USA}
    \affiliation{Department of Physics, University of Colorado, Boulder, Colorado 80309, USA}
    
    \author{S. Kotler}
    \altaffiliation{\textit{Current Address}: Racah Institute of Physics, The Hebrew University of Jerusalem, Jerusalem, 91904, Israel}
    \affiliation{Associate of the National Institute of Standards and Technology, Boulder, Colorado 80305, USA}
    \affiliation{Department of Physics, University of Colorado, Boulder, Colorado 80309, USA}
    
    \author{F. Lecocq}
    \affiliation{National Institute of Standards and Technology, Boulder, Colorado 80305, USA}
    
    \author{J. Teufel}
    \affiliation{National Institute of Standards and Technology, Boulder, Colorado 80305, USA}
    
    \author{J. Aumentado}
    \affiliation{National Institute of Standards and Technology, Boulder, Colorado 80305, USA}

    \author{E. Kapit}
    \affiliation{Colorado School of Mines, Dept. of Physics, 1500 Illinois St., Golden, CO 80401}
    
    \author{R. W. Simmonds}
    \email{E-mail: raymond.simmonds@nist.gov}
    \affiliation{National Institute of Standards and Technology, Boulder, Colorado 80305, USA}

\begin{abstract}
Future quantum information processors require tunable coupling architectures that can produce high fidelity logical gates between two or more qubits. Parametric coupling is a powerful technique for generating tunable interactions between many qubits.\cite{monroe2021programmable} Here, we present a highly flexible parametric coupling scheme with superconducting qubits that provides complete removal of residual $ZZ$ coupling and the implementation of driven SWAP or SWAP-free controlled-$Z$ (c$Z$) gates. Our fully integrated, 2D on-chip coupler design is only weakly flux tunable, cancels static linear coupling between the qubits, avoids internal coupler dynamics or excitations, and is extensible to multi-qubit circuit-QED systems. Exploring gate fidelity versus gate duration allows us to maximize two-qubit gate fidelity, while providing insights into possible error sources for these gates. Randomized benchmarking over several hours reveals that the parametric SWAP c$Z$ gate achieves an average fidelity of $99.44\pm 0.09$\% in a gate duration of 70~ns and a dispersively driven parametric SWAP-free c$Z$ gate attains an average fidelity of $99.47\pm 0.07$\% in only 30~ns. The fidelity remained above this value for over 8~hours and peaked twice with a maximum of $99.67\pm 0.14$\%. Overall, our parametric approach combines versatility, precision, speed, and high performance in one compact coupler design.
\end{abstract} 

    \title{Fast, tunable, high fidelity cZ-gates between superconducting qubits with parametric microwave control of ZZ-coupling} 
    \maketitle 
    
Quantum computers hold the potential to outperform classical computers by solving specific problems that are unattainable for classical systems.\cite{shor1994algorithms, nielsen2010quantum, arute2019supramacy}
To achieve quantum advantage in computation, two-qubit gates must be fast to maintain high fidelities within the entire architecture to allow for quantum error correction.\cite{Gaebler2016} 
So far, it is not clear what the optimal coupling architecture should be, but it is clear that a larger connectivity graph between elements combined with well-controlled quantum evolutions is advantageous for achieving both useful quantum computations and quantum simulations.\cite{monroe2021programmable}
For superconducting architectures with finite coherence times, two of the most challenging problems that remain are:
(i) the ability to perform quantum gates over short timescales\cite{barends2014superconducting,kelly2015state,Sung2021,stehlik2021tunable} with minimal uncontrolled or stray couplings to ensure high fidelity operations, and
(ii) to achieve this beyond just nearest neighbor pairs and possibly between any two arbitrarily chosen qubits within a $N$-qubit architecture\cite{monroe2021programmable,Bravyi2024}.
Thus far, there has been considerable progress in improving gate fidelities with tunable couplers, between nearest-neighbor qubits or even extended nearest-neighbor qubits.\cite{xu2020high,stehlik2021tunable,Sung2021,moskalenko2022high,marxer2023long,Ding2023,Zhang2024}
Gate fidelities beyond 99.9\% have been achieved with fluxonium qubits\cite{Ding2023,Zhang2024,Lin2024} with all coherence times beyond $100\,\mu$s, with one example\cite{Zhang2024} utilizing parametric gates.
As a way to move towards achieving all-to-all coupling and a more modular architecture, various approaches have been pursued to enable parametric interactions with superconducting circuits\cite{Rigetti2018,Noh2023,Zhou2021,Brown2022,Paolo2022,Marinelli2023}, analogous to some trapped-ion systems\cite{monroe2021programmable}.
In fact, parametric coupling has emerged as an essential tool for producing efficient and high-fidelity quantum operations with various types of superconducting systems, including planar cavities\cite{ZakkaBajjani2011}, cavity-QED\cite{Allman2014}, 3D cavities\cite{Lu2023}, transmon qubits\cite{didier2018parametric,Li2018,Zurich_CZ_2020,Hong2020,Nakamura_2QGates_PRA,Li2022}, fluxonium qubits\cite{Zhang2024}, and its application has been expanded beyond two-qubit gates to include higher levels or multi-qubit and cavity QED systems\cite{Li2018,Noh2023,Zhou2021,Roy2022,Marinelli2023}.
With growing interest in parametric coupling for quantum computing, researchers continue to develop new techniques and explore innovative applications, driving new advancements in the field.\cite{Paolo2022,Petrescu2023,Zhang2024}
In this work, we show a 2D architecture on-chip that uses parametric drives to control various resonant and non-resonant coupling dynamics providing a flexible coupling architecture\cite{Lu2017,Noh2023,Brown2022,Perez2023,Li2023a,Li2023b} that can achieve fast, high fidelity two-qubit gates with lower coherence transmons.
%
    \begin{figure*}[t] 
    {\includegraphics[width=1\hsize]{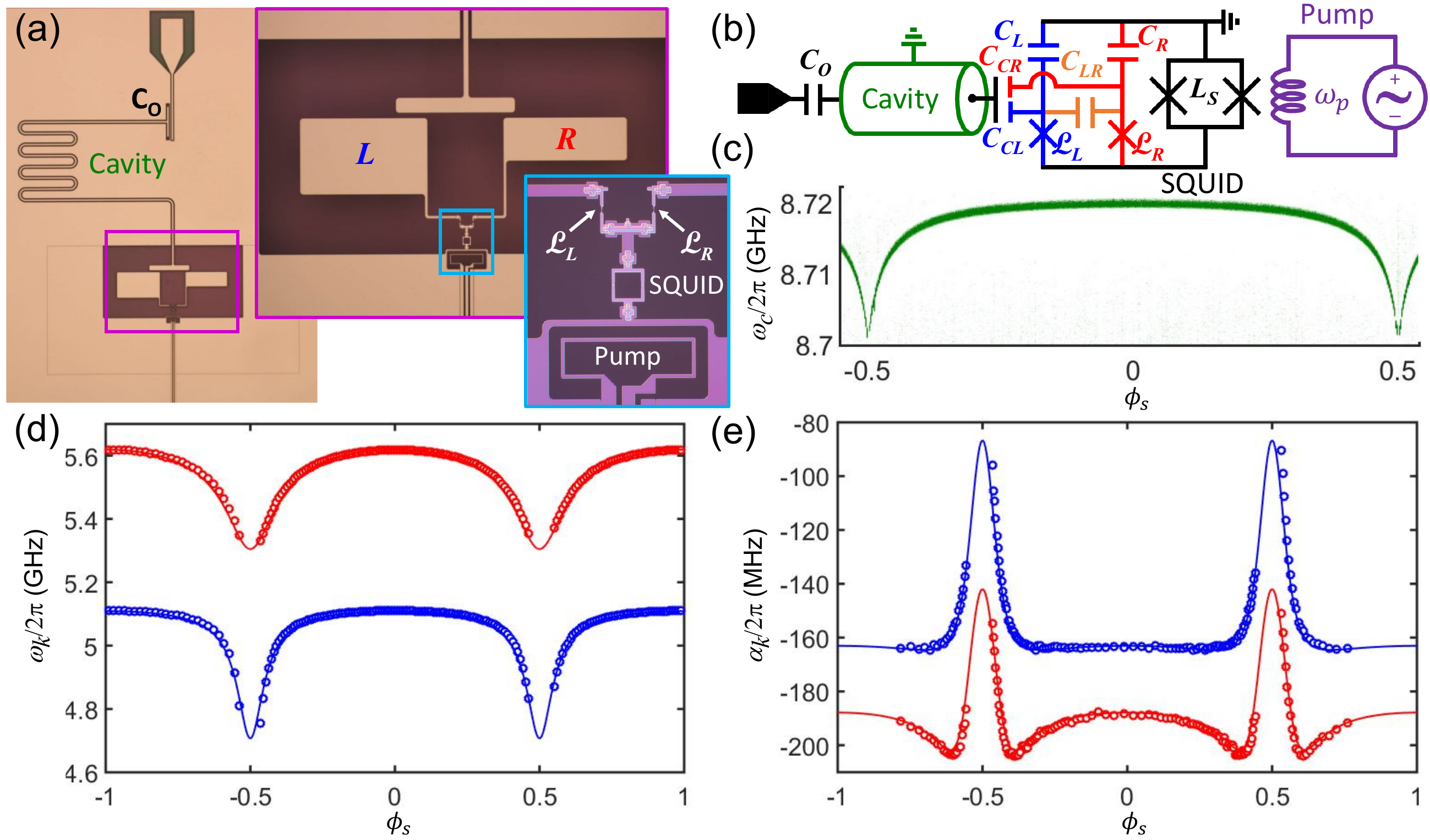}}
    \caption{{\textbf {A flux and parametrically tunable two-qubit system.}} (a) Optical micrograph of a tunable coupled qubit system with zoom-in views to show more details. (b) Schematic circuit diagram of the system including the Left transmon ($L$) and Right transmon ($R$) plus the readout cavity. The parametric flux drive is applied through a shared SQUID coupler. (c) Cavity spectrum as a function of $\phi_s = \Phi_s/\Phi_0$. (d) Plot of the $\phi_s$-periodic $L$ and $R$ transmon frequencies. (e) Plot of the $\phi_s$-periodic $L$ and $R$ transmon anharmonicities.}
    \label{Fig_1} 
    \end{figure*}

%
Parametric techniques rely on modulating a controlled parameter at a specific frequency and phase with a variable amplitude, such as an electrically or mechanically variable capacitance or a current or flux dependent inductance. \cite{yurke1989observation,Tian2008,Teufel2011,ZakkaBajjani2011,Allman2014}
A single parametric coupler is versatile, providing either single or dual mode amplification\cite{yurke1989observation,Teufel2011,Ranzani2015}, frequency conversion and dispersive interactions\cite{Noh2023,Perez2023}, or non-reciprocal directional dynamics\cite{Ranzani2015}.
In this study, we present a parametric coupler design strategy that minimizes the number of on-chip control and readout ports and is capable of eliminating residual $ZZ$ coupling between two transmon qubits, while performing various types of two-qubit gates: a parametric parametric iSWAP, a SWAP-based controlled-$Z$ (p-SWAP c$Z$), and a dispersive parametric SWAP-free controlled-$Z$ (p-SWAP-free c$Z$) gates.
%
%
%
%
We find that the gate errors are a function of gate duration, the shape of the pulse envelope, and are vulnerable to very large amplitude parametric drives that can rectify the qubit frequencies or induce single-qubit rotations from $n$-photon absorption at sub-harmonics of the individual qubit transition frequencies (see Supplemental Material for more details). 
Identifying these issues provides us with a path to further improve parametric-based gates.
Ultimately, this design can be extended to include more qubits and cavities\cite{Noh2023}, such integrated multi-qubit modules\cite{Zhou2021} could help facilitate necessary schemes for error correction\cite{gambetta2017building,preskill2018quantum,Bravyi2024} 

Our two-qubit system of study is displayed in Fig.~1(a) and (b) with two transmons, labelled ``Left" ($L$) and ``Right" ($R$), each having a main Josephson junction and a shared two-junction SQUID parametric coupler with total inductance $L_s$.
Here, $\mathcal{L}_k=\Phi_o/2\pi I_k$ represents the (nonlinear) Josephson inductance for $k\in{L,R}$ with nominally equal critical currents $I_o = I_L\approx I_R$ for the main qubit junctions, $E_{J} = I_o\Phi_o/2\pi$ is the Josephson energy and $\Phi_o = h/2e$ is the magnetic flux quantum.
Each transmon has capacitance $C_k$ (for $k\in{L,R}$), with mutual capacitance $C_{LR}$.
The two qubits also share the same readout cavity, each capacitively coupled through $C_{ck}$ (for $k\in{L,R}$), which also serves as the qubit drive line through the cavity coupling capacitor $C_o$ to the feedline.
This leads to a very weakly tunable cavity frequency that is periodic in the coupler's static flux bias $\phi_s=\Phi_s/\Phi_o$.
Our compact layout only requires two on-chip ports for the whole system, one for the single element excitations plus readout, and one for setting all static operating frequencies (static coupler bias) and for generating all coupled interactions (r.f. coupler ``pump" drive).
This type of architecture can be extended by adding more qubits or cavities to the coupler, without increasing the number of on-chip control and readout ports.\cite{Noh2023,Brown2022}
In Fig.~1(d), we see that the two qubits are quite closely packed in frequency space and tune with $\phi_s$ in unison, so their frequencies will never collide. 
Here, we denote their frequencies as $\omega_L$ and $\omega_R$, which represents the transition from the ground state $|gg\rangle$ to the joint two-qubit energy levels $|eg\rangle$ and $|ge\rangle$, respectively (see Fig.~2(c) for more details).
The shared SQUID provides only a mild tunability of the two transmons, where the two nominally equal SQUID junctions are each about $20\times E_{J}$, resulting in an inductive participation ratio of only a few percent at zero flux bias.
The system is designed to operate at a fixed static flux, where the two qubit frequencies are fixed, as well as the static coupling between them.
This mild tunability balances our ability to provide strong parametric interactions ($\lesssim 100$~MHz), allowing us to perform two-qubit gates quickly, while still minimizing either qubit's susceptibility ($\lesssim1$GHz/$\Phi_o$) to flux noise in the SQUID (see Supplemental Material for more details).\cite{Hutchings2017,Garcia2022,Noh2023}
In Fig.~1(e), we see that the anharmonicity $\alpha_k$ (for $k\in{L,R}$) of the two transmons is roughly constant except for a sharp decrease as $L_s$ increases rapidly near $\phi_s=0.5$.
Our device design differs considerably from most current coupler designs that usually incorporate a capacitively coupled SQUID that acts more like a tunable qubit or resonator between the qubits.
In contrast, our SQUID coupler is fully integrated galvanically into the qubits themselves and its self-resonance frequency is $>20$~GHz for all $\phi_s$, too high to be excited during coupled operations (see Supplemental Material for more details).
This also means that the coupler's resonant frequency will not collide with nearby qubits or cavities and any additional degrees of freedom, internal coupler dynamics, or decoherence channels are unlikely to perturb its operation.
%
    \begin{figure*}[t] 
    {\includegraphics[width=1\hsize]{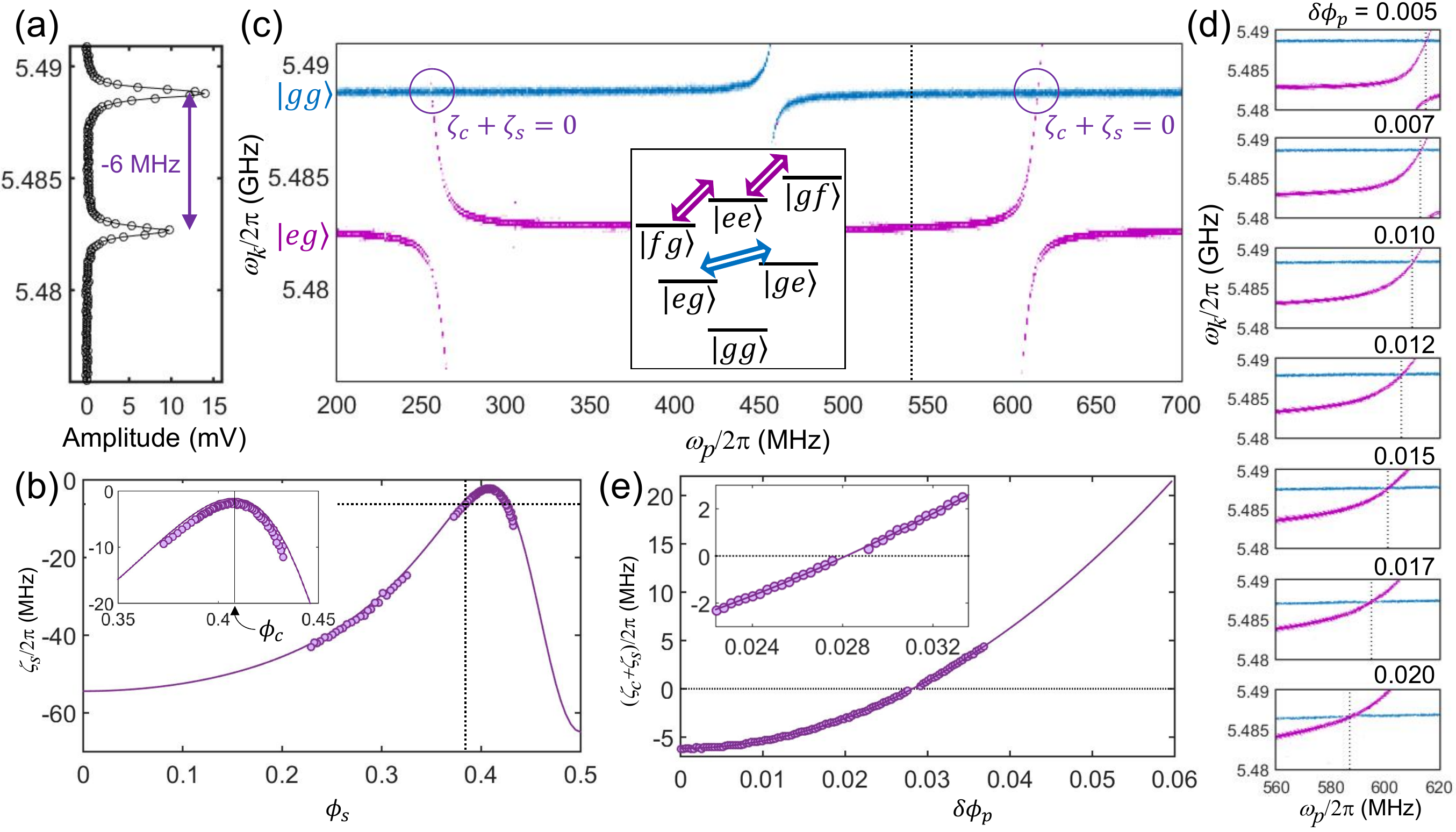}}
    \caption{{\textbf {Residual $ZZ$ coupling and cancellation.}} (a) Spectroscopic scan showing $ZZ$ coupling when $\phi_s\approx 0.386$. (b) Static $ZZ$ coupling $\zeta_s$ as a function of $\phi_s$, dashed vertical line indicates when $\phi_s\approx 0.386$, horizontal line indicates when $\zeta_s\approx -6$~MHz. Inset shows zoom-in near $\phi_c$ (vertical line). (c) Spectroscopic scan in the presence of the parametric pump when $\phi_s\approx 0.386$. Circles indicate cancellation of the total $ZZ$-coupling. Inset shows two-qubit energy level diagram with arrows indicating parametric transitions between the one and two excitation state manifolds. (d) Spectroscopic scan for increasing pump amplitude (see text for details). (e) $\zeta_c+\zeta_s$ as a function of pump amplitude showing parametric cancellation of the $ZZ$ coupling. Inset shows zoom-in near $\zeta_c+\zeta_s=0$.}
    \label{Fig_2} 
    \end{figure*}

%
Our device design also takes advantage of the relatively large mutual capacitance $C_{LR}$ between the two closely spaced transmons to largely cancel out the mutual inductive coupling due to the SQUID coupler.
The $L$ and $R$ qubit electrodes are designed in size, orientation, and separation so that the negative capacitive coupling equals the positive inductive coupling of the SQUID at the static ``cancellation flux'' bias, $\phi_c=\Phi_c/\Phi_o\approx 0.41$, where parametric coupling strengths will be strong.
At a flux bias far from $\phi_c$, the residual static coupling $g_s$ between the qubits is predominantly capacitive when $\phi_s<\phi_c$ or inductive when $\phi_s>\phi_c$.
Because the SQUID coupler is fully integrated, when the two qubits (with $\alpha_k<0$, for $k\in{L,R}$) are very strongly coupled ($g_s\gg 0$) with a static detuning of $\Delta_s = \omega_R-\omega_L\approx 0.5$~GHz, there is a very strong \emph{static} $ZZ$-coupling  between them\cite{barends2014superconducting,Long2021}, 
\begin{equation}
 \zeta_s=2g_s^2\frac{(\alpha_L+\alpha_R)}{(\Delta_s-\alpha_L)(\Delta_s+\alpha_R)}. 
 \label{Eq_1}
\end{equation}
In Fig.~2(a), we show an example of weak $ZZ$ coupling near $\phi_c$ at $\phi_s\approx 0.386$, where $\zeta_s/2\pi\approx -6$~MHz, and in (b), we show $\zeta_s$ as a function of $\phi_s$, where $|\zeta_s|/2\pi>50$~MHz at $\phi_s=0$ or $0.5$.
Notice that Eq.~\ref{Eq_1} only captures the induced \emph{nonlinear} interactions from the static \emph{linear} coupling between the two detuned qubits, which should go to zero at $\phi_c$.
However, we see that at $\phi_c$ we have $|\rm{min}(\zeta_s)|/2\pi\approx 2$~MHz.
This is due to the purely nonlinear coupling that results from a small fraction of the $L$ transmon's current passing through the $R$ transmon's main qubit junction and vise-versa.
Using our design layout, Ansys Q3D, and the ``scQubits" package\cite{Groszkowski2021,Chitta2022}, we find very good agreement between our numerical predictions (solid lines) and the experimental data (circles) in Fig.~1(d),(e) and Fig.~2(b), including the residual $\rm{min}(\zeta_s)$ at $\phi_c$ (see Supplemental Material for more details).  
%

With the qubits detuned by $\Delta_s$, energy exchange between two-qubit energy levels with an equal number of excitations is suppressed. 
The joint two-transmon levels are labelled $|LR\rangle$, denoting the individual $L$ and $R$ transmon levels as $g$, $e$, $f$, and so on (see Fig.~2(c)).  
However, if in addition to the static flux bias $\phi_s$, we drive the SQUID coupler with a cosinusoidal flux modulation,  $\phi(t)=\delta\phi_p\cos(\omega_p t)+\phi_s$, then we can generate new coupling terms between any two of the two-qubit states $|b\rangle$ and $|t\rangle$ written in an interaction frame with respect to the free Hamiltonian\cite{Allman2014},
\begin{equation}
 H_{Ip} = \hbar g_{p}\left(e^{i\omega_{p}t} + e^{-i\omega_{p}t}\right)(q_L q_R^\dagger e^{-i\Delta_{bt}t} + {\rm h.c.}),
 \label{Eq_2}
\end{equation}
where we will ignore sum frequency or counter-rotating terms, $q_k^\dagger$ ($q_k$) is the creation (annihilation) operator of the $k$th transmon mode, and $\Delta_{bt}$ is the difference in frequency or detuning between the higher `top' level $|t\rangle$ and the lower `bottom' level $|b\rangle$.
Here, we focus on $\omega_p=\Delta_{bt}+\Delta_{p}$, where $\Delta_p$ is the detuning from the difference frequency $\Delta_{bt}$. 
The parametric coupling strength is related to the flux dependent static coupling strength\cite{Allman2014} through $2g_{p} = (dg_{s}(\phi_s)/d\phi)\delta\phi_p$ (see Supplemental Material for more details). 
Of course, under ideal conditions $g_{s}(\phi_c) = 0$, while $dg_{s}(\phi_c)/d\phi\neq 0$. 
It is clear from Eq.~(\ref{Eq_2}) that when $\Delta_p = 0$, $\omega_p = \Delta_{bt}$ so that $H_{Ip} = \hbar g_{p}(q_L q_R^\dagger + q_L^\dagger q_R)$ provides a SWAP interaction between the qubits.
In general, the pump frequency can be detuned by $\Delta_{p}$ in order to sweep between parametrically-induced resonant-type ($\Delta_{p} \ll g_{p}$) and dispersive-type ($\Delta_{p} \gg g_{p}$) interactions near the transition between the states $|b\rangle$ and $|t\rangle$.
This behavior can been seen in the continuous frequency domain (see Fig.~2(c)) by measuring the spectrum of the $R$ qubit.
Starting from $|gg\rangle$, when the spectroscopic tone is resonant with the $R$ qubit transition, we move to state $|ge\rangle$.
For a fixed pump amplitude of $\delta_p=0.005$, we see an avoided crossing in this spectrum with a coupling strength $2g_{p}/2\pi\approx 5$~MHz centered at $\omega_p/2\pi\approx450$~MHz.
This occurs when the two parametrically interacting states are $|b\rangle = |eg\rangle$ and $|t\rangle = |ge\rangle$ and $\omega_p = \Delta_{bt} = \omega_{ge}-\omega_{eg}$, exchanging the single excitation manifold of states, as depicted in the inset of Fig.~2(c).
After $\pi$-pulsing the $L$ qubit, we start from $|eg\rangle$ and the $R$ qubit spectroscopic tone moves the state to $|ee\rangle$ with a resonance frequency  shifted by $\zeta_s/2\pi = -6$~MHz (see the line-cut of both peaks shown in Fig.~2(a)).
As the pump tone increases above $200$~MHz, an avoided crossing appears when $\omega_p = \omega_{gf}-\omega_{ee}$ and the two parametrically connected states are $|b\rangle =|ee\rangle$ and $|t\rangle =|gf\rangle$.
And, lastly when $\omega_p = \omega_{ee}-\omega_{fg}$, the interaction is between states $|b\rangle =|fg\rangle$ and $|t\rangle =|ee\rangle$.
These two splittings occur for parametric interactions within the two-photon excitation manifold of states, as depicted in the inset of Fig.~2(c).
%

The presence of the residual $ZZ$ coupling near $\phi_c$ gives us an opportunity to show the versatility of our parametric coupling scheme by allowing us to use parametric dispersive shifts to cancel $\zeta_s$.\cite{Noh2023,Perez2023}
Recall that the presence of $\zeta_s$ shifts the two-qubit energy level spectrum (shown in the inset of Fig.~2(c)), so that the position of $|ee\rangle$ satisfies the relation $\omega_{ee} = \omega_{eg}+\omega_{ge}+\zeta_s$.
Notice, in Fig.~2(c), that there are two values for $\omega_p$ where the spectroscopic peaks for initial states $|gg\rangle$ and $|eg\rangle$ cross, as denoted by the circles. 
This occurs when the total $ZZ$ coupling $\zeta_c+\zeta_s=0$.  
The parametrically induced dispersive shifts $\zeta_c$, analogous to the static case, follow from Eq.~(\ref{Eq_1}) with the reassignments: $g_{s}\rightarrow g_{p}$ and $\Delta_s\rightarrow\Delta_{p}$. 
And, according to Eq.~(\ref{Eq_1}), as we increase the pump amplitude $\delta\phi_p$, $g_p$ increases and the value for the detuning $\Delta_p$ that cancels $\zeta_s$ must also increase as well. 
This is observed in Fig.~2(d), as the cancellation point (dashed vertical line) moves increasingly to the left of $\Delta_{bt}$.
In addition, careful observation shows a clear decrease in the frequency of the spectroscopic line corresponding to the initial state $|gg\rangle$ as the pump amplitude increases due to a rectification effect from the curvature of the $\omega_k(\phi_s)$ curves (see Supplemental Material for more details).
If we choose an $\omega_p$ that is maximally detuned from both parametric transitions from the single ($\omega_{ge}-\omega_{eg}$) and double ($\omega_{fg}-\omega_{ee}$) manifold of states, we can avoid unwanted SWAP interactions (as discussed below) and ensure purely dispersive shifts that can fully cancel $\zeta_s$.
An example is shown in Fig.~2(e), where we plot the total effective $ZZ$ coupling $\zeta_s+\zeta_c$ as a function of pump amplitude for $\omega_p/2\pi=545$~MHz, vertical dashed line in Fig.~2(c).
Here, we begin at $-6$~MHz, then increase as $g_p^2$ until the spectroscopic peaks merge and we cross zero $ZZ$ coupling.
If we continue to increase the pump amplitude $\delta\phi_p$, $\zeta_c$ continues to increase as $g_p^2$ and dominates, leading to progressively stronger positive $ZZ$ coupling.
%
    \begin{figure*}[t]
    {\includegraphics[width=1\hsize]{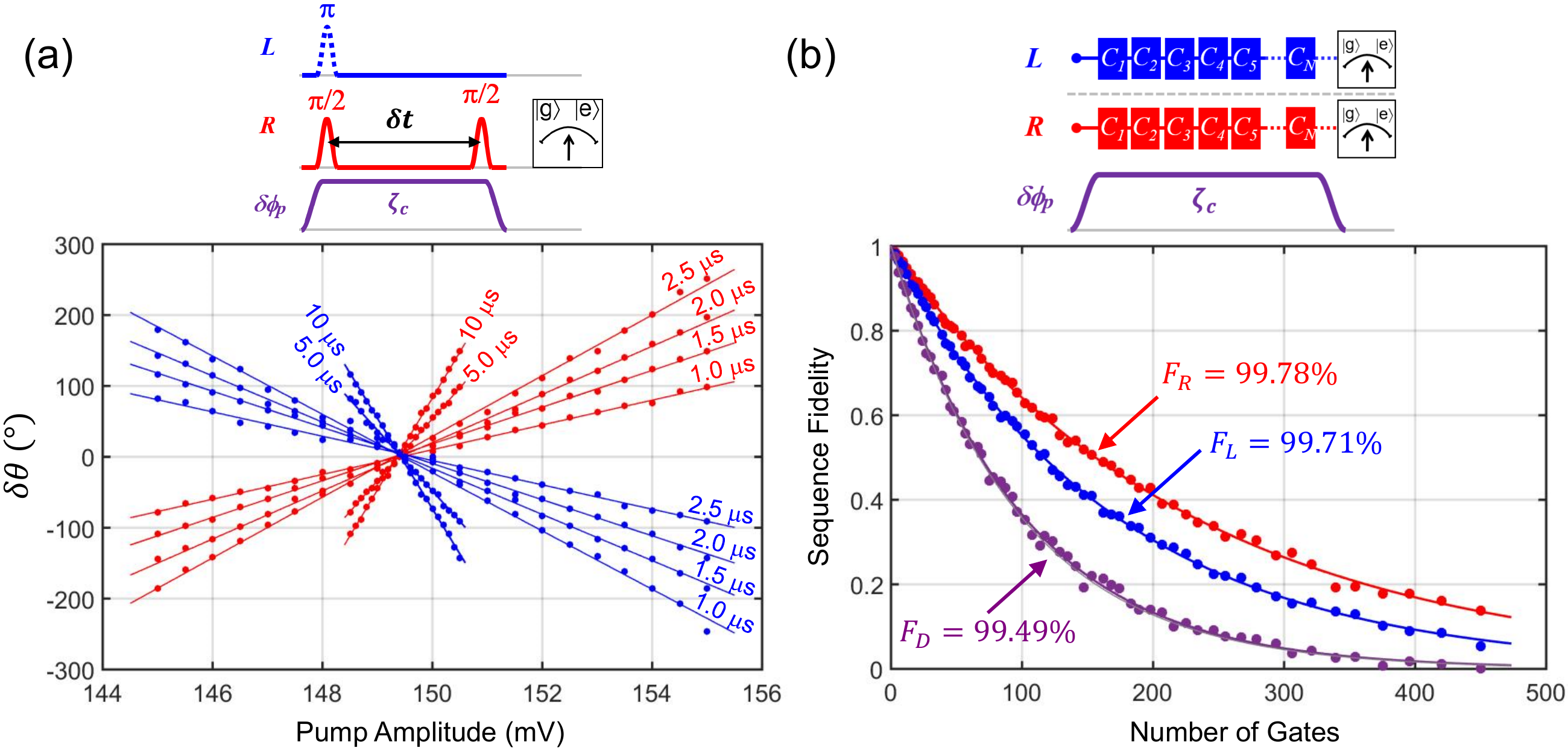}}
    \caption{{\textbf {Fine cancellation of $ZZ$ coupling.}} (a) Pulse diagram  of the cross-Ramsey measurement and data showing $\delta\theta$ as a function of pump amplitude (see text for details). For each value of $\delta t$, all the red (blue) linear fit lines for the $R$ ($L$) qubit data cross zero at the same pump amplitude where the total $ZZ$ coupling is eliminated. (b) Pulse diagram  and data for single qubit and simultaneous two-qubit RB showing that $\zeta_c+\zeta_s = 0$ with no additional correlated gate errors. Solids lines are fits to the data. The solid gray line is the expectation based on combining the single qubit errors.}
    \label{Fig_3} 
    \end{figure*}

%
Due to the scatter in the spectroscopic data, the quadratic fit allows us to identify zero $ZZ$ coupling with an uncertainty of about $\pm 50$~kHz.
In order to improve this by an order-of-magnitude, we use cross-Ramsey-measurements\cite{Houck_PRA} on the $R$ ($L$) qubit, whose phase evolution depends on the state of the $L$ ($R$) qubit and the size of the residual $ZZ$ coupling. 
In this experiment, the delay time between the two $\pi/2$ pulses on the $R$ qubit are fixed at $\delta t$, while the phase difference on the second pulse spans $2\pi$ producing a sine-wave.
We perform this experiment on the $R$ ($L$) qubit with and without the $\pi$ pulse on the $L$ ($R$) qubit and measure the phase difference between these two measurements as a function of the pump amplitude $\delta\phi_p$ around the $\zeta_s$ cancellation point predicted by the spectroscopic data in Fig.~2(e).
As seen in Fig.~3(a), the resulting phase difference, $\delta\theta = (\zeta_c+\zeta_s)\delta t$, shows a linear dependence on pump amplitude and passes through zero.
These more sensitive cross-Ramsey measurements were performed on both qubits for several $\delta t$ showing all curves passing through a single pump amplitude that ensures $(\zeta_c+\zeta_s)/2\pi=0\pm 5.5$~kHz. 
As a further verification, we performed randomized benchmarking\cite{Emerson2005,Knill2008} (RB) on both qubits simultaneously, as another sensitive test of residual $ZZ$-coupling.\cite{Houck_PRA}
As shown in Fig.~3(b), we find that the error per gate for simultaneous RB, where we measure the probability that we properly returned to $|gg\rangle$, is equal to the sum of the individual error rates found from single qubit RB, which indicates that additional correlated errors from residual $ZZ$ coupling is negligible.
%

Next, we show parametric interactions between the two qubits in the time domain, which leads us to performing specific gate operations.
By applying a coherent pump signal in the vicinity of $\Delta_{bt}$, parametric coupling induces transitions between the states $|b\rangle$ and $|t\rangle$ in a similar way to standard Rabi oscillations.\cite{ZakkaBajjani2011,Allman2014}
The probability of being in the initially empty state over time is given by,
\begin{equation}
 P(t) = \frac{4g_p^2}{\Omega^2}\sin^2\left(\frac{\Omega t}{2}\right),
 \label{Eq_3}
\end{equation}
where $\Omega = \sqrt{4g_p^2 + \Delta_p^2}$ is the parametric SWAP frequency when $\omega_p=\Delta_{bt}+\Delta_p$. 
When $\Delta_p=0$, the on-resonance (in the rotating frame) SWAP frequency $\Omega$ is $2g_p$, which is equal to the total mode splitting size seen in the spectroscopic data.
Also, notice that the probability amplitude of these oscillations reaches unity only when $\Delta_p=0$.
This will be important later when we discuss sources of gate errors.
An example of these parametric oscillations is shown in Fig.~4(a) and (b) between states $|b\rangle=|eg\rangle$ and $|t\rangle=|ge\rangle$ and $|b\rangle=|fg\rangle$ and $|t\rangle=|ee\rangle$, respectively. 
This data was taken with a similarly designed two-qubit device to that shown in Fig.~\ref{Fig_1}, but made from all aluminum on a high purity silicon substrate.
The characteristic shape due to Eq.~(\ref{Eq_3}) is most visible in Fig.~4(b). 
Line-cuts with a higher density of points were taken through the center of these data when $\Delta_p=0$.
Fig.~4(a) shows a maximum SWAP frequency of $\Omega/2\pi\approx 100$~MHz and Fig.~4(b) shows $\Omega/2\pi\approx 25$~MHz.
This provides the speed necessary to perform fast two-qubit iSWAP or c$Z$ gates in about 10 -- 40~ns.
Although we performed iSWAP gates utilizing states $|b\rangle=|eg\rangle$ and $|t\rangle=|ge\rangle$ with an unoptimized fidelity of 95\%, we choose to focus our attention on the c$Z$ gates described below.
%
    \begin{figure*}[t] 
    {\includegraphics[width=1\hsize]{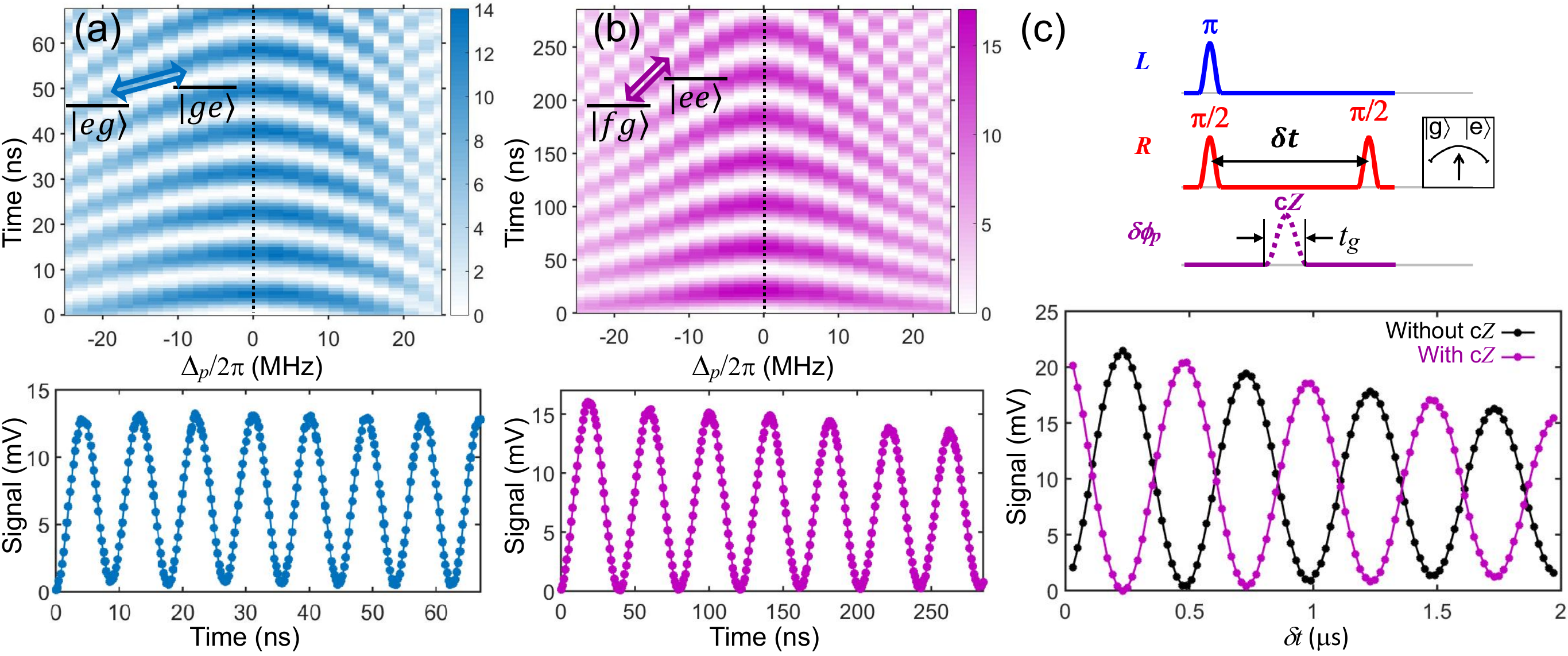}}
    \caption{{\textbf {Parametric SWAP oscillations and demonstration of a p-SWAP c$Z$ gate.}} (a) Top Panel: Parametric SWAPs between states $|eg\rangle$ and $|ge\rangle$ as a function of $\Delta_p$. Bottom panel: Rabi oscillations with a maximum rate of about 100~MHz when $\Delta_p=0$ (vertical dashed line in top panel). (b) Top Panel: Parametric SWAPs between states $|fg\rangle$ and $|ee\rangle$ as a function of $\Delta_p$. Bottom panel: Rabi oscillations with a maximum rate of about 25~MHz when $\Delta_p=0$ (vertical dashed line in top panel). (c) Pulse sequence for the cross-Ramsey experiment to verify operation of the p-SWAP c$Z$ gate. Adding a c$Z$ gate shifts the phase of the Ramsey oscillations by $\pi$ as expected.}
    \label{Fig_4} 
    \end{figure*}

Now, let us consider a SWAP-induced parametric controlled-Z (p-SWAP c$Z$) gate\cite{Strauch_PRL}, which can be realized via a complete rotation from the state $|ee\rangle$ to one of the other two-photon excited manifold of states $|fg\rangle$ or $|gf\rangle$ and back to $|ee\rangle$. 
Parametric two-qubit gates using similar principles have been realized with other systems\cite{Nakamura_2QGates_PRA, Zurich_CZ_2020, Rigetti2018}, but with either limited parametric coupling strength \cite{Zurich_CZ_2020, Rigetti2018} or reduced coherence times \cite{Nakamura_2QGates_PRA}.
As we will see, our parametric coupling strategy can take advantage of high gate speeds to achieve a large ratio between coherence time and gate duration to minimize gate errors in the presence of qubit decoherence.
We show in Fig.~4(c) a simple demonstration of a p-SWAP c$Z$ gate, again with the all aluminum device.
Here, we perform a cross-Ramsey experiment on the $R$ qubit with a $\pi$-pulse on the $L$ qubit with and without a c$Z$ gate applied half-way between the cross-Ramsey sequence. 
After accounting for rectification of the qubit frequencies, we observe a $\pi$ phase shift between the two possible Ramsey oscillations. 
Cross-Ramsey experiments are helpful in the calibration procedures for optimizing the c$Z$ gates because the rectification of the qubit frequencies is dependent on $\omega_p$ and $\delta\phi_p$ (see Supplemental Material for more details).
To quantify the p-SWAP c$Z$ gate fidelity, we perform standard interleaved RB repeatedly over many hours.\cite{Corcoles2013,barends2014superconducting}
In Fig.~5(a), we show the measured total error per gate $\varepsilon_t$ as a function of the gate duration $t_g$ (see pulse diagram in Fig.~4(c), as $t_g$ is defined by the edge-to-edge pulse width). 
First, we focus on pulse envelopes made from two fixed 20~ns Hann function rise and fall pulse edges plus a variable length plateau, where each RB data point was averaged for about 9~hours.
For each gate duration, we have performed a suite of calibrations and use RB\cite{Kelly2014} plus a covariance matrix adaptation evolution strategy (CMA-ES) for numerical optimization\cite{Zurich_leakage_2020} of the gate performance (see Supplemental Material for more details).
In general, the total error per gate $\varepsilon_t = \varepsilon_{\rm dec} + \varepsilon_{\rm con}$ can be broken into decoherence-induced errors $\varepsilon_{\rm dec}$ and any control errors $\varepsilon_{\rm con}$, which could include noise in the control electronics or coherent evolutions from unintended Hamiltonian dynamics. 
We observe that when the gate duration is long ($t_g>150$~ns), the error rate is dominated by the individual qubit coherences, where $\varepsilon_{\rm dec}=(2/5)(\Gamma_{1,{\rm eff}}+2\Gamma_{2,{\rm eff}})t_g$, with $\Gamma_{1,{\rm eff}}=(1/2)(1/T_{1,L}+1/T_{1,R})$ and $\Gamma_{2,{\rm eff}}=(1/2)(1/T_{2,L}+1/T_{2,R})$.\cite{abad2022,marxer2023long} 
The dashed lines in Fig.~5(a) show the prediction when $T_{1,{\rm eff}}=1/\Gamma_{1,{\rm eff}}=16.3\,\mu$s and $T_{2,{\rm eff}}=1/\Gamma_{2,{\rm eff}}=22.7\,\mu$s, consistent with measured qubit coherences (see Supplemental Material for more details).
We also plot the individual contributions from $\Gamma_{1,{\rm eff}}$ (dotted line) and $\Gamma_{2,{\rm eff}}$ (dash-dot line), which shows the comparative roles energy relaxation and phase coherence play in determining gate errors.
This suggests that as $t_g$ decreases with respect to decoherence rates, the gate error should decrease steadily towards zero.
However, our experimental results show that as $t_g$ decreases, the observed $\varepsilon_t$ deviates from the decoherence line when $\varepsilon_{\rm con}$ increases. 
As a result, $\varepsilon_t$ reaches a minimum value at $t_g=70$~ns where we achieve our best p-SWAP c$Z$ gate fidelity of $99.44\pm 0.09$\%, as shown in Fig.~5(b).
If we compare this to the fidelity of an idle gate in place of the p-SWAP c$Z$ gate, we find a much improved value of $99.88$\%, which suggests that $\varepsilon_{\rm con} = 1.2\times 10^{-3}<\varepsilon_{\rm dec}$. 
This shows that simply idling during the gate in the presence of decoherence produces $\varepsilon_{\rm dec} = 4.4\times 10^{-3}$, close the theoretical decoherence limit line when $t_g=70$~ns ($\varepsilon_{\rm dec} = 4.2\times 10^{-3}$). 
%
%
As gate durations decrease further, $\varepsilon_{\rm con}>\varepsilon_{\rm dec}$ dominating the total error.
%

In order to investigate sources for $\varepsilon_{\rm con}$, we carried out a series of gate fidelity measurements with variations in the shape of the gate's pump pulse envelope.
Let us first consider that when $t_g=80$~ns, the pulse envelope is made from equal durations of Hann function pulse edges and a plateau. 
As the pulse width decreases further towards $t_g=40$~ns, the spectrum of these pulses in frequency space becomes wider, following more closely the Fourier transform of a pure Hann pulse, which is a broadened version of a rectangular window function.
Thus, if a narrowing spectral width (decreasing $t_g$) leads to leakage errors as a limiting factor (increasing $\varepsilon_{\rm con}$), then the pure Hann pulses would improve our gate fidelity.
However, experiments show the opposite.
First, we independently verified that leakage to higher transmon levels was not a significant source of error.\cite{barends2014superconducting}
Next, we carried out c$Z$ gate experiments where the pump pulse envelope was a pure Hann function of different widths, as shown in Fig.~5(a).
Notice that this group of data has a total error per gate systematically higher than the experiments with plateaued pulses for similar qubit coherence.
%
    \begin{figure*}[t] 
    {\includegraphics[width=1\hsize]{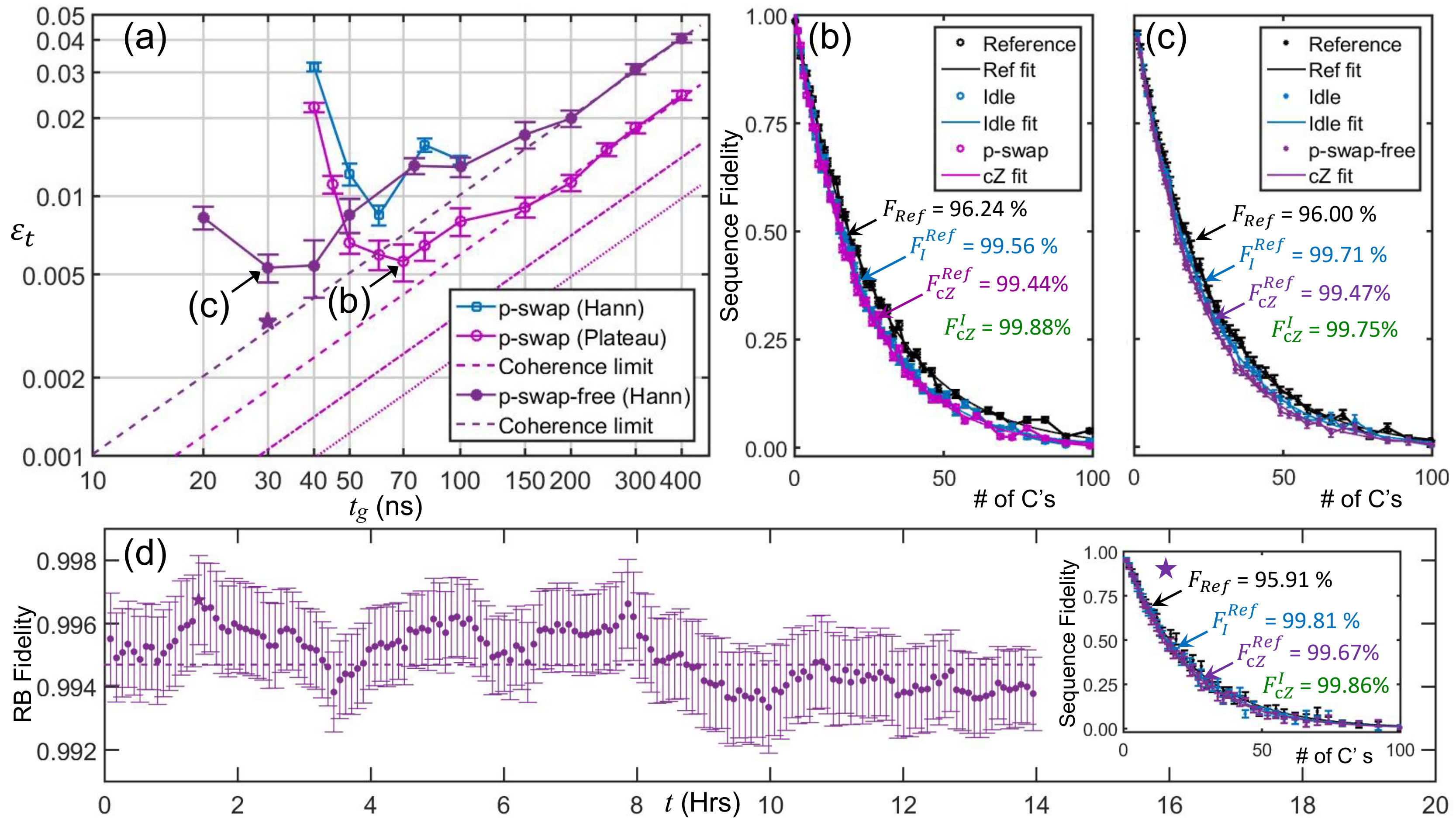}}
    \caption{{\textbf {Demonstration of parametric controlled-Z gates.}} (a) RB for various parametric c$Z$ gates as a function of the gate duration $t_g$. Dashed lines represent coherence limits. The individual contributions from decoherence are shown for energy relaxation ($\Gamma_{1,{\rm eff}}$, dotted line) and phase coherence ($\Gamma_{2,{\rm eff}}$, dash-dot line) for the plateau pulses. $\star$ shows peak fidelity, see (d). (b) RB showing the best result for the p-SWAP c$Z$ gate with fidelity $99.44\pm 0.09$\% for $t_g=70$~ns when averaged over 9~hours. (c) RB showing the result for the p-SWAP-free c$Z$ gate with fidelity $99.47\pm 0.07$\% for $t_g=30$~ns when averaged over about 18~hours. (d) A running average of the p-SWAP-free c$Z$ gate of duration 30~ns over 14~hours with a peak fidelity of $99.67\pm 0.14$\% ($\star$). Dashed horizontal line shows average value (see (c)). Inset shows RB results for the peak fidelity ($\star$). All thin solid lines are RB fits to the data (see main text). The axis label ``\# of C’s” is shorthand for ``number of Clifford gates”.}
    \label{Fig_5} 
    \end{figure*}

%
We can understand the source of these errors by returning to the concept of how this gate functions.
Ideally, starting from the $|ee\rangle$ state, the parametric pump pulse operates at exactly $\omega_p=\Delta_{bt}=\omega_{ee}-\omega_{fg}$, so that the state evolution under the Hamiltonian (from Eq.~(\ref{Eq_2})) occurs at the center of the Rabi curve (from Eq.~(\ref{Eq_3}) shown in Fig.~4(a)) returning all of the state population back to $|ee\rangle$ with a final phase accumulation of exactly $\pi$.
In reality, because the frequencies of both qubits, their anharmonicities, and $\zeta_s$ all experience a slight adjustment due to the rectification effect that depends on pump amplitude $\delta\phi_p$ and frequency $\omega_p$, it is difficult to maintain a perfect evolution of the states and cancellation of $\zeta_s$ (see Supplemental Material for more details).
For our experiments, we choose a fixed cancellation tone for $\zeta_c$ to ensure proper single qubit gates in the absence of any gate tones.
For the gate tone, we use a constant frequency $\omega_p=\tilde{\omega}_{ee}-\tilde{\omega}_{fg}$ set according to the rectified values of the two-qubit states at the maximum pump amplitude, as measured during parametric Rabi oscillations as we sweep $\Delta_p$ (see for example Fig.~4 and the Supplemental Material). 
Thus, during the rise and fall of the Hann pulse envelope, the frequency $\omega_p$ is slightly detuned from optimal.  
For low amplitude pump pulses mostly comprised of a pulse plateau (larger $t_g$), the evolution of the gate matches closely with the idealized case and the residual $ZZ$ coupling is still negligible.
On the contrary, for pure Hann pulses, the pump frequency only matches at the maximum amplitude of the pulse, modifying the Hamiltonian evolution of the gate (recall Eq.~(\ref{Eq_3})).  
In addition, as the pulse amplitude increases for shorter $t_g$, the static $\zeta_c$ tone is not completely cancelling $\zeta_s$ during the gate, although this effect is compensated for by the reduced gate duration over which it occurs.
These problems could be properly addressed by varying the drive frequency $\omega_p(t)$ and cancellation tone $\zeta_c(t)$ over the duration of the gate, which we plan to explore in future experiments.

In order to remedy these issues, we next turn to a parametric c$Z$ gate that is SWAP-free.
A nice feature of the p-SWAP-free c$Z$ gate is that there is no requirement to optimize a single parametric drive frequency, making this gate more resilient to rectification-induced errors.
Recall in Fig.~2(e), we could cancel $\zeta_s$ or choose to drive harder so that $\zeta_c+\zeta_s>0$, producing strong positive $ZZ$ coupling. 
This is a much more natural way to realize a c$Z$ gate\cite{stehlik2021tunable}, where a net frequency shift $\zeta_{cZ}(t)$ of the state $|ee\rangle$ over time $t_g$ can accumulate the necessary conditional phase,
\begin{equation}
 \pi = \int_0^{t_g}\zeta_{cZ}(t)dt.
 \label{Eq_4}
\end{equation}
To perform the gate, we add an additional tone $\zeta_g(t)$ along with the constant cancellation tone $\zeta_c$, so $\zeta_{cZ}(t)=\zeta_g(t)+\zeta_c+\zeta_s$.
Any slight deviations over time in $\Delta_{bt}$ during the gate pulse due to rectification may lead to distortions so that $\zeta_{cZ}(t)$ is not optimal, but we only require the integral in Eq.~(\ref{Eq_4}) to be satisfied. 
As a first approximation, this can be compensated by a small increase or decrease in the gate duration.
This means that we can use pure Hann pulses without any plateau and not suffer from additional errors from pump-induced rectification. 
After performing a suite of calibrations and using CMA-ES (see Supplemental Material for more details), we measured $\varepsilon_t$ for various $t_g$, in a similar way as for the p-SWAP c$Z$ gate.
We find that for long gate durations, the p-SWAP-free c$Z$ gate is limited by the combined decoherence of the two qubits, with $T_{1,{\rm eff}}=11.6\,\mu$s and $T_{2,{\rm eff}}=12.0\,\mu$s for this particular cool-down (see dashed curve in Fig.~\ref{Fig_5}(a) and Supplemental Material for more details).
Again, as $t_g$ is reduced the error rate decreases following the decoherence prediction and then begins to deviate.
The minimum gate error is observed to be $99.47\pm 0.07$\% with a pure Hann pulse with total duration (rise and fall) of 30~ns, as is shown in Fig.~5(c).
Again, we can also compare this to the fidelity of an idle gate in place of the p-SWAP-free c$Z$ gate, which improves to $99.76$\%, showing that $\varepsilon_{\rm con} = 2.4\times 10^{-3}<\varepsilon_{\rm dec}$.
Here, we find $\varepsilon_{\rm dec} = 2.9\times 10^{-3}$ is close to the theoretical decoherence limit line when $t_g=30$~ns ($\varepsilon_{\rm dec} = 3.0\times 10^{-3}$). 
In Fig.~\ref{Fig_5}(d), we plot the p-SWAP-free gate fidelity over time with each point representing a 4~hour moving average of a roughly 18~hour long repeating dataset.
We see that the maximum gate fidelity peaks at $99.67\pm 0.14$\% with respect to the reference (see $\star$ and inset).
At this optimal point, with respect to the idle, the fidelity is $99.86$\% so that $\varepsilon_{\rm con} = 1.4\times 10^{-3}$ and $\varepsilon_{\rm dec} = 1.9\times 10^{-3}$, better than the theoretical decoherence limit line when $t_g=30$~ns ($\varepsilon_{\rm dec} = 3.0\times 10^{-3}$).
Recall that this limit is assessed as an average over long periods of time, showing that coherence fluctuations also produce variations in gate fidelity over time, most likely due to two-level system defects (see Supplemental Material for more details). 
That said, the p-SWAP-free gate fidelity peaked twice and remained almost entirely above the mean value for over 8~hours (see dashed line and Fig.~\ref{Fig_5}(c)).
%
%

%
For both the p-SWAP and p-SWAP-free c$Z$ gates, the gate error rises rapidly for the shortest gate durations.
To investigate this we performed some high amplitude pump drive tests, where the initial two-qubit state was $|gg\rangle$ (see Supplemental Material for more details).
We found that it is possible for the parametric pump to drive multi-photon induced single qubit transitions when $\omega_p\approx\omega_k/n$, where $n$ is the number of photons absorbed.
For the highest pump amplitudes, if $\omega_p$ was in the vicinity of these frequencies, the pump can cause unwanted single qubit Rabi rotations.
In principle, these coherent single qubit errors could be nulled out with additional deterministic single qubit drives that could counteract these stray rotations.\cite{Sung2021,Ding2023}
In future designs, we plan to significantly decrease the capacitive coupling between the bias coil and the transmons in order to reduce any direct driving and resultant multi-photon absorption.\cite{Li2023a}

%
To conclude, we have demonstrated a versatile, hardware-efficient, fast, and high-performance parametric coupler between two mildly tunable transmon qubits.
We are able to eliminate stray $ZZ$ coupling between the two qubits and performed three types of two-qubit gates: p-iSWAP, p-SWAP c$Z$, and p-SWAP-free c$Z$.
We measured an average gate error over many hours as a function of gate duration and found maximum gate fidelities above $99$\% for all parametric gates.
%
The p-SWAP-free c$Z$ gate reached a peak fidelity of $99.67\pm 0.14$\% for a gate duration of 30 ns and remained above $99.47$ for over 8~hours.
For longer gate durations above 150~ns, the gate error is almost entirely limited by qubit decoherence, while the shortest gates are limited mostly by erroneous Hamiltonian evolutions.
Exploring different pulse shapes indicates that a significant portion of the control errors in the p-SWAP c$Z$ gate can be ascribed to a rectification effect, which can be avoided when performing p-SWAP-free c$Z$ gates.
Our results suggest that modest improvements in qubit coherence in next generation devices should lead to parametric two-qubit gates with fidelities above 99.9\%. 
In addition, future circuit designs with transmons should reduce residual static $ZZ$ coupling to $<300$~kHz, whereas introducing fluxonium qubits provides exact cancellation of all residual static $ZZ$ coupling. 
These concepts can be scaled into integrated multi-qubit cavity QED systems to provide in-situ parity readout useful for error correction.\cite{Bravyi2024,Noh2023}
%
 
\section*{Acknowledgements}
This work was partially performed under financial assistance from U.S. Department of Commerce, National Institute of Standards and Technology. Eliot Kapit's work was supported by NSF grant PHY-1653820 and ARO grant No. W911NF-18-1-0125. We thank Frederick W. Strauch for useful discussions and suggestions for improving gate fidelity in the experiments. We thank Manuel C. Castellanos-Beltran and Joshua Combes for commenting on the manuscript.

\section*{Author contributions}
X.Y.J. conducted the experiment. X.Y.J., Z.P. and R.W.S. analyzed and modelled the data. R.W.S. designed and K.C. fabricated the device. S.K. contributed to the measurement set-up and software. E.K. provided theoretical support and suggestions for the experiment. X.Y.J. and R.W.S. wrote the manuscript and supplemental material. R.W.S. conceived the experiment and supervised the project. All authors contributed to the preparation of the manuscript.

\section*{Methods}
The samples were fabricated on 76~mm sapphire wafers. The main wiring layer is formed by a fluorine-based reactive ion, dry gas etch of a 100~nm thick sputtered niobium film. Next, Josephson junctions are fabricated using the well known ``Dolan bridge technique'' that relies on a double-angle ($\pm 14.3^o$) evaporation of aluminum using a bi-layer resist stack consisting of MMA/PMMA (550~nm/240~nm), with an intermediate step for oxidization (100~mTorr for 300 seconds) of the first layer of aluminum (35~nm). The second deposition provides the top aluminum layer (75~nm) and forms a tunnel barrier across a narrow overlap region forming the Josephson junctions (qubit junction areas were approximately 100~nm~$\times$~120~nm, while SQUID junctions were approximately $1.8~\mu\mathrm{m}\times 0.1~\mu\mathrm{m}$). This was achieved in a custom deposition chamber with a load-lock that allows computer controlled oxidization for creating reproducible junction current densities. The final step requires connecting the Josephson junctions to the niobium wiring layer through an aluminum patch layer that is e-beam evaporated in the same chamber that was used for making junctions and then lifted off. To ensure superconducting contacts between all layers, we use an ion mill cleaning step (beam = 300 V, accelerator = 950 V, filament = 3.47 A) for 30 seconds before depositing aluminum. 

Once diced, the chips are mounted into a sample box made of aluminum and bonded directly to non-magnetic microwave SMA bulkhead feedthrus for wiring connections, with additional bond wires between the ground plane on the chip and the inside of the box. This aluminum box is then placed inside another copper box of cylindrical shape. A niobium tube is then placed over this cylinder and is thermalized by a copper tube that covers it. A $\mu$-metal tube is then placed over this cylinder with another copper tube over that for thermalization. Finally, a copper can lined with black carbonized foam encloses all these layers. The coaxial cables are routed out of the ends of the cylinders and into non-magnetic SMA feedthrus that pass through the copper can. The entire shielded enclosure is mounted to the bottom of a dilution refrigerator and cooled to 10~mK. For more details on the experimental setup, see Supplementary Section I.

\newpage 
\section{Supplementary Information: \\Fast, tunable, high fidelity cZ-gates between superconducting qubits with parametric microwave control of ZZ-coupling}

\section{The experimental setup}

\begin{figure}[h!]
\centering
\includegraphics[width=5in]{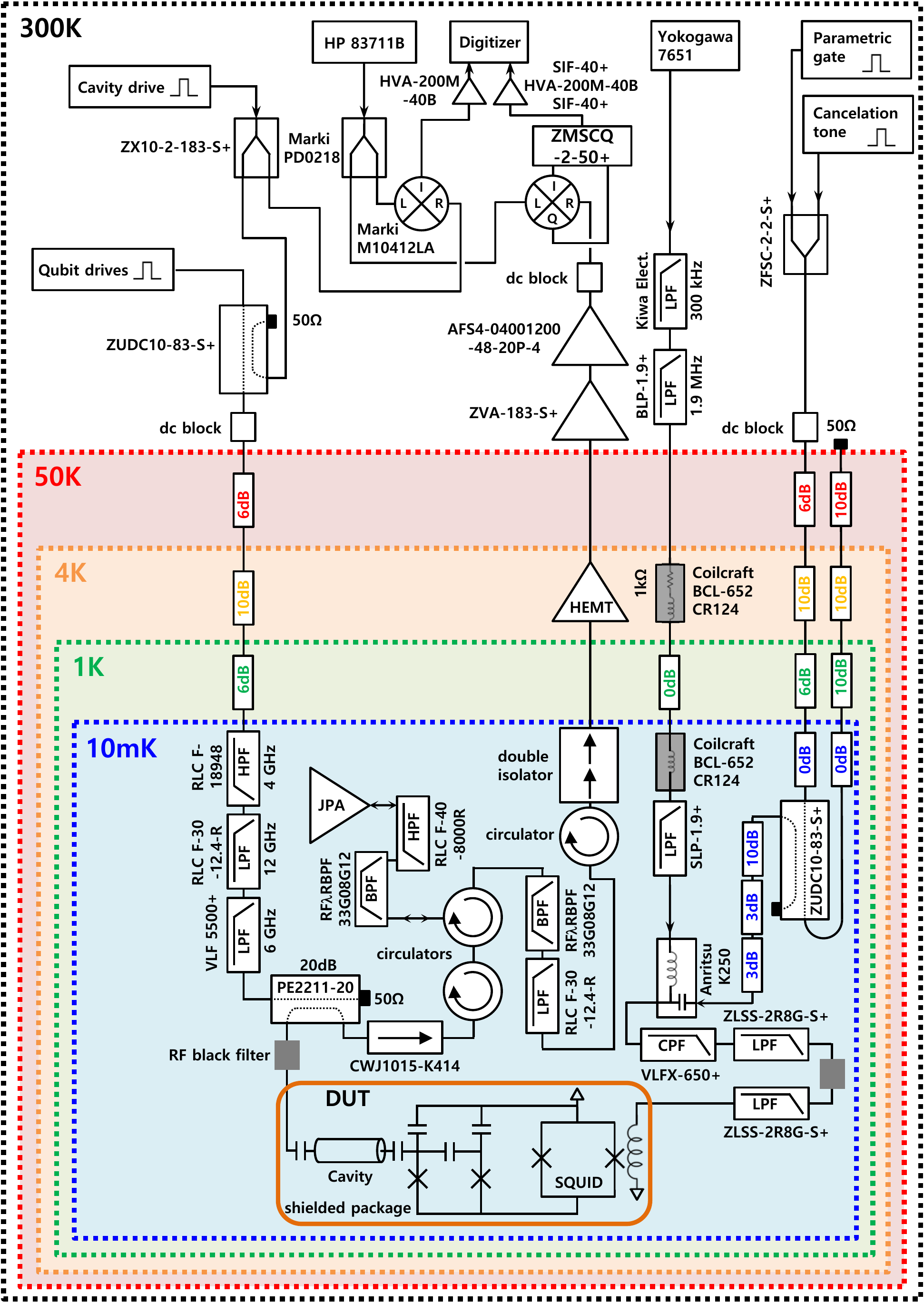}
\caption{\textbf{Schematic of the experimental setup.}}
\label{sFig1}
\end{figure}

\begin{figure}[h!]
\centering
\includegraphics[width=5in]{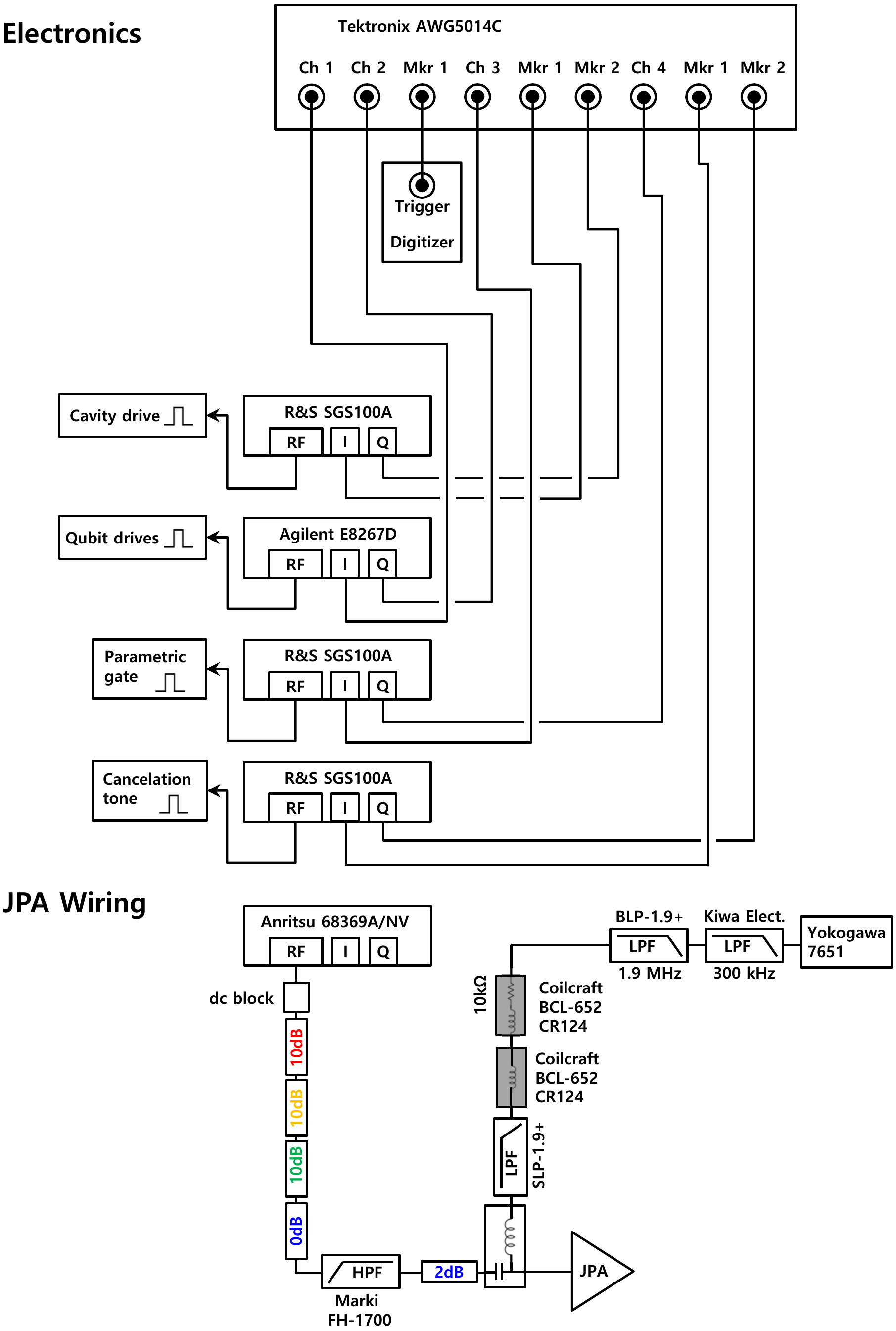}
\caption{\textbf{Wiring diagrams.}}
\label{sFig2}
\end{figure}

\begin{figure}[h!]
\centering
\includegraphics[width=5in]{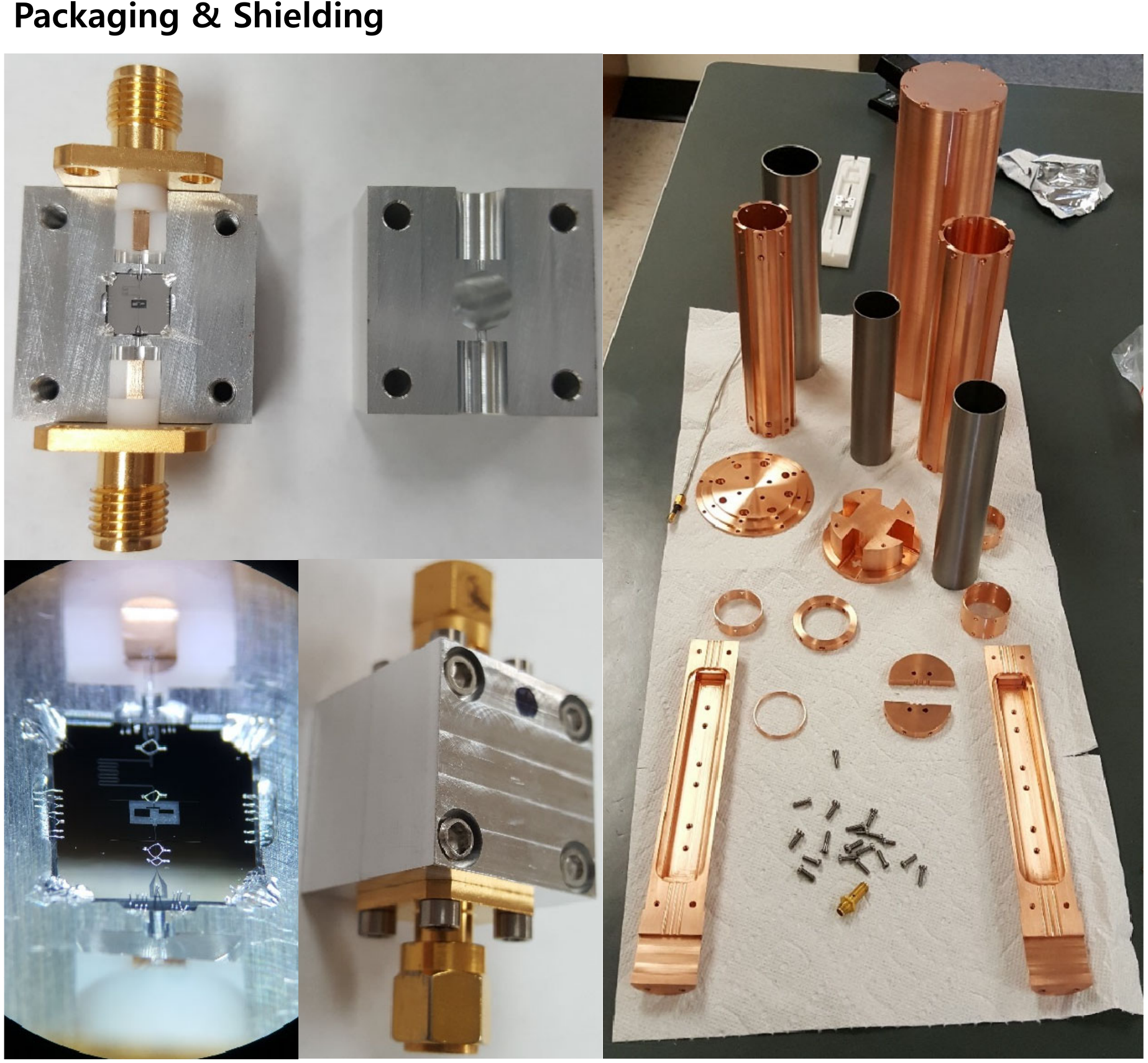}
\caption{\textbf{Packaging and Shielding.}}
\label{sFig3}
\end{figure}

\begin{figure}[t!]
\centering
\includegraphics[width=4in]{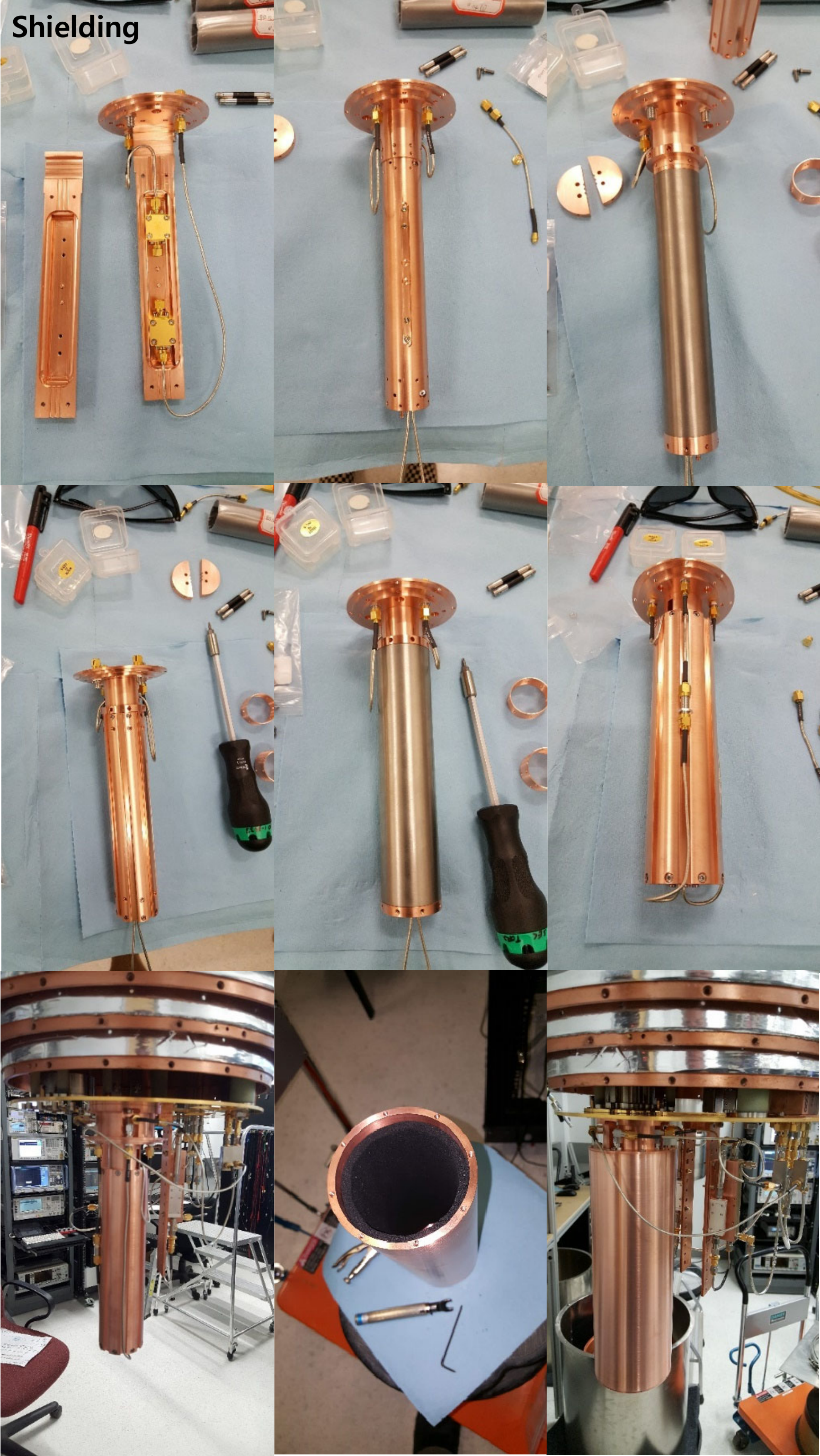}
\caption{\textbf{Step-by-step shielding assembly.}}
\label{sFig4}
\end{figure}

A detailed schematic of the instrumentation and cryogenic setup can be seen in Fig.~\ref{sFig1}. The device under test (DUT) is connected to the experimental setup through two SMA ports: a cavity port that applies both cavity and qubit drive tones and a flux bias port that provides both a dc-flux bias and an rf-flux bias for adding parametric drive tones. All control and readout microwave tones were independently pulsed on two separate coaxial ports. The cavity and qubit drive signals were combined at room temperature and sent to the cavity, while the parametric microwave tones were sent through a bias-T to the pump port (see also Fig.~\ref{sFig2}). These separate tones also required a pulse timing calibrating so that the tones could be properly aligned in-time. The cavity pulses are constructed from one microwave source gated by a main arbitrary waveform generator (AWG), shown in Fig.~\ref{sFig2}. The cavity signal is attenuated and filtered as it is sent to the DUT and is then recovered after passing through cryogenic circulators at the mK-stage and then a number of amplification stages (a HEMT at 4~K and following amplifiers at room temperature). The original cavity tone, as well as the response from the cavity, are mixed with a detuned local oscillator tone so that the heterodyne signal can be amplified (HVA-200M-40B preamplifier) and digitized (Gage) for analysis. The qubit drive tones are generated by independent synthesized rf sources that are gated in time with the main AWG. The dc-flux bias is provided by a low noise Yokogawa 5760 and passes through multiple filter sections on the way to the DUT, passing through an Anritsu K250 bias-T at the mK-stage. The rf-flux bias tone is generated in a similar fashion as the qubit pulses, although at lower frequency. These parametric pulse tones are attenuated and pass through their own appropriate filter configuration on their way to the DUT, also passing through the Anritsu bias-T. Filters on the rf flux line are configured to create a pass band between approximately 0 - 1 GHz, which allows parametric modulation at the difference frequency between the qubits and the cavity, while fully isolating the cavity and the qubits at their resonance frequencies.

Our sample box is machined from high purity aluminum in a 3D-cavity like fashion, where the chip is secured by indium squished into the corners of a mounting ledge in the center of a cylindrical cavity with a lowest fundamental near 30~GHz, shown in Fig.~\ref{sFig2}. The bond wire connections to the bias and cavity leads on-chip pass through small cylindrical tunnels that lead directly to SMA connectors, without any additional circuit boards. This box is then mounted inside a cylindrical copper box that is then sequentially enclosed by alternating layers of magnetic shielding and copper tubes to increase thermalization, as shown in Fig.~\ref{sFig3} and Fig.~\ref{sFig4}. The first shield tube is made from pure niobium and the second is cryoperm. The outermost copper shield can was lined with black carbonized foam. All of this was mounted to the bottom of a dilution refrigerator that had a mixing chamber shield that was tin-plated and acted as an additional layer of thermal and magnetic shielding. 

\section{Qubit rectification and calibration of the parametric pump drive}
\label{calib}
The data shown in Fig.~2(c) in the main text requires measuring the qubit spectrum while sweeping the frequency $\omega_p$ of the parametric pump. During these measurements with a parametric drive, it is crucial to maintain a constant pump amplitude for all pump frequencies. In this section, we discuss a systematic calibration procedure to compensate for the standing waves in the pump line that produce an oscillatory dependence of the pump amplitude with frequency (see Fig.~\ref{sFig6}) and to extract the pump's flux modulation amplitude $\delta\phi_p$ at the device. In principle, the pump calibration can be performed at any static flux bias $\phi_s$, however it is most convenient to perform this calibration at $\phi_s=0$, where rectification of the qubit frequencies is a maximum and first order parametric interactions are suppressed.

\begin{figure}[b!]
\centering
\includegraphics[width=\textwidth]{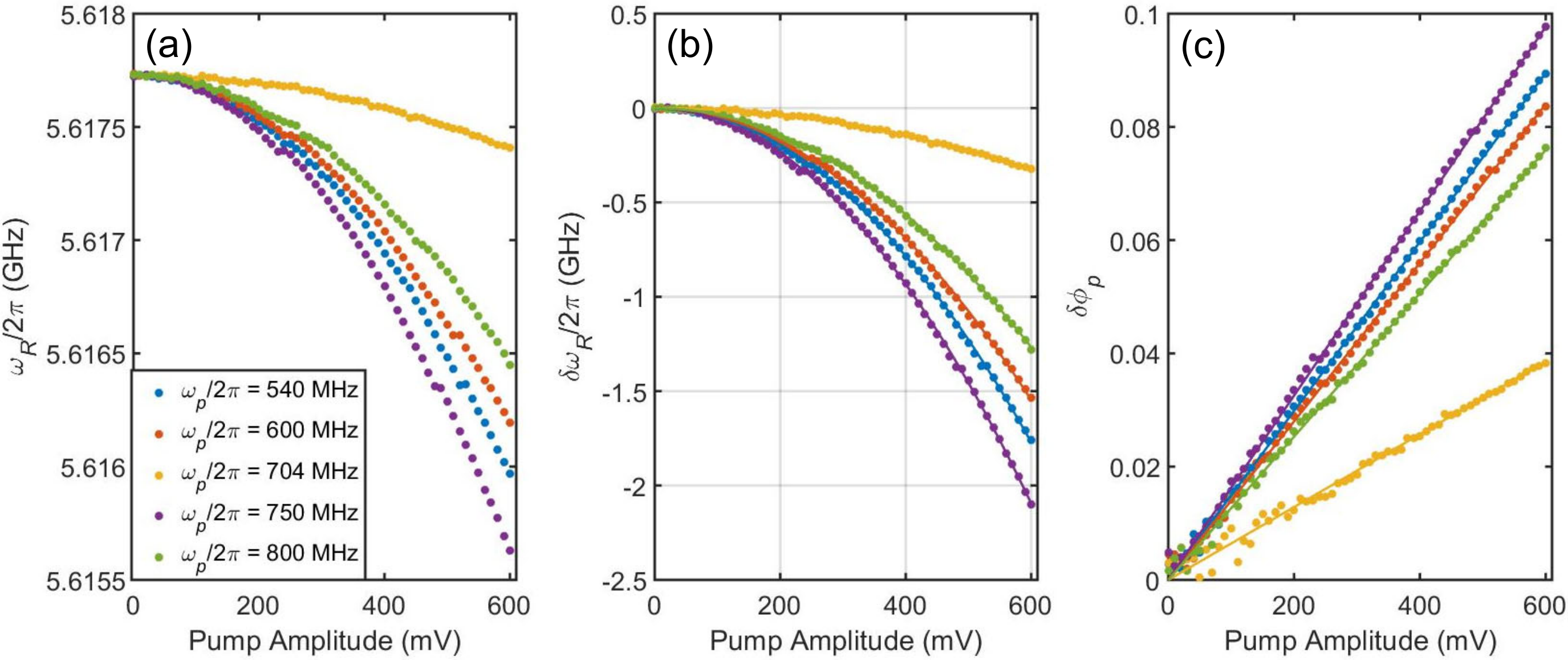}
\caption{\textbf{Rectification of the resonant frequency of the $R$ transmon.} (a) Examples showing the rectification of $\omega_R$ (the $R$ transmon) as a function of signal generator pump amplitude for five pump frequencies ($i=1,2,\dots 5$ for $\omega_p/2\pi =$ 540, 600, 704, 750, \& 800~MHz). (b) Extracted $\delta\omega_R$ as a function of signal generator pump amplitude for the five pump frequencies. The solid lines represent a quadratic fit to the data. (c) The calibrated values for $\delta\phi_p$ as a function of pump amplitude for each pump frequency. The solid lines are linear fits to the data with slopes $S_i$ and intercept 0.}
\label{sFig5}
\end{figure}

\begin{figure}[b!]
\centering
\includegraphics[width=5in]{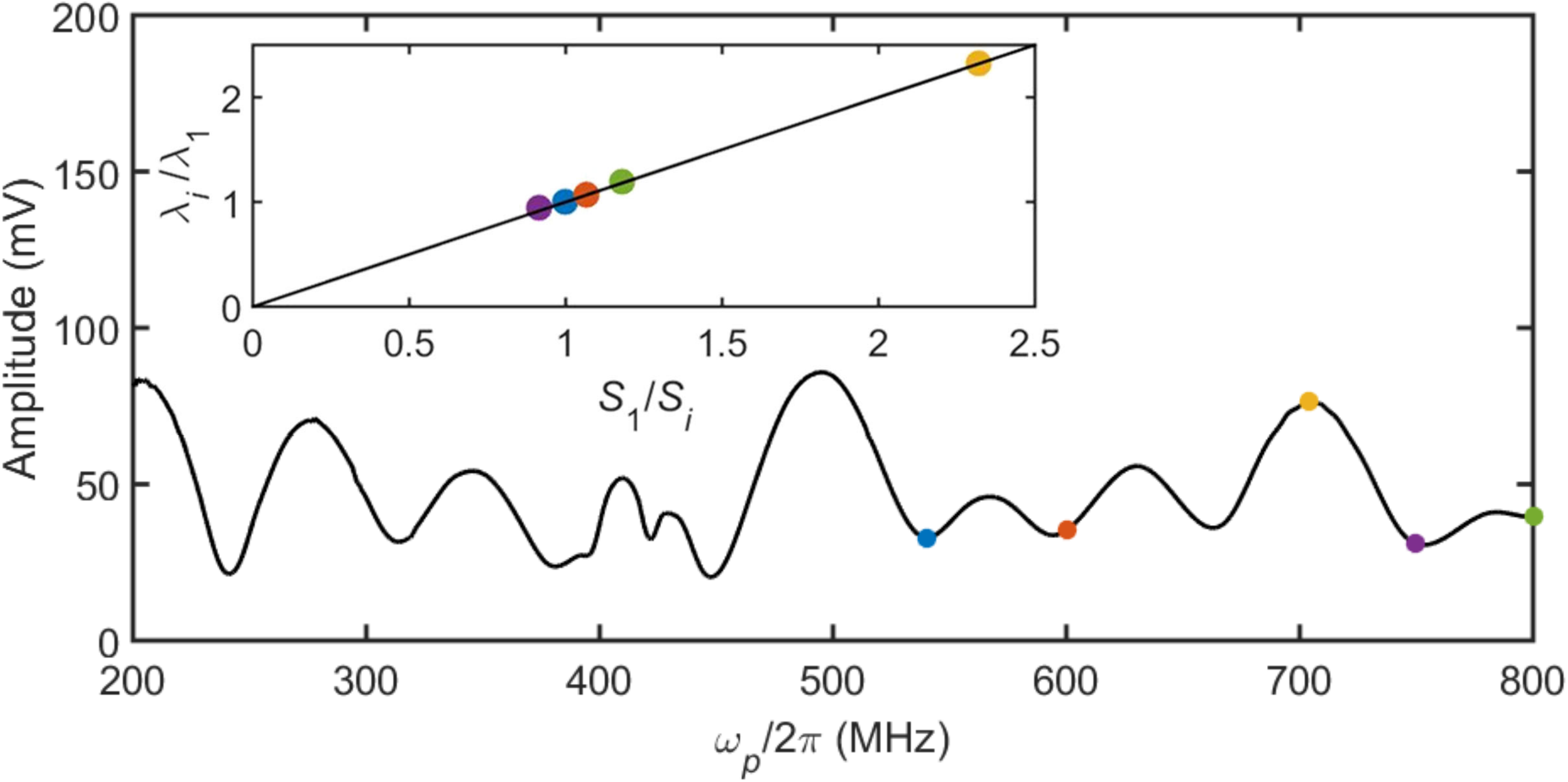}
\caption{\textbf{Signal generator amplitude as a function of frequency after pump calibration.} The signal generator's amplitude is found by multiplying the target constant amplitude by the calibration factor $\lambda(\omega_p)$. The five pump frequencies ($i=1,2,\dots 5$ for $\omega_p/2\pi =$ 540, 600, 704, 750, \& 800~MHz) where rectification was measured directly in Fig.~\ref{sFig5} are identified with corresponding colored dots at each $\lambda_iV_n$. The inset shows the ratio $\lambda_i/\lambda_1$ as a function of the inverse normalized slopes $S_1/S_i$, for the five pump frequencies. The solid line has a slope of 1 and intercept 0.}
\label{sFig6}
\end{figure}

In the presence of a parametric pump, the resonant frequency of the transmon $\omega_k(\phi_s)$ ($k\in\{L,R\}$) at a given static flux bias $\phi_s = \Phi_s/\Phi_0$ is shifted or rectified to a new, average value $\overline{\omega_k (\phi_s)}$ due to any curvature in its flux dependence. If the flux applied to the SQUID loop is modulated such that $\phi(t)=\delta\phi_p\cos(\omega_p t)+\phi_s$, then $\omega_k(\phi(t))$ can be averaged over one period $\tau_p=2\pi/\omega_p$ through a Taylor expansion giving
\begin{eqnarray}
\overline{\omega_k (\phi_s)} &=& \frac{1}{\tau_p}\int^{\tau_p}_{0} \omega_k (\phi(t)) dt \\ \nonumber
&=& \frac{1}{\tau_p}\int^{\tau_p}_{0} [\omega_k(\phi_s) + \omega_k'(\phi_s)\delta\phi_p\cos(w_p t) + 
\frac{1}{2}\omega_k''(\phi_s)\delta\phi_p^2\cos^2(w_p t) + \cdots]dt \\ \nonumber
&=& \omega_k(\phi_s) + \frac{1}{4}\omega_k''(\phi_s)\delta\phi_p^2 + O(\delta\phi_p^4) \cdots
\end{eqnarray}
where $\omega_k''(\phi_s) = d^2\omega_k/d\phi^2$, evaluated at $\phi=\phi_s$. From this expression, it is clear that the rectified shift of the qubit frequency $\delta\omega_k = \overline{\omega_k (\phi_s)}-\omega_k(\phi_s)\approx\omega_k''(\phi_s)\delta\phi_p^2/4$ is dominated by contributions coming from the second derivative of $\omega_k(\phi)$ at $\phi_s$ or the curvature of the qubit's modulation with flux. We can use the rectification of (either) qubit frequency $\delta\omega_k$ to quantify $\delta\phi_p$ as a function of the amplitude and frequency settings on the microwave signal generator for the pump,
\begin{equation}
\delta\phi_p\approx 2\sqrt{\frac{\delta\omega_k(\phi_s)}{\omega_k''(\phi_s)}}
\label{sEq}
\end{equation}
A clear example of the rectification of the $R$ transmon with signal generator pump drive amplitude is shown in Fig.~\ref{sFig5} for five color-coded pump frequencies (540, 600, 704, 750, \& 800~MHz) with $\phi_s=0$. Fig.~\ref{sFig5}(a),(b) shows that the resonant frequency of the $R$ transmon is quadratically dependent on the amplitude of the parametric pump with differing strengths depending on the pump frequency. As shown in Fig.~\ref{sFig5}(c), using $\delta\omega_R$ and $\omega_R''(0)$ from fits to the qubit spectrum's flux modulation curve (see Fig.~1 in the main text), we can use Eq.~(\ref{sEq}) to calibrate $\delta\phi_p$ as a function of the pump amplitude for each pump frequency.

\begin{figure}[b!]
\centering
\includegraphics[width=5in]{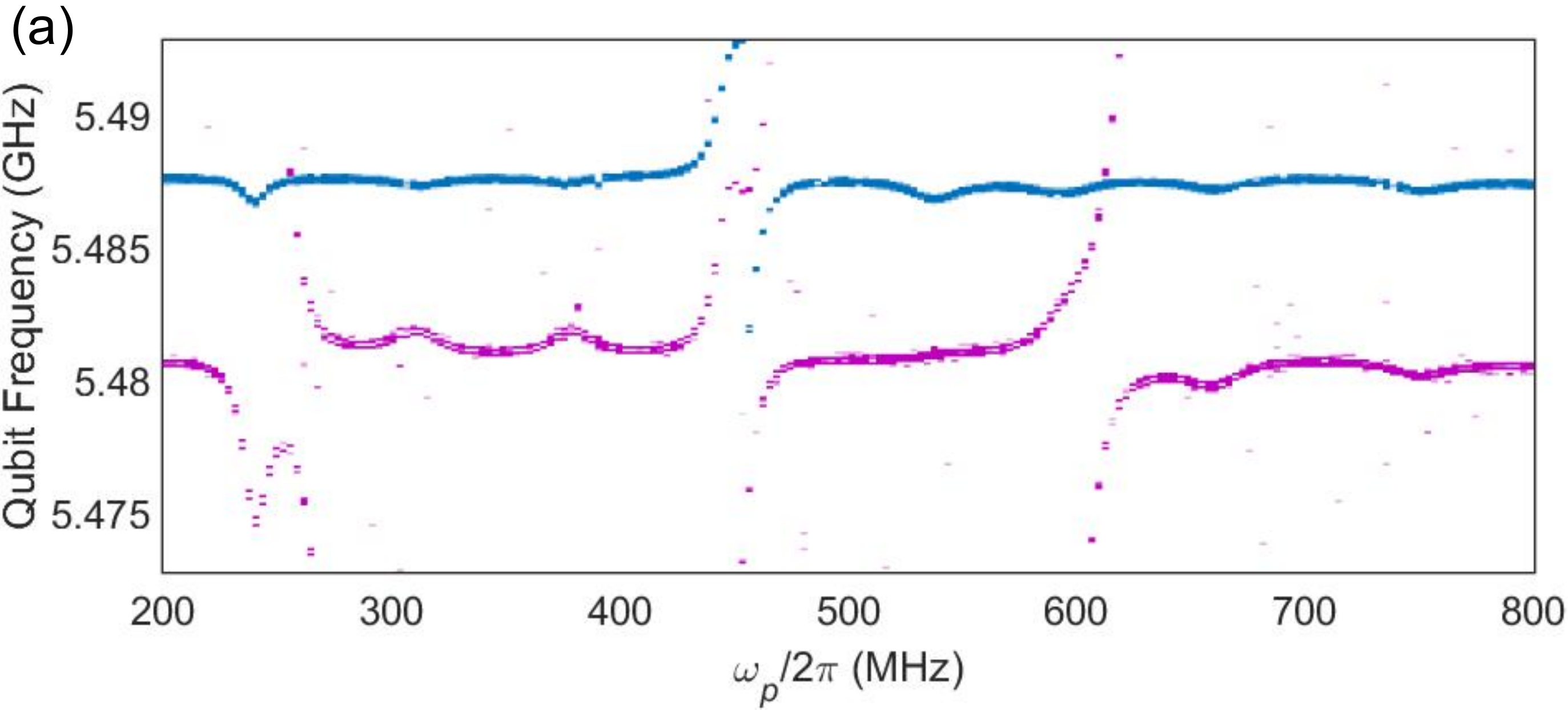}
\includegraphics[width=5in]{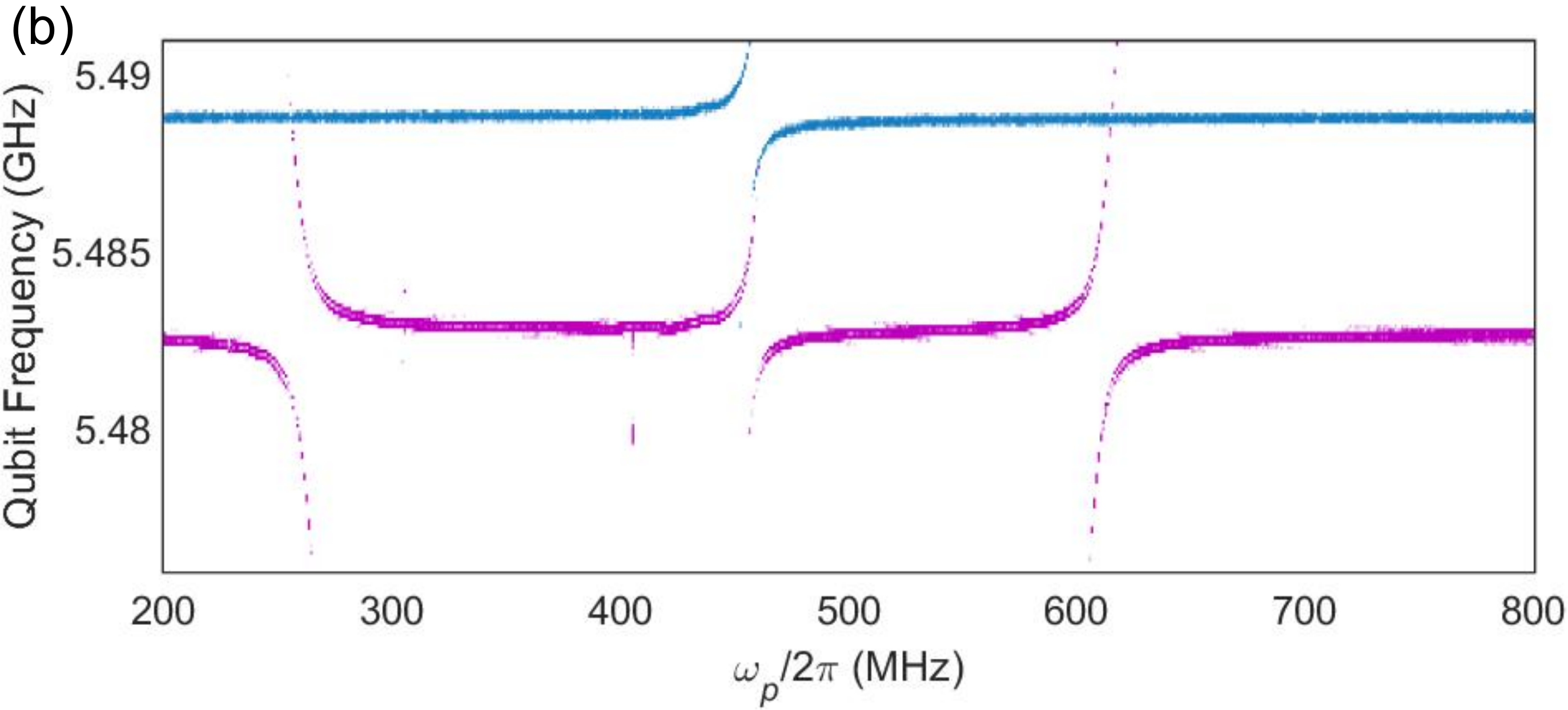}
\caption{\textbf{Before and after pump calibration.} (a) Without compensation. (b) With compensation (see Fig.~2(c) of the main text).}
\label{sFig7}
\end{figure}

To perform the full calibration and compensation, we measure the resonant frequency of the $R$ transmon while sweeping the pump frequency between 200~MHz and 800~MHz at a fixed amplitude setting for the microwave signal generator for the pump. We choose to perform this calibration at the symmetric point $\phi_s = 0$, where $\omega_R'(0)=0$ (higher order, odd derivatives also vanish) in order to suppress strong parametric couplings between the transmons, which is known (to first-order) to be proportional to $\sqrt{\omega_R'(\phi)}$ (see section \ref{Basic_model} below). In this way, any shifts $\delta\omega_R$ of the qubit frequency should be due primarily to rectification and proportional to $\delta\phi_p^2$ (as we saw in Fig.~\ref{sFig5}). In order to maintain a constant $\delta\phi_p$ at the SQUID loop of the device for each $\omega_p$, the amplitude setting on the microwave signal generator must vary to compensate for the frequency dependent attenuation of the entire pump line. We use qubit rectification data across this full frequency range at constant pump generator amplitude (not shown) to calculate a multiplication factor as a function of pump frequency, 
\begin{equation}
\lambda(\omega_p) = \frac{\lambda_0}{\sqrt{|\delta\omega_R(\omega_p)|}}    
\end{equation}
where the normalization constant $\lambda_0$ is chosen so that $\lambda=1$ at a given pump frequency. 

In Fig.~\ref{sFig6}, we show an example of this calibration result when measuring the spectrum of the $R$ transmon in the presence of the parametric pump when $\phi_s = 0.386$, (as shown in Fig.~2(c) of the main text and Fig.~\ref{sFig7} below). Here, we have adjusted the amplitude setting on the microwave signal generator for the pump by multiplying our {\it nominal} amplitude $V_n=33$~mV by $\lambda(\omega_p)$. The five pump frequencies ($i=1,2,\dots 5$ for $\omega_p/2\pi =$ 540, 600, 704, 750, \& 800~MHz) where rectification was measured directly in Fig.~\ref{sFig5} are identified with corresponding colored dots at each $\lambda_iV_n$. The inset shows the ratio $\lambda_i/\lambda_1$ as a function of the inverse normalized slopes $S_1/S_i$, for the five pump frequencies. The solid line has a slope of 1 and intercept 0. This shows that $\lambda$ can give the relative size of the inverse slopes $1/S$ for any $\omega_p$. Therefore, after measuring $\lambda(\omega_p)$ and one single rectification curve measured at one pump frequency (see Fig.~\ref{sFig5}), we can find any $S(\omega_p)$ and calibrate the size of $\delta\phi_p$ at any $\omega_p$. 

This calibration and compensation procedure was important for taking the data presented in this work. As an example, Fig.~\ref{sFig7} shows the $R$ transmon spectrum as a function of the parametric pump frequency, with and without the compensation adjustment of the pump signal generator amplitude (shown in Fig.~\ref{sFig6}). Without the compensation, as shown in Fig.~\ref{sFig7}(a), multiple undulations appear as the pump frequency changes making it difficult to clearly distinguish the avoided crossings due to the parametric mode couplings between the $L$ and $R$ transmon. Again, this indicates that the amplitude of the parametric pump is not uniform over the frequency range swept. Fig.~\ref{sFig7}(b) shows the result of the same measurement as in (a), but with the compensating numerical factor $\lambda(\omega_p)$ taken into account. Notice that the resonant frequency of the $R$ transmon is now mostly constant over this range of pump frequencies, which clearly demonstrates a proper compensation for the unwanted frequency dependence of the standing waves in the pump line. Moreover, the undulations have disappeared leaving a clear signature of the parametrically generated avoided crossings on top of a uniform background. Clear measurements of the two-qubit parametric interactions with pump detuning both in the frequency and time-domain require a constant pump amplitude for all pump frequencies, ensuring the removal of these artifacts. 

\section{Basic circuit model}
\label{Basic_model}
We can approximate each transmon frequency by,
\begin{equation}
    \omega_k \approx \frac{1}{\sqrt{(\mathcal{L}_k+L_s)C_k}} - E_{ck}/\hbar,
    \label{Eq_freq_k}
\end{equation}
where $C_k$ represents the total capacitance (including any additional mutual capacitance) for each transmon, for $k\in\{L,R\}$, and $E_{ck}=(2\pi\hbar/\Phi_o)^2/(8 C_k)$ is the charging energy for each transmon. At zero flux, where the SQUID inductance contributes about $4\%$ of the transmon's total inductance, we can neglect the SQUID's contribution completely and extract the transmon junction inductance $\mathcal{L}_k$ and capacitance from measurements of the transmon transition frequency and anharmonicity. Here we have,
\begin{eqnarray}
         \mathcal{L}_k &=& \frac{1}{C_k}\frac{1}{(\omega_k+E_{ck}/\hbar)^2} \\
        C_q &=& \frac{(2\pi\hbar/\Phi_o)^2}{8 E_{ck}}
\end{eqnarray}
Each transmon's $\mathcal{L}_k$ was made with nominally the same critical current Josephson junction. The SQUID inductor was made from two nominally identical SQUID junctions within a loop, each with a critical current that is $\eta_s$-times larger than the main transmon junctions. As seen in the next section with a more detailed model, fits give $\eta_s = 16.3$, which is close to the design's as-drawn value of $15.2$ for the area ratio of the SQUID and the transmon junctions. 

Because transmons are nearly linear oscillators, we can use a simple model for linear coupled oscillators. In general, the two transmons share the dc-SQUID inductor, which leads to a sharing of current that provides an inductive coupling strength $g_{L}$. In addition, the large capacitors of the two transmons being relatively close together lead to a capacitive coupling with strength $g_{C}$. These coupling strengths are roughly given by,
\begin{eqnarray}
    g_{L} &=& \frac{1}{2}\frac{L_s}{\sqrt{(\mathcal{L}_L+L_s)(\mathcal{L}_R+L_s)}}\sqrt{\omega_L\omega_R} \label{Eq_g_L}\\
    g_{C} &=& \frac{1}{2}\frac{C_{LR}}{\sqrt{(C_L+C_{LR})(C_R+C_{LR})}}\sqrt{\omega_L\omega_R}\label{Eq_g_C}
\end{eqnarray}
where $L_s$ and $\omega_k$ are flux dependent. The total static coupling is the difference of the inductive and capacitive contributions, $g_{s}=g_{L}-g_{C}$. As discussed in the main text, at a particular flux, there is a SQUID inductance $L_s$ where $g_{L}=g_{C}$ and $g_{s}=0$. 

From these relations, we can first investigate the relative strength of the parametric coupling at each flux bias\cite{EandF2011,Allman2014} by calculating the derivative of $g_{s}$ with respect to $\phi$, which contains both the inductive and capacitive contributions,
\begin{eqnarray}
    \frac{\partial g_{s}}{\partial\phi} &=& \\ \nonumber
    &=& \left[\frac{1}{L_s}\frac{\partial L_s}{\partial\phi} + \frac{3}{2}\frac{1}{\omega_L}\frac{\partial \omega_L}{\partial\phi} +
    \frac{3}{2}\frac{1}{\omega_R}\frac{\partial \omega_R}{\partial\phi}
    \right]g_{L} - 
    \left[\frac{1}{2}\frac{1}{\omega_L}\frac{\partial \omega_L}{\partial\phi}+
    \frac{1}{2}\frac{1}{\omega_R}\frac{\partial \omega_R}{\partial\phi}
    \right]g_{C} \\
    &\approx& \left[\frac{1}{L_s}\frac{\partial L_s}{\partial\phi}\right]g_{L} 
\end{eqnarray}
since $(1/\omega_k)(\partial \omega_k/\partial\phi)\ll 
    (1/L_s)(\partial L_s/\partial\phi)$, so that,

\begin{equation}
    \frac{\partial g_{s}}{\partial\phi}\approx\frac{1}{2}\frac{\partial L_s}{\partial\phi}\frac{1}{\sqrt{(\mathcal{L}_L+L_s)(\mathcal{L}_R+L_s)}}\sqrt{\omega_L\omega_R}
\end{equation}
and we know from Eq.~(\ref{Eq_freq_k}) above, $\partial\omega_k/\partial\phi\approx-(\omega_k/2)(\partial L_s/\partial\phi)/(\mathcal{L}_k+L_s)$ for $k\in\{L,R\}$, giving,
\begin{equation}
\left|\frac{\partial g_{s}}{\partial\phi}\right|\approx\sqrt{\left|\frac{\partial \omega_L}{\partial\phi}\right|\left|\frac{\partial \omega_R}{\partial\phi}\right|}
\end{equation}
Then the parametric coupling strength can be estimated from\cite{Allman2014}
\begin{equation}
g_p = \frac{1}{2}\left|\frac{\partial g_{s}}{\partial\phi}\right|\delta\phi_p\approx\frac{1}{2}\sqrt{\left|\frac{\partial \omega_L}{\partial\phi}\right|\left|\frac{\partial \omega_R}{\partial\phi}\right|}\delta\phi_p
\end{equation}
This shows that the parametric coupling strength is directly related to the size of the frequency modulation of both qubits with flux. This is important when considering that the phase coherence of the qubits is inversely proportional to the same factors in the expression above for the parametric coupling strength. Thus, there is a trade-off between achieving large parametric coupling strengths for relatively small flux modulations and maintaining high phase coherence in the presence of flux noise in the SQUID coupler $-$see Section~\ref{coherence} below. 

\section{Determination of full circuit model and parameters}
\begin{figure}[b!]
\centering
\includegraphics[width=5in]{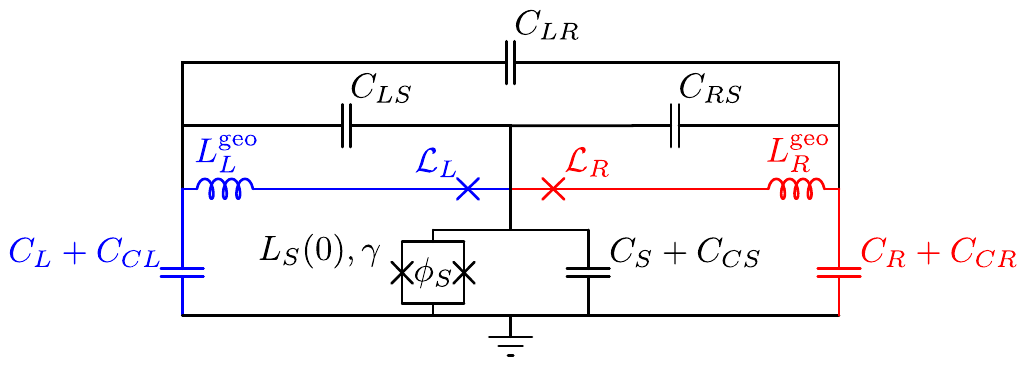}%
\caption{{\textbf {Lumped circuit model for parameter extraction.}} Extended circuit schematic diagram of Fig.~1(b) (see main text) with the addition of parasitic geometric inductances $L^{\mathrm{geo}}_L$ and $L^{\mathrm{geo}}_R$. Final fitting circuit element parameters are listed in Table \ref{tab:circparams}.}
\label{sFig_circ}
\end{figure}

The main circuit parameters were determined by considering the lumped circuit model shown in Fig.~1(b) in the absence of the coplanar waveguide readout cavity, apart from unavoidable capacitive loading to both the transmons and their mutual capacitive coupling. These effects simply are absorbed into the remaining capacitive parameter values used in the final circuit model. Here, the coplanar waveguide cavity coupling capacitances ($C_{CL}$ and $C_{CR}$) are explicitly accounted for in the two corresponding transmon shunting capacitances ($C_{L}$ and $C_{R}$). To get a more accurate fitting of the qubit transition frequencies, it is important to consider the additional geometric inductance of the 8~$\mu$m wide wires connecting each transmon’s Josephson junction to the roughly rectangular capacitor electrodes. A more detailed lumped circuit schematic is shown in Fig. \ref{sFig_circ}.  

In this model, although not explicitly shown in Fig.~\ref{sFig_circ}, Josephson junction capacitances are initially assumed to be 1~fF for the qubit junctions and 18~fF for the SQUID junctions based on their relative areas. Their precise values do not significantly effect the fit, rather they are set to reasonable values and included in the circuit model, which is necessary for treating the junctions in the charge basis. The remaining capacitance values are determined from the device layout. Using Ansys Q3D Extractor we can extract the capacitance matrix of the system from the design layout GDS file for the metal wiring layer up to a scaling factor for the sapphire relative permittivity. This restricts the number of free parameters in the circuit model and fixes the resultant circuit values to be consistent with the proportional geometries of the design. Thus, all the capacitances scale with this single factor corresponding to the sapphire substrate relative permittivity used in the Q3D simulation. While Ansys HFSS can take into account an anisotropic relative permittivity reflective of sapphire, Ansys Q3D only allows a simple isotropic permittivity. For both microstrip or coplanar waveguide structures, analytic derivations show an effective isotropic value can be taken as the geometric mean of the parallel and perpendicular components. Although we receive a room temperature value for sapphire's relative permittivity from the vendor supplying our source material, there are limited reports of low temperature values in the current literature. Actual values can also be skewed by differences in the sapphire growth mechanisms, wafer cutting and surface polishing techniques. To fit the experimental data, we are left to determine the capacitive scaling factor and the inductances: $L$ and $R$ transmon junctions, SQUID junctions with potential asymmetry, and the geometric inductances. The geometric inductances are kept in comparable ratio to that of their different lengths.

\begin{table}[b!]
\centering
\begin{tabular}{|l|c|c|l|c|c|} 
 \hline
 Param. & value & units & Param. & value & units \\ [0.5ex] 
 \hline\hline
 $\mathcal{L}_L$ & 7.37 & nH & $\mathcal{L}_R$ & 7.64 & nH \\
 \hline
 $L^{\text{geo}}_L$ & 0.16 & nH & $L^{\text{geo}}_R$ & 0.22 & nH \\
 \hline
 $C_L$ & 83.48 & fF & $C_R$ & 65.15 & fF \\
 \hline
 $C_{Lj}$ & 1.12 & fF & $C_{Rj}$ & 1.12 & fF \\
 \hline
 $L_S(0)$ & 0.23 & nH & $\gamma$ & 0.75 & \\
 \hline
 $C_S$ & 4.30 & fF & $C_{CS}$ & 0.14 & fF \\
 \hline
 $C_{CL}$ & 18.34 & fF & $C_{CR}$ & 12.44 & fF \\
 \hline 
 $C_{LS}$ & 1.68 & fF & $C_{RS}$ & 1.55 & fF \\
  \hline
 $C_{Sj1}$ & 23.04 & fF & $C_{Sj2}$ & 17.28 & fF \\
 \hline
 $C_{LR}$ & 7.34 & fF  & & &\\
 \hline
\end{tabular}
\caption{{\textbf {Fitted circuit parameter values.}}}
\label{tab:circparams}
\end{table}
 
The lumped circuit model shown in Fig.~\ref{sFig_circ} is implemented in scQubits using the custom circuit class.\cite{Groszkowski2021, Chitta2022} In order to diagonalize the system in a reasonable time with an exponentially large computational Hilbert space, hierarchical diagonalization is utilized.\cite{Kerman2020} This consists of a change of basis and then identification of subsystems within the circuit. The subsystems correspond to the SQUID coupler and the two transmons without the SQUID inductance. This returns the eigenenergies of the circuit evaluated at each value of external flux bias $\phi$. The levels can be identified in increasing energy: $E_{gg}$, $E_{eg}$, $E_{ge}$, $E_{fg}$, $E_{ee}$, $E_{gf}$. From this the qubit frequencies $\omega_{k}$, anharmonicities $\alpha_{k}$, and the $ZZ$ coupling $\zeta_s$ can be calculated: 
 \begin{align}
     \hbar\omega_L &= E_{eg} - E_{gg} \\
     \hbar\omega_R &= E_{ge} - E_{gg} \\
     \hbar\alpha_L &= (E_{fg} - E_{eg}) - (E_{eg} - E_{gg}) \\
     \hbar\alpha_R &= (E_{gf} - E_{ge}) - (E_{ge} - E_{gg}) \\
     \hbar\zeta_s &= (E_{ee} - E_{ge}) - (E_{eg} - E_{gg}) 
 \end{align}

\begin{figure}[t!]
\centering
\includegraphics[width=\textwidth]{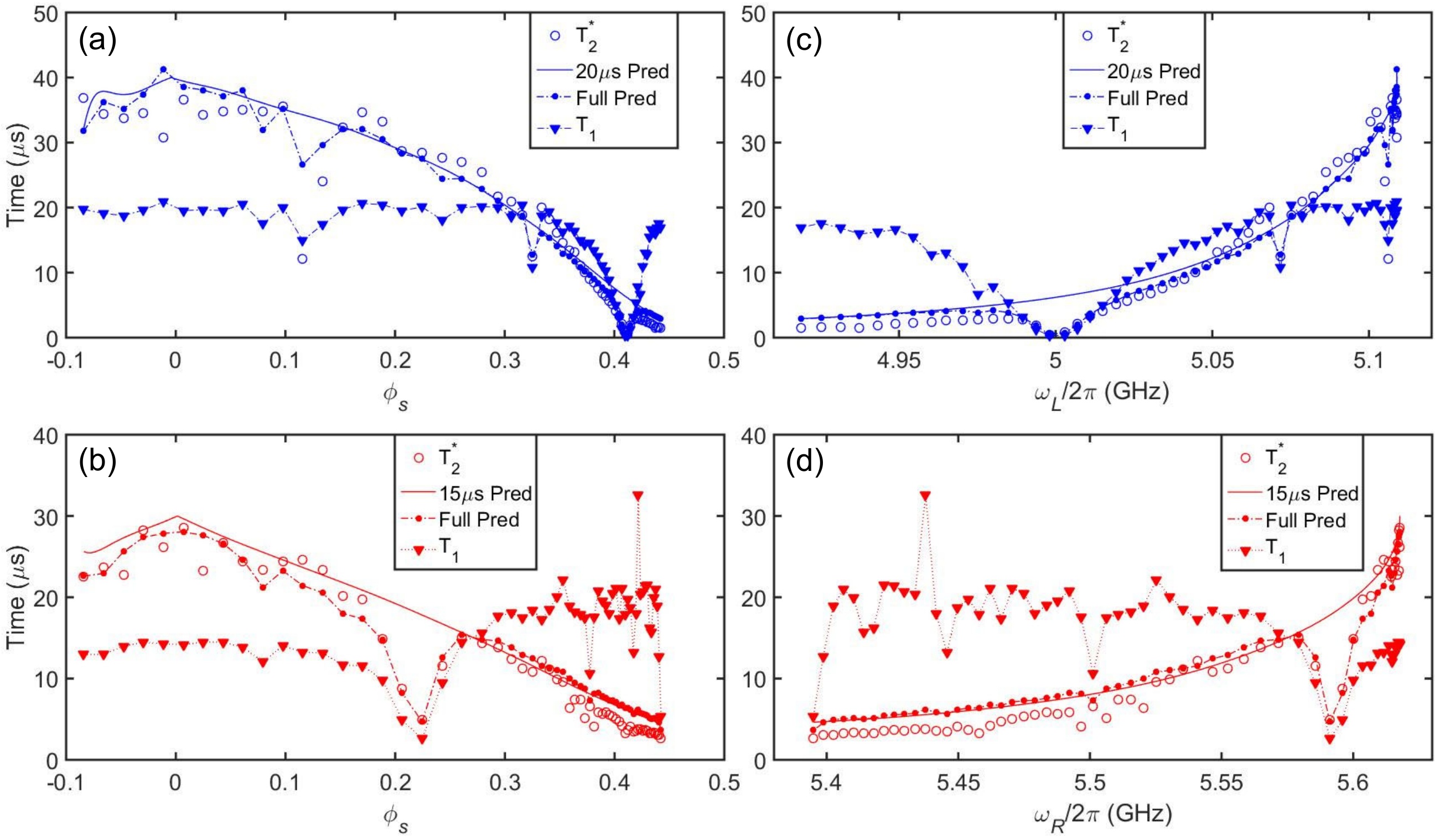}
\caption{\textbf{Device coherence.} Coherence for the $L$ qubit as a function of  with bias flux (a) and qubit frequency (c). Coherence for the $R$ qubit as a function of  with bias flux (b) and qubit frequency (d). The solid and dashed-dot lines show predictions for dephasing coming from flux noise with $\sqrt{A_\Phi} = 4.1\,\mu\Phi_o$ using and average $T_1$ or the full $T_1$ data set respectively. Notice that each qubit has a ``dead-zone'' where the overall coherence is reduced significantly. When plotted against qubit frequency, these regions look like a single-mode Purcell effect from a mode at 5~GHz and 5.59~GHz for the $L$ and $R$ qubits respectively.}
\label{sFig9}
\end{figure} 

The extracted circuit parameters are determined by fitting the simulated qubit frequencies, anharmonicities, and static $ZZ$ coupling to the experimentally measured data. The resulting fits are shown in the main text in Fig.~1(d), Fig.~1(e) and Fig.~2(b). Our fits correspond to a geometric mean relative permittivity of sapphire of $\epsilon_r = 11.49$. Additionally, the bare resonance of the SQUID coupler mode can be computed (not shown) and tunes from approximately 50~GHz to 20~GHz over the entire range of external flux $\phi$. These values are far enough above any of the qubit transition frequencies of interest so that these modes can be completely neglected. This is an additional advantage of our circuit design strategy. 

\section{Sample Coherence}
\label{coherence}
\subsection{Coherence versus flux and frequency}
We measured the coherence of our samples as a function of the static flux bias $\phi_s$. This includes measuring the energy relaxation $T_1$ and the Ramsey dephasing time $T_2^*$ of each qubit. Data for the main device used to produce the fast parametric gates is shown in Fig.~\ref{sFig9}. Coherence of the qubit can be plotted as a function of bias flux, as shown in Fig.~\ref{sFig9}(a),(b) ($L$, $R$), or as a function of qubit frequency, as shown in Fig.~\ref{sFig9}(c),(d) ($L$, $R$). Notice that each qubit has a ``dead-zone'' where the overall coherence is reduced significantly. When plotted against qubit frequency, these regions look like a single-mode Purcell effect from a mode at 5~GHz and 5.59~GHz for the $L$ and $R$ transmons respectively. These modes are \emph{not} indicative of two-level system defects because they survive multiple temperature cycles to room temperature, however they do move and sometimes disappear and return. They are \emph{not} consistent with the packaging (shown in Fig.~\ref{sFig3}) that encloses the chip in a cylindrical cavity with a lowest fundamental mode of $\sim 36$~GHz. Needless to say, these modes are still under investigation. Notice that at $\phi_s=0$, and the highest qubit frequencies, both qubits show Ramsey coherence times with $T_2^*\approx 2T_1$, where they are first-order insensitive to flux bias noise. As discussed below, the solid and dashed-dot lines show predictions accounting for flux noise dephasing with $\sqrt{A_\Phi} = 4.1\,\mu\Phi_o$ using the average $T_1$ or the full $T_1$ data set respectively.

\begin{figure}[b!]
\centering
\includegraphics[width=5in]{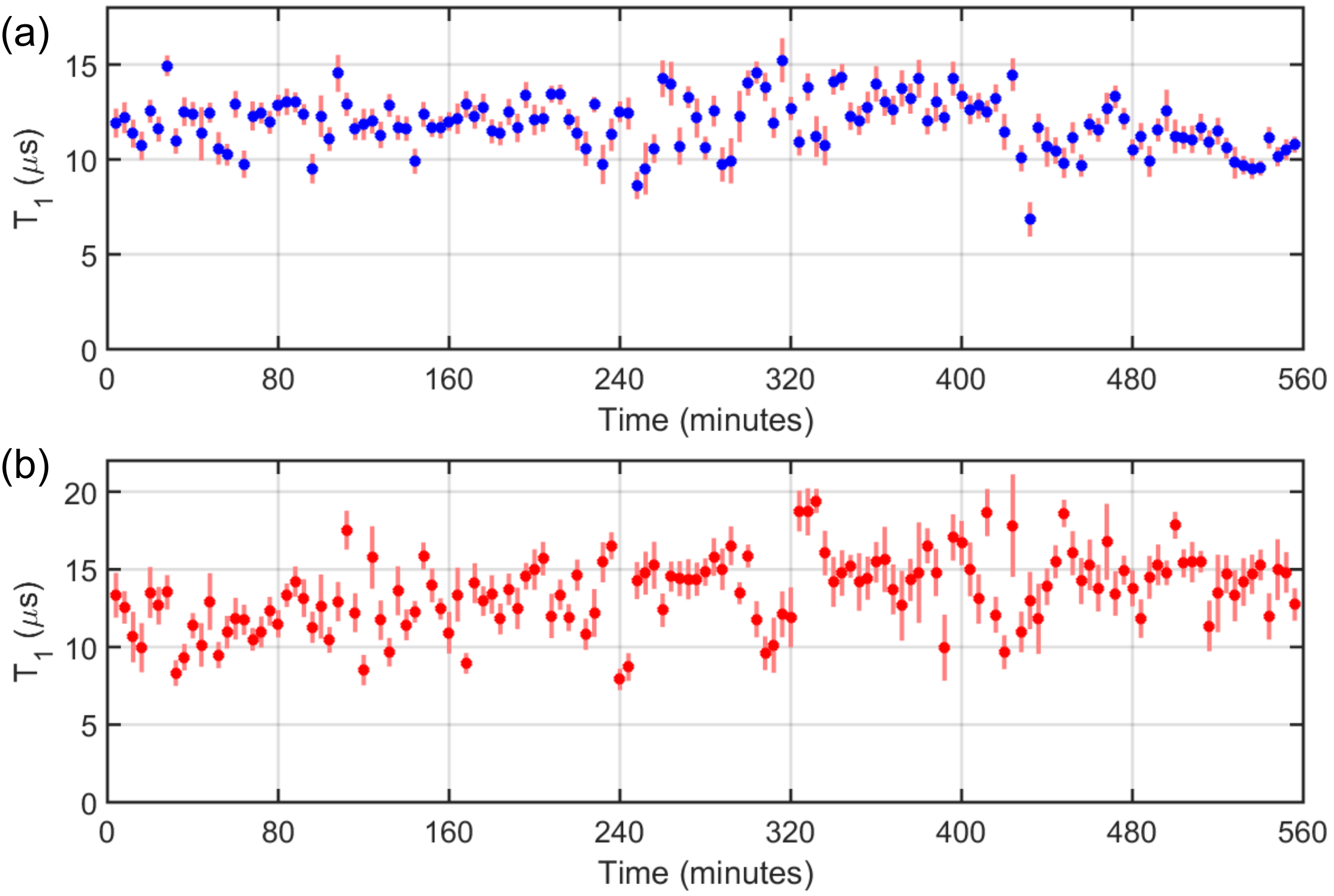}
\caption{\textbf{Measuring relaxation times repeatedly.} (a) $T_1$ as a function of repetition time for the $L$ qubit. (b) $T_1$ as a function of repetition time for the $R$ qubit. This data was taken over the full time scale for each qubit at separate times, not interleaved. This data came from a two-qubit device of similar design but fabricated with all aluminum transmons on a silicon substrate.}
\label{sFig10}
\end{figure}

\begin{figure}[b!]
\centering
\includegraphics[width=5in]{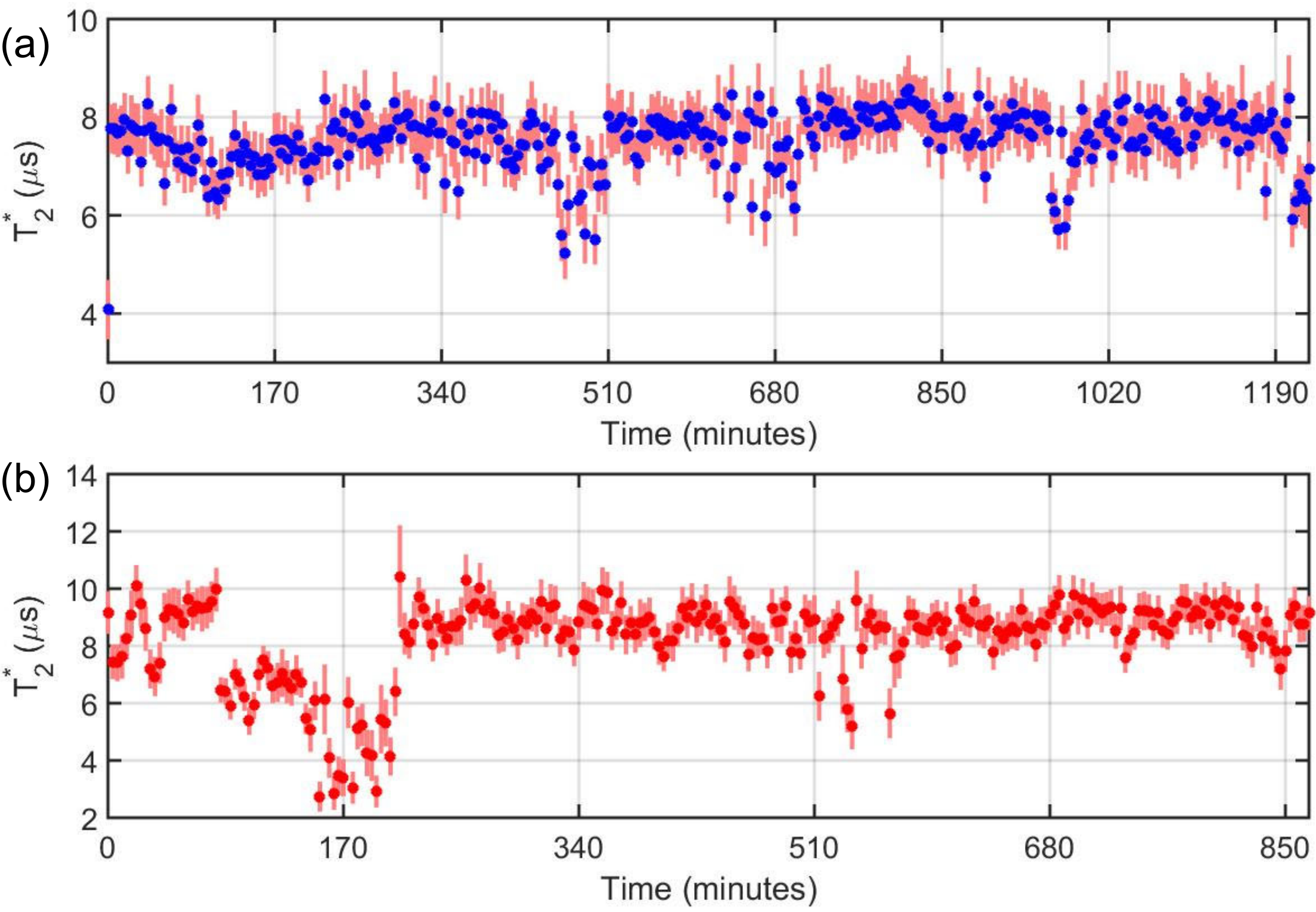}
\caption{\textbf{Measuring Ramsey dephasing times repeatedly.} (a) $T_2^*$ as a function of repetition time for the $L$ qubit. (b) $T_2^*$ as a function of repetition time for the $R$ qubit. This data was taken over the full time scale for each qubit at separate times, not interleaved. This data came from a two-qubit device of similar design but fabricated with all aluminum transmons on a silicon substrate.}
\label{sFig11}
\end{figure}

\subsection{Coherence, flux sensitivity, and circuit design}
Our design strategy balances strong, tunable parametric interactions with significant protection from qubit dephasing from bias flux noise. Much of this discussion follows that found in the supplemental material for T.~Noh~{\it et al}\cite{Noh2023}. In our design, we reduce the slopes of the qubit frequencies with flux at the cancellation flux $\phi_c$ by relying on the weak tunability provided only by the shared SQUID coupler, which then reduces $\partial g_{s}/\partial\phi$ as well. However, the parametric coupling strength also depends on the size of the parametric pump amplitude $\delta\phi_p$. This means that it is possible to compensate a weak slope (lower flux noise induced dephasing rates) with a strong drive to obtain sizable parametric coupling strengths. We only need modulations of order 100~MHz in order to achieve about 100~MHz of coupling. As a benefit, our circuit design strategy has far weaker sensitivity to flux noise than fully tunable transmons based on two-junction SQUIDs, allowing us to minimize dephasing across most of a flux quantum.   

In M. D.~Hutchings~\textit{et al.}\cite{Hutchings2017}, they describe optimizing a balance between flux tunability, a desirable feature for overcoming issues with critical current deviations from fabrication spread, and maintaining qubit phase coherence. In that work, they showed that transmons with an asymmetry of 15:1 (the ratio of the left and right SQUID junction critical currents) with a sensitivity to flux ($|d\omega_q/d\Phi|/2\pi$) less than 1~GHz$/\Phi_o$ did not show significant additional dephasing when the flux noise was about $1.3~\mu\Phi_o/\sqrt{\mathrm{Hz}}$. Following their technique, we can calculate the flux sensitivity of our design from knowing the measured qubit frequency as a function of flux. At the cancellation flux $\phi_c$ the two qubits have nearly the same slope of approximately $1.5$~GHz$/\Phi_o$ and $\lesssim 1.2$~GHz$/\Phi_o$ where we operated the two-qubit gates. These values are comparable to the levels found in M.D.~Hutchings~\textit{et al.}\cite{Hutchings2017} and can be reduced further as well, as found in T.~Noh~\textit{et al.}\cite{Noh2023}.

We characterize the flux noise in the same way that was done in M. D. Hutchings \textit{et al.}\cite{Hutchings2017}, where $\Gamma_{\phi}=\Delta\phi|\partial\omega_k/\partial\phi|$ ($k\in\{L,R\}$) represents the dephasing rate from low-frequency flux noise with a $1/f$ power spectrum $S_\Phi(f)=A_\Phi/|f|$ and $\Delta\phi = \sqrt{A_{\Phi}|\ln(\omega_{IR}\tau)|}/\Phi_o$, where $\omega_{IR}/2\pi = 1$~Hz is the infrared cutoff frequency and $\tau=10~\mu$s, comparable to $1/\Gamma_{\phi}$. The lines in Fig.~\ref{sFig9} are given by $1/T_2^*=1/(2T_1)+\Gamma_{\phi}+\Gamma_{bg}$, where we use the average measured $T_1$, $\omega_k(\phi)$ as a function of flux to calculate $|\partial\omega_k/\partial\phi|$ across the qubit's frequency range and $\Delta\phi$ is the best fit value to the $T_2^*$ data. Here, $\Gamma_{bg} = 1/T_2^*-1/(2T_1)$ when $|\partial\omega_k/\partial\phi|=0$ and represents the background dephasing when $\Gamma_{\phi}=0$. In our case, $1/\Gamma_{bg}>500\,\mu$s for both transmons because $T_2^*\approx 2T_1$ at the maximum qubit frequencies. From our fit, we find one value, $\sqrt{A_{\Phi}}=4.1~\mu\Phi_o$, slightly larger than more typical values of $1\--3~\mu\Phi_o$ found in other experiments\cite{Hutchings2017}. In the future, the maximum $T_1$ and $T_2^*$ values can significantly be improved with better materials, cleaner fabrication, and improved filtering, shielding, and packaging to reduce further flux noise and infrared radiation from impinging on the sample. 

\begin{figure}[b!]
\centering
\includegraphics[width=5in]{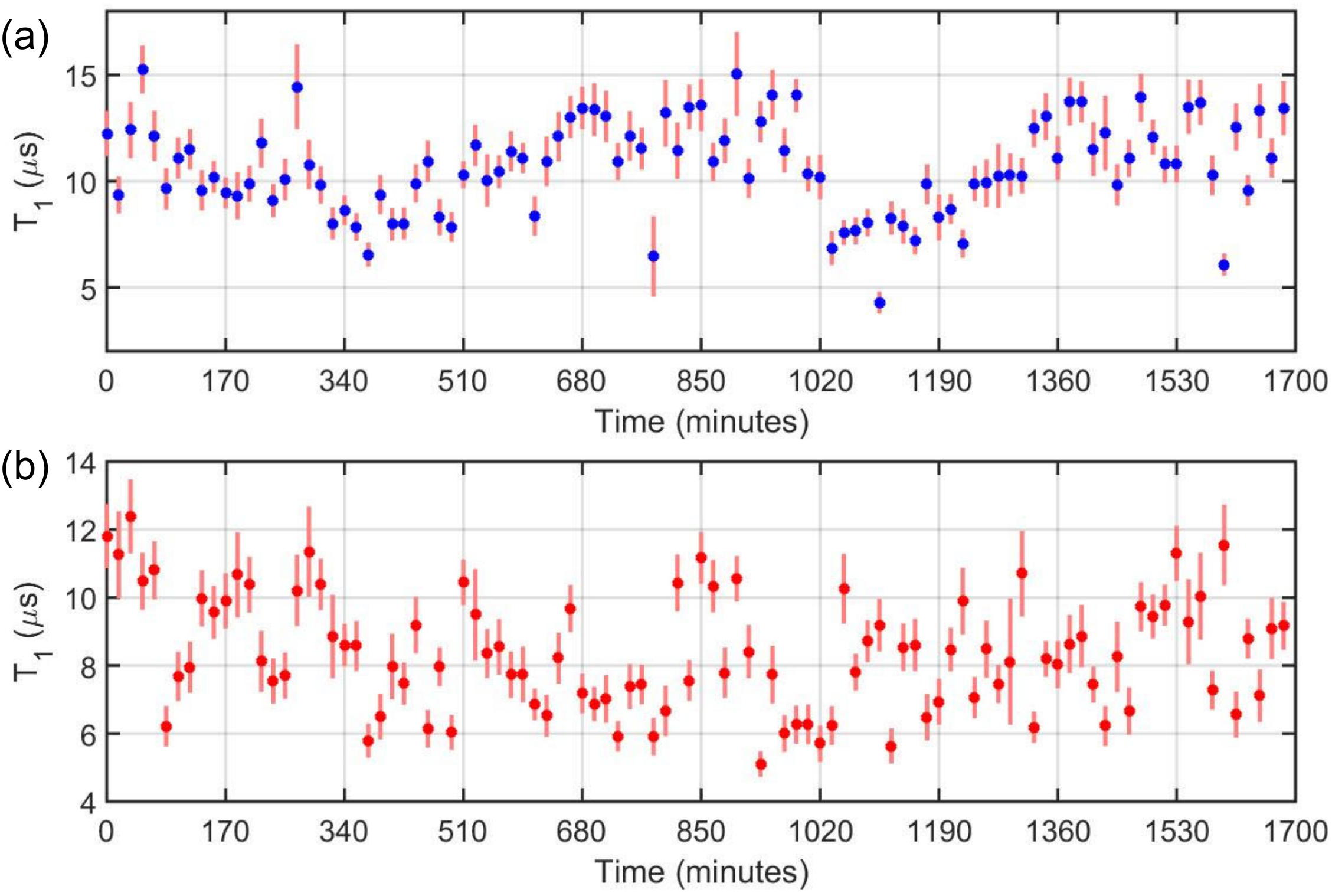}
\caption{\textbf{Measuring relaxation repeatedly interleaved over time.} (a) $T_1$ as a function of repetition time for the $L$ qubit. (b) $T_1$ as a function of repetition time for the $R$ qubit. This data came from the main two-qubit device fabricated with niobium wiring and aluminum Josephson junctions, as described in the Methods Section of the main text.}
\label{sFig12}
\end{figure}

\begin{figure}[b!]
\centering
\includegraphics[width=5in]{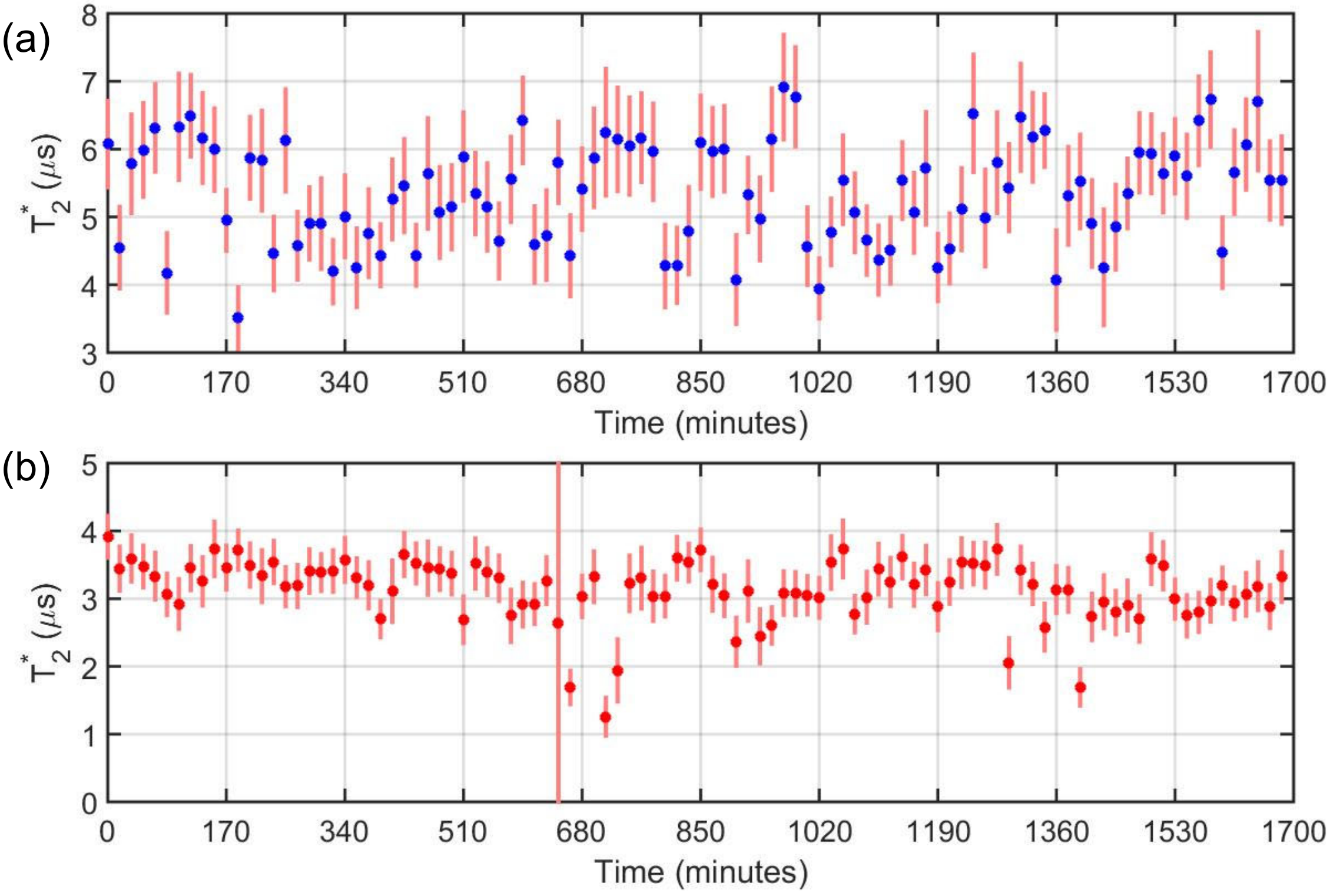}
\caption{\textbf{Measuring Ramsey dephasing repeatedly interleaved over time.} (a) $T_2^*$ as a function of repetition time for the $L$ qubit. (b) $T_2^*$ as a function of repetition time for the $R$ qubit. This data came from the main two-qubit device fabricated with niobium wiring and aluminum Josephson junctions, as described in Methods Section of the main text.}
\label{sFig13}
\end{figure}

\subsection{Coherence over time}
Next, we plot the coherence times for both qubits over time by repeating measurements for long periods as shown in Fig.~\ref{sFig10} and Fig.~\ref{sFig11}. Here, we focused on separately measuring one quantity on one qubit for many minutes, then moving on to another quantity. These results shows fluctuations over time with sometimes dramatic changes or jumps over relatively short periods of time. This data was taken on a similar two-qubit design but made from all aluminum wiring on high purity silicon substrates. We found that the coherence times were comparable to our niobium on sapphire samples. 

Finally, we plot in Fig.~\ref{sFig12} and Fig.~\ref{sFig13} repeated \emph{interleaved} coherence time measurements for both qubits over time for the niobium wiring device used for fast parametric gates. This data was taken near the operation flux for performing the gates. It does not represent the best coherence times we found at this location or per cool-down and was not taken during the time we made our final gate measurements. Here, we focused on interleaving the measurements of each quantity on each qubit in order to look for correlations in the coherence fluctuations between the qubits. For this data, we measured $T_1$ and $T_2^*$ on the $L$ qubit then the same on the $R$ qubit and repeated this over time for approximately 28 hours. Although we see fluctuations over time, there is no clear correlations between the two qubits.

\begin{figure}[b!]
\centering
\includegraphics[width=\textwidth]{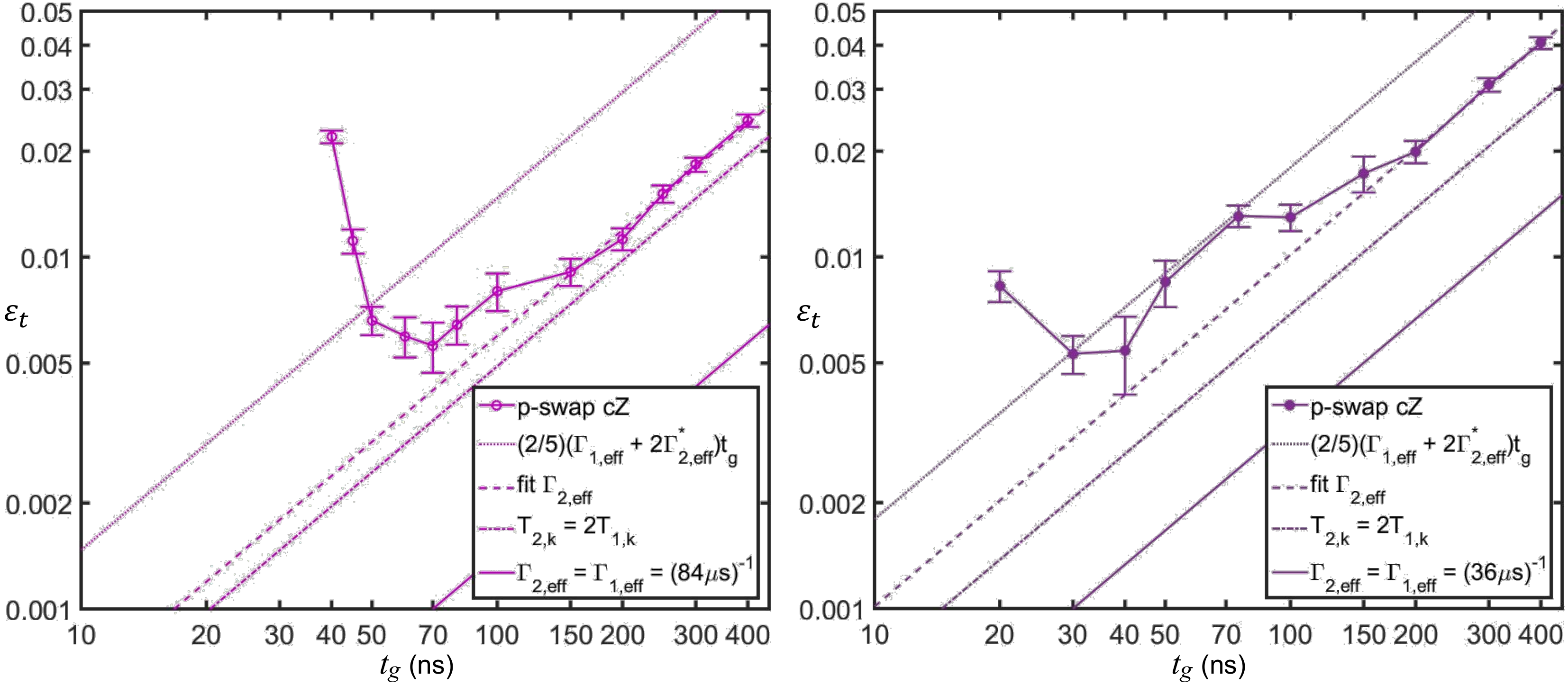}
\caption{\textbf{Coherence limits on gate fidelity.} Total error per gate $\varepsilon_t$ of gate duration $t_g$ for: (a) the p-SWAP c$Z$ gate or (b) the p-SWAP-free c$Z$ gate.}
\label{sFig14}
\end{figure}

\subsection{Coherence limits on gate fidelity}
In order to calculate the limits on gate fidelity based on the individual qubit coherences in the main text, we use the $T_1$ (energy decay) values to obtain $\Gamma_{1,{\rm eff}}=(1/2)(1/T_{1,L}+1/T_{1,R})$ and we use the $T_2^*$ (Ramsey decay) values to obtain an estimate of $\Gamma_{2,{\rm eff}}^*=(1/2)(1/T_{2,L}^*+1/T_{2,R}^*)$. We measured both $T_1$ and $T_2^*$ before we performed randomized benchmarking (RB). For the p-SWAP c$Z$ data using plateau pulses, we measured $T_{1,L}=13.93\,\mu$s and $T_{1,R}=19.78\,\mu$s and $T_{2,L}^*=7.75\,\mu$s and $T_{2,R}^*=5.64\,\mu$s. This then gives $1/\Gamma_{1,{\rm eff}}=16.35\,\mu$s and $1/\Gamma_{2,{\rm eff}}^*=6.53\,\mu$s. For the p-SWAP-free c$Z$ data on a different cool-down, the measured values were a bit lower with $T_{1,L}=10.82\,\mu$s and $T_{1,R}=12.56\,\mu$s and $T_{2,L}^*=6.45\,\mu$s and $T_{2,R}^*=4.81\,\mu$s. This then gives $1/\Gamma_{1,{\rm eff}}=11.63\,\mu$s and $1/\Gamma_{2,{\rm eff}}^*=5.51\,\mu$s. If we simply plot $\varepsilon_{\rm dec}^*=(2/5)(\Gamma_{1,{\rm eff}}+2\Gamma_{2,{\rm eff}}^*)t_g$ derived from the Ramsey data for these values we find a much larger error than measured, as shown in Fig.~\ref{sFig14} by the dotted lines. Using the Ramsey measurements overestimates the error per gate. This seems reasonable because RB tends to have a natural echoing effect that reduces low frequency noise, which is why we require $T_2$ values that are related to the phase coherence of the qubits. If we assume $\Gamma_{1,{\rm eff}}$ is an accurate effective mean value, recall above that $T_1$ values vary over time, then we can estimate the best possible performance when the phase coherence is a maximum or $T_{2,k} = 2T_{1,k}$ for $k\in{L,R}$. This then leads to $\varepsilon_{\rm dec}^{\rm min}=(4/5)\Gamma_{1,{\rm eff}}t_g$, which outperforms our measured data, as shown in Fig.~\ref{sFig14} by the dash-dot lines. In order to find the effective phase coherence decay rate $\Gamma_{2,{\rm eff}}$ that explains the data, we choose to use this as a fit parameter for data with $t_g\geq 200$~ns. The result is plotted as dashed lines in Fig.\Ref{sFig14} and in Fig.~5(a) of the main text. The fit gives $T_{2,{\rm eff}}=1/\Gamma_{2,{\rm eff}}=22.7\,\mu$s or $T_{2,{\rm eff}}/T_{1,{\rm eff}}\approx 1.4$ for the p-SWAP c$Z$ data and $T_{2,{\rm eff}}=1/\Gamma_{2,{\rm eff}}=12.0\,\mu$s or $T_{2,{\rm eff}}/T_{1,{\rm eff}}\approx 1.0$ for the p-SWAP-free c$Z$ data. These are very reasonable effective values that are valid over the many hours the data was taken. if we continue to expect modest coherence from transmon qubits, where $T_{2,{\rm eff}}=T_{1,{\rm eff}}$ or $\Gamma_{2,{\rm eff}}=\Gamma_{1,{\rm eff}}$, then $\varepsilon_{\rm dec}^{2=1}=(6/5)\Gamma_{1,{\rm eff}}t_g$. We can then ask: What coherence values for $T_{1,{\rm eff}}$ limit the gate fidelity to 99.9\% when $t_g=70$~ns for the p-SWAP c$Z$ or when $t_g=30$~ns for the p-SWAP-free c$Z$ gates? It turns out that we only require $T_{1,{\rm eff}}=84\,\mu$s for the p-SWAP c$Z$ and $T_{1,{\rm eff}}=36\,\mu$s for the p-SWAP-free c$Z$ gate. These results are shown as the solid curves in Fig.~\ref{sFig14}. This is at least a factor of 2 improvement over this sample's coherence. Coherence values beyond $100\,\mu$s have been achieved in the literature, which should lead to fast parametric gates with gate fidelities beyond 99.9\%. This represents our goal for future devices.   

\subsection{Coherence reductions due to two-level system defects}
One major difficulty we have had with our qubit samples is two-level system (TLS) defects. Although during a given cool-down the spectrum of the qubits may look clean across an entire flux quantum, we have seen TLS appear from no where. Usually, if our laboratory loses power (for way too many reasons and for no good reason at all!) and the devices warm up above 100~K, there is a chance that we will have a TLS waiting for us. Usually, we do not see them as strongly resonant TLS within the spectrum of the two-qubits. Rarely, they may make one of the qubits have telegraph like fluctuations in its frequency. More often, we just see the coherence drop or fluctuate more than usual. It is likely that the stability of our two-qubit gates (shown in Fig.~5(d) of the main text) are influenced by these types of fluctuations. More work is needed to carefully characterize and quantify these effects. 

\begin{figure}[t!]
\centering
\includegraphics[width=5in]{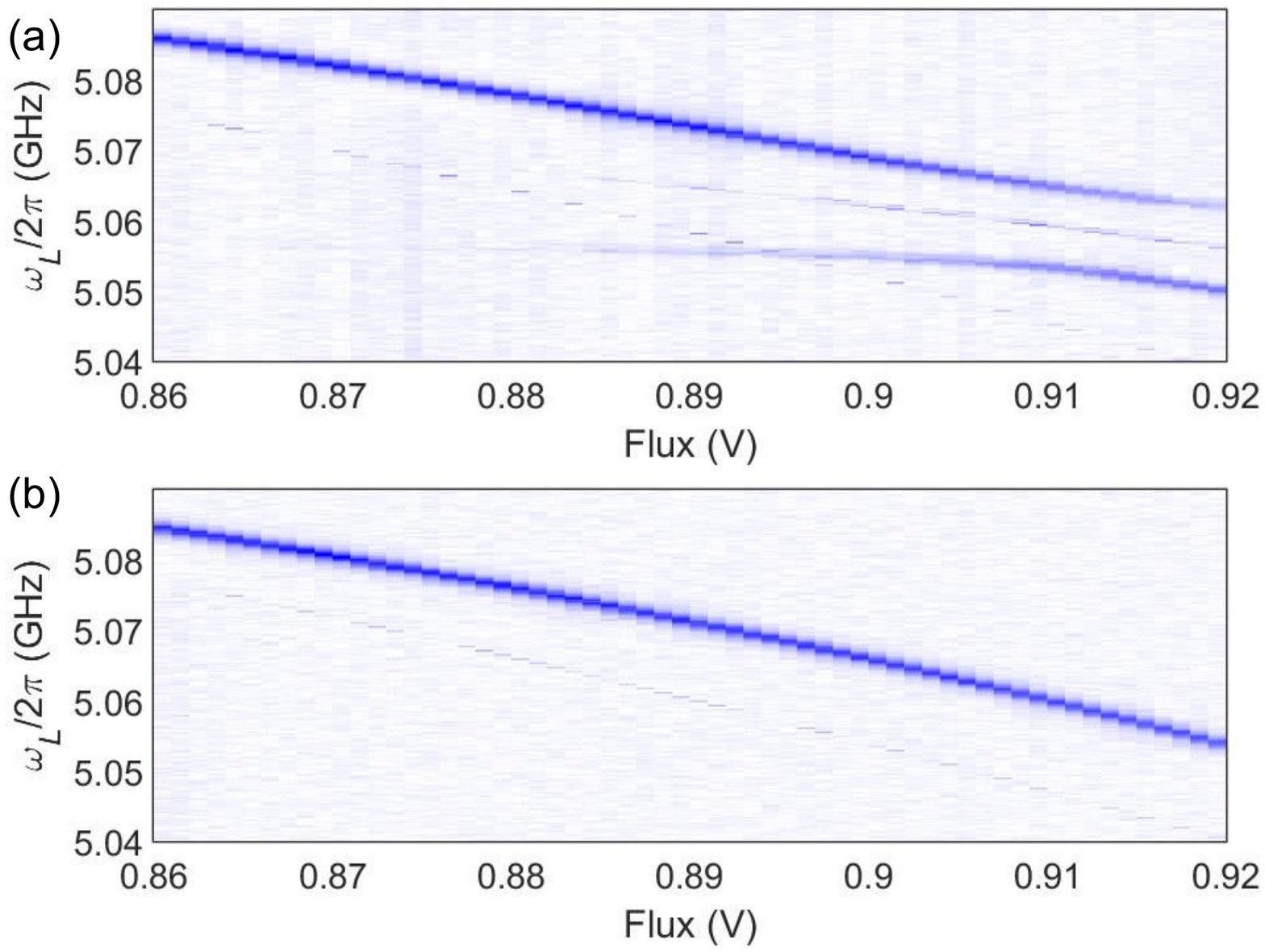}
\caption{\textbf{$L$ qubit spectroscopy.} (a) This spectrum shows a clear approximately 10~MHz mode splitting, signaling the presence of a two-level system. (b) After this device was cycled in temperature above 100~K then re-cooled the two-level system disappeared leaving a clear spectrum.}
\label{sFig15}
\end{figure}

\begin{figure}[t!]
\centering
\includegraphics[width=5in]{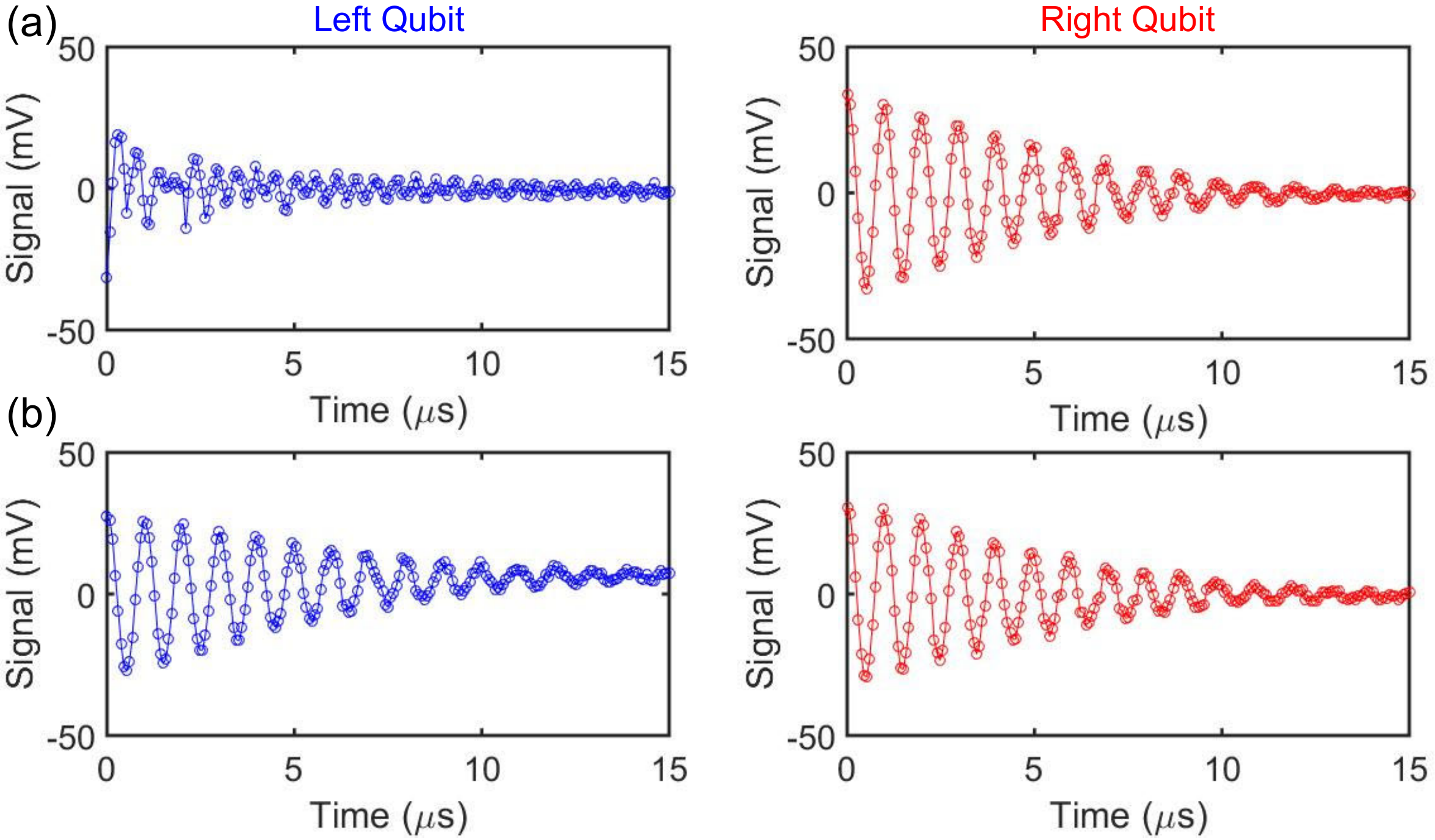}
\caption{\textbf{$L$ and $R$ qubit Ramsey oscillations.} (a) In the presence of a strongly coupled two-level system, the Ramsey oscillations at the operation flux are severely altered in the $L$ qubit. (b) After this device was cycled in temperature, the clear Ramsey oscillations at the operation flux are restored.}
\label{sFig16}
\end{figure}

The strongest example of a TLS defect in our system is shown in Fig.~\ref{sFig15} and Fig.~\ref{sFig16}. In Fig.~\ref{sFig15}(a), we clearly see a mode splitting in the spectrum of the $L$ transmon, which indicates the presence of a strongly coupled TLS. Here, the coupling strength is about 10~MHz $-$quite significant! In this case, when taking measurements to tune-up the two-qubit parametric gates, we suddenly saw a drastic drop in the gate fidelity. Upon further inspection its was clear that the $L$ qubit had experienced a significant coherence drop. This was made clearer by measuring the Ramsey oscillations of both qubits, as shown in Fig.~\ref{sFig16}(a). Notice that the Ramsey oscillation amplitude is very low for the $L$ transmon (and the frequency is too high. It was after this that we actually took the spectrum shown in Fig.~\ref{sFig16}(a)). Notice that the $R$ qubit is not affected by the TLS. Our circuit design keeps the qubits in relatively close proximity, so it seems that the TLS is mostly likely within the $L$ qubit's small Josephson junction.

We have tried many strategies to try to remove TLS to improve coherence. We have moved the flux bias to extremely large positive and negative values, as well as modulated those fluxes with ramps over time. We have driven both ports with strong rf-power in order to ``rattle'' the circuit. Some of these techniques worked, but not every time. We have also tried to warm up the device above the 1~K critical temperature of the aluminum Josephson junctions without much success. The most reliable technique is increasing the temperature of the device above about 100~K. In Fig.~\ref{sFig15}(b) and Fig.~\ref{sFig16}(b), we show that a temperature cycling of the device removed the strongly coupled TLS completely. The fabrication of our qubit Josephson junctions produces junction sizes that are relatively small, with junction areas near 130$\times$130~nm$^2$. Although we may be avoiding most of the TLS within the qubit junctions, it's still no guarantee. We believe more materials research will be needed to produce nearly defect-free Josephson junctions that will ensure more stable devices and two-qubit gates. 

\section{Two-qubit readout and fidelity}
\label{readout}
\begin{figure}[t!]
\centering
\includegraphics[width=\textwidth]{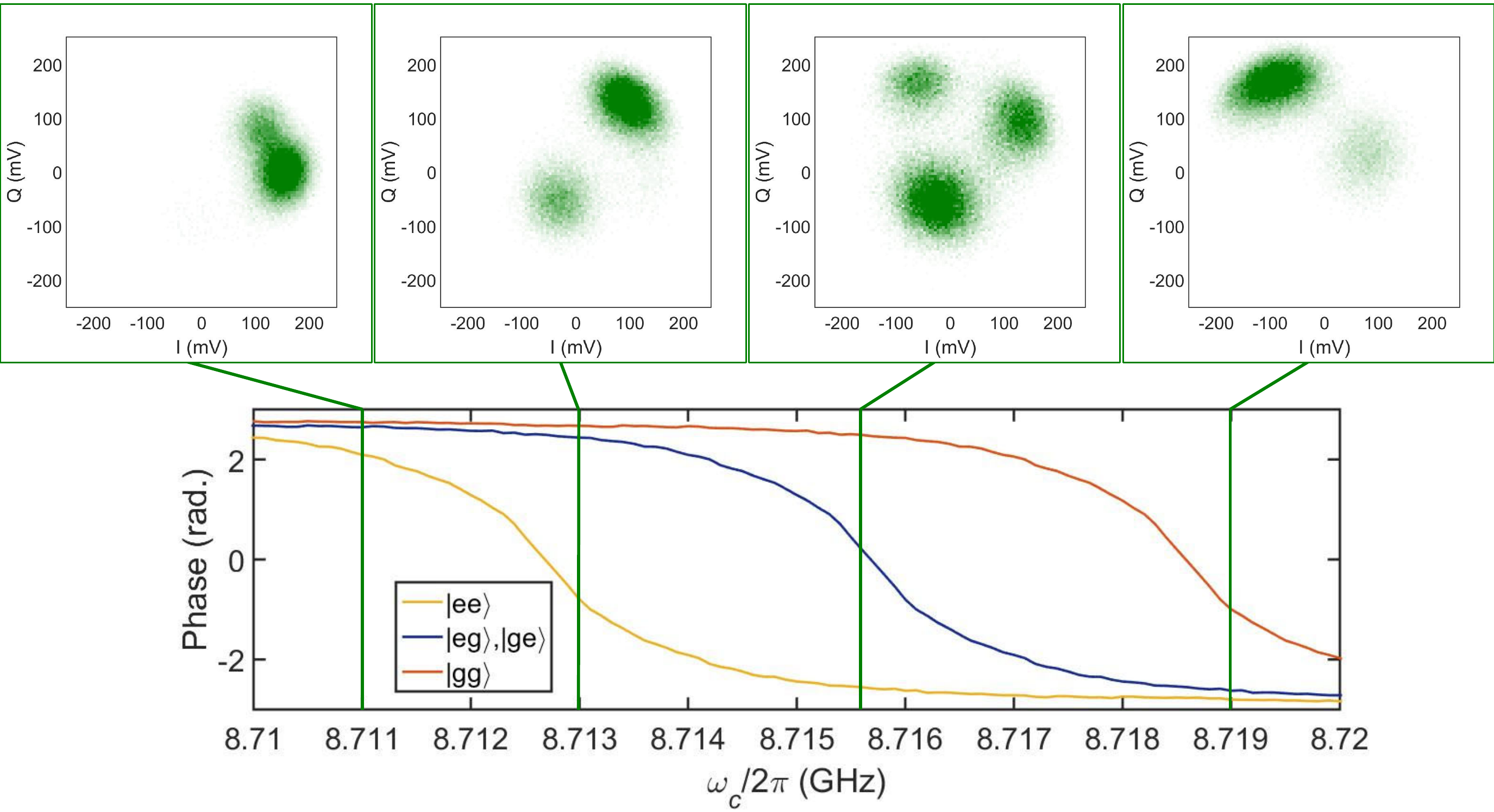}
\caption{\textbf{Readout cavity measurements as a function of drive frequency.} The main plot shows the average phase of the readout cavity as a function of the drive frequency when prepared in a mixture of two-qubit states. We also show $I-Q$ phase portraits at several cavity frequencies. Notice that between 8.715 and 8.716~GHz there are three well separated regions that can distinguish three possibilities: $|gg\rangle$, $|ee\rangle$, and the overlapping response of $|eg\rangle$ and $|ge\rangle$.}
\label{sFig17}
\end{figure}

\begin{figure}[t!]
\centering
\includegraphics[width=4.5in]{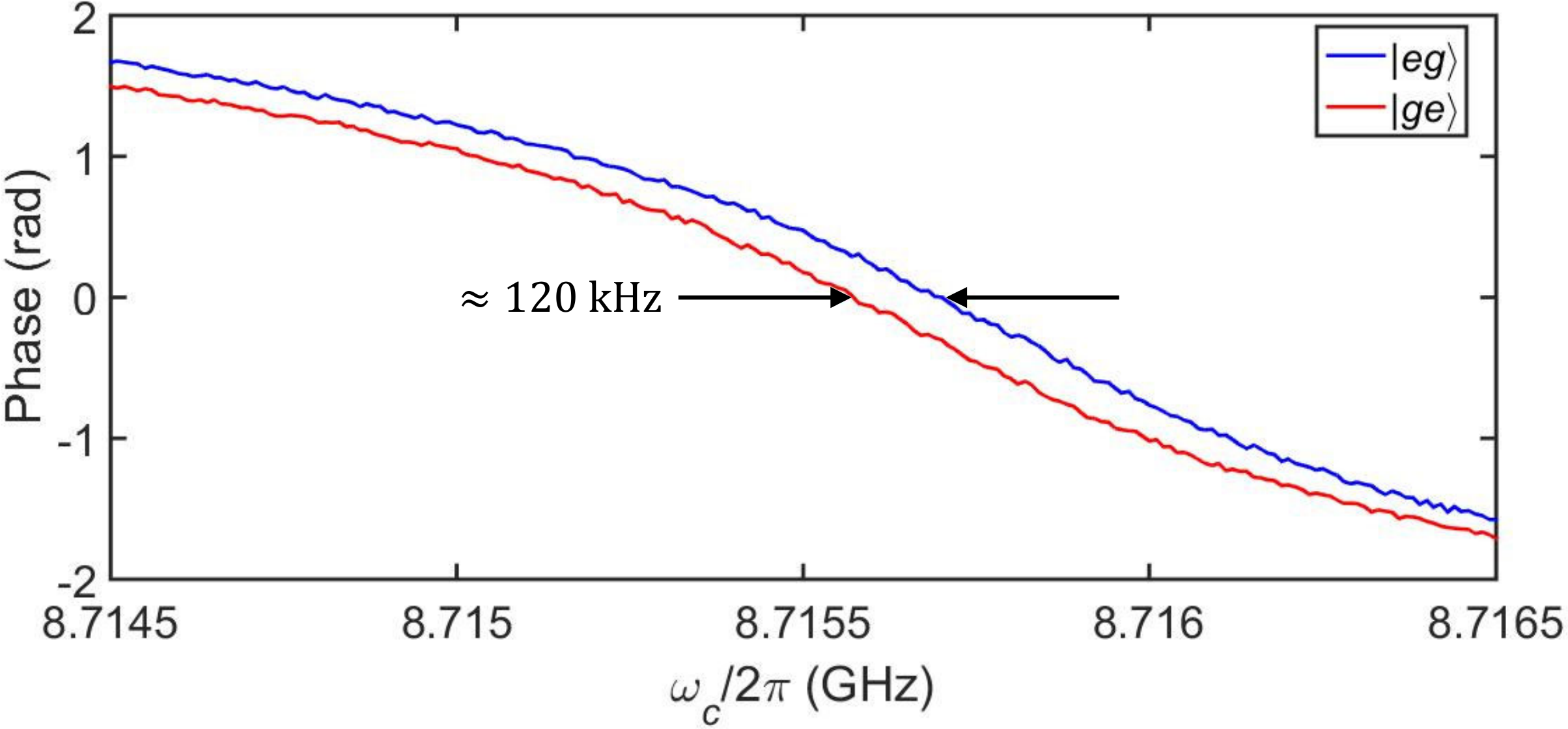}
\caption{\textbf{Fine measurements of the readout cavity for states $|eg\rangle$ and $|ge\rangle$.} By measuring the readout cavity response when preparing either $|eg\rangle$ or $|ge\rangle$, we can see that both qubits produce nearly equal dispersive shifts on the cavity. The total shifts are $2\chi_L/2\pi\approx2.8$~MHz and $2\chi_R/2\pi\approx2.9$~MHz, with a difference of 120~kHz.}
\label{sFig18}
\end{figure}

For our circuit design, we use a joint two-qubit readout with one coplanar waveguide cavity coupled to both qubits. This simplifies the layout and requires only one cavity response to be monitored with a Josephson Parameteric Amplifier (JPA) to achieve single-shot readout. The JPA we use was designed and fabricated by our group using a niobium trilayer fabrication process.\cite{Gabe2020} This simplifies the measurement strategy to some degree, but also requires more effort to calibrate the readout properties in order to distinguish all four computational states of the two-qubit system. Because both qubits impart a disperseive shift onto the cavity, there are four cavity frequencies for the four states $|gg\rangle$, $|eg\rangle$, $|ge\rangle$, $|ee\rangle$. For our system, the dispersive shifts from each qubit (for one excitation) are nearly equal, $\chi_L\approx\chi_R$, so that the states $|eg\rangle$ and $|ge\rangle$ are nearly indistinguishable.\cite{Noh2023} We can see this by monitoring the readout cavity as a function of the drive frequency when the qubits are prepared in a mixture of all four two-qubit states. This leads to three possible phase curves with drive frequency as shown in Fig.~\ref{sFig17}. At the lowest and highest drive frequencies, the phase response is nearly the same for all states, but between 8.715 and 8.716~GHz there are three well separated phase values that can distinguish three possibilities: $|gg\rangle$, $|ee\rangle$, and the overlapping response of $|eg\rangle$ and $|ge\rangle$. If we carefully measure the readout cavity response when preparing either $|eg\rangle$ or $|ge\rangle$, we can measure the phase response for both qubits separately (see Fig.~\ref{sFig18}), revealing that $2\chi_L/2\pi\approx2.8$~MHz and $2\chi_R/2\pi\approx2.9$~MHz, with a difference of about 120~kHz.

Also in Fig.~\ref{sFig17}, we show the single-shot histograms in the $I-Q$ plane at several cavity drive frequencies. At the lowest (highest) drive frequency we can distinguish only one (two) regions in $I-Q$-space. This response splits into three regions at the central drive frequency allowing us to clearly distinguish three ``balls'' for $|gg\rangle$, $|ee\rangle$, and the overlapping response of $|eg\rangle$ and $|ge\rangle$. To maximize the two-qubit readout fidelity we must calibrate the readout drive frequency and amplitude in order to maximally separate these balls in $I-Q$-space. This can be quantified by measuring the three angles $\theta_i$ and three distances $L_i$ that characterize the triangle formed by connecting the centroids of the three balls. 

\begin{figure}[t!]
\centering
\includegraphics[width=\textwidth]{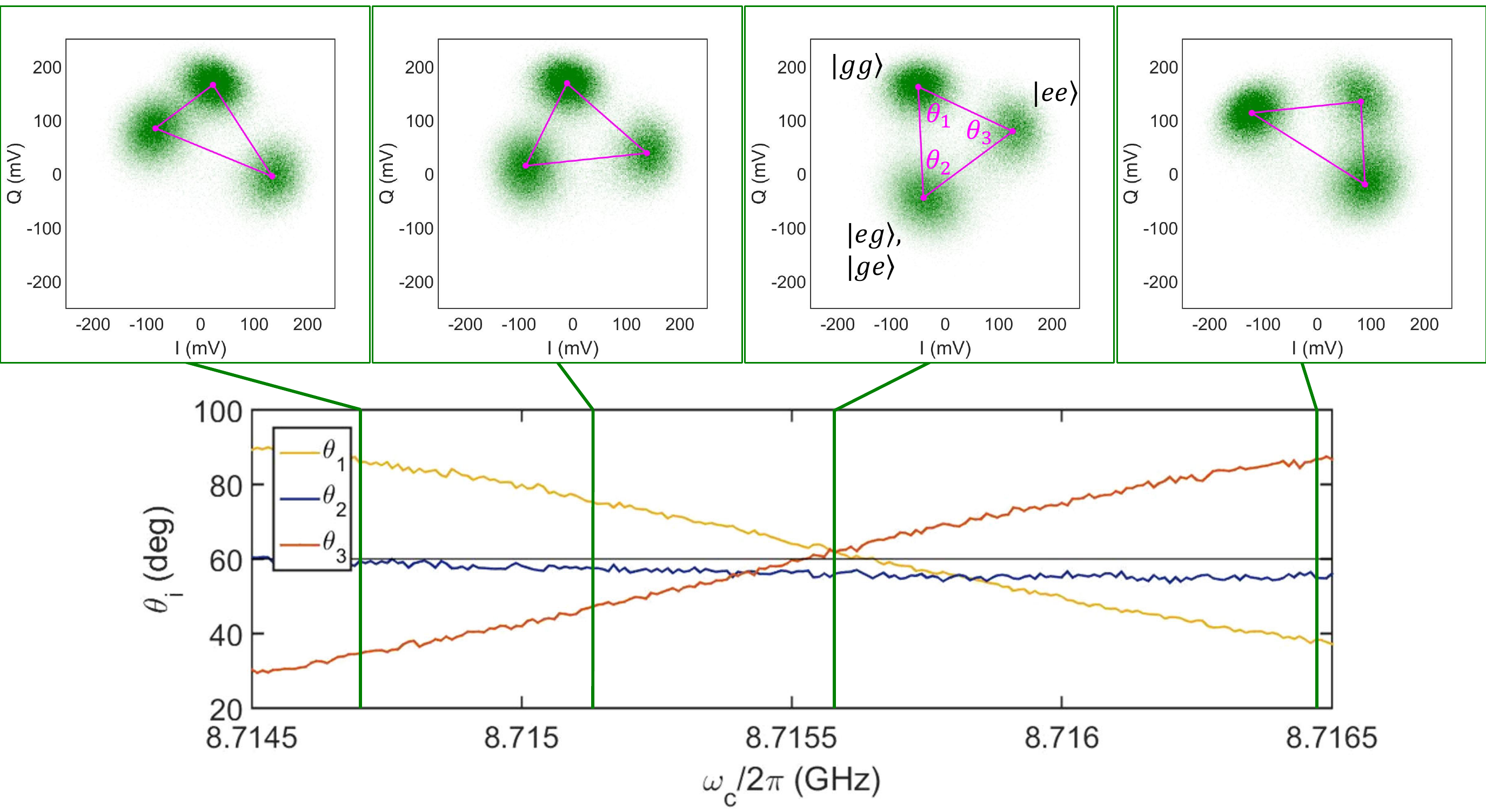}
\caption{\textbf{Readout cavity measurements as a function of drive frequency at drive amplitude of 1150~mV.} The main plot shows the three angles $\theta_i$ that characterize the triangle formed by connecting the center of the three response regions in the $I-Q$ plane as a function of the drive frequency when prepared in a mixture of two-qubit states. We can see that near 8.7155~GHz the three regions are maximally separated mostly closely, forming an equilateral triangle.}
\label{sFig19}
\end{figure}

\begin{figure}[t!]
\centering
\includegraphics[width=\textwidth]{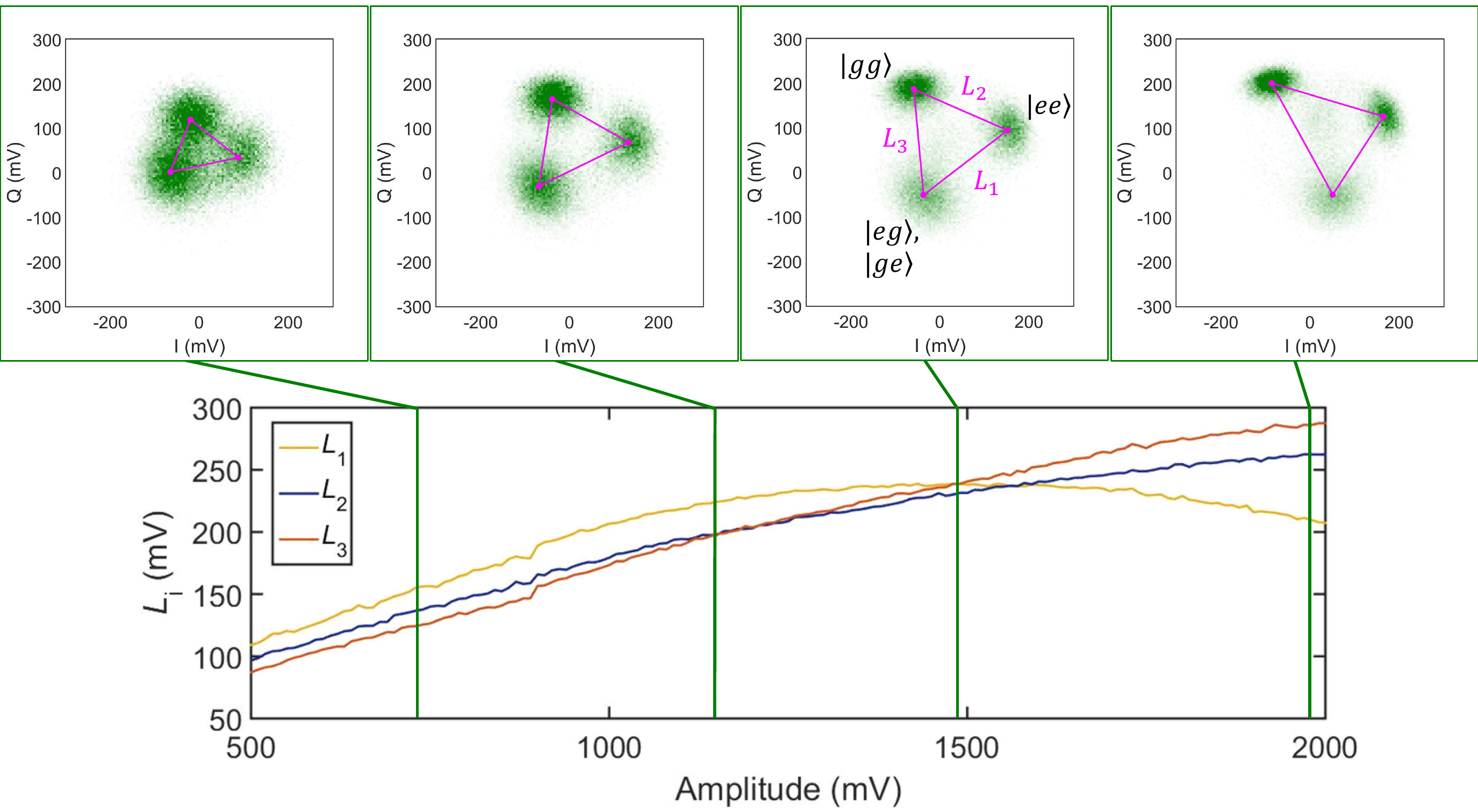}
\caption{\textbf{Readout cavity measurements as a function of drive amplitude at drive frequency 8.7154~GHz.} The main plot shows the three lengths $L_i$ that characterizes the distances connecting the center of the three response regions in the $I-Q$ plane as a function of the drive amplitude when prepared in four two-qubit states $|gg\rangle$, $|eg\rangle$, $|ge\rangle$, and $|ee\rangle$. We can see that near 1500~mV, the three distances are equally and maximally separated. Notice that for large amplitudes, we can see major distortions and point-spreading of the three regions.}
\label{sFig20}
\end{figure}

In Fig.~\ref{sFig19}, we show the three angles $\theta_i$ as a function of the drive frequency. In the $I-Q$ portraits, we can see that above and below the central frequency the scalene triangles have an acute angles where two of the balls have poor distinguishability. However, it is clear that near 8.7155~GHz the three balls are maximally separated forming an equilateral triangle with all $\theta_i\approx 60^o$. In Fig.~\ref{sFig20}, we show the three lengths $L_i$ as a function of the cavity drive amplitude when driven near the optimal cavity drive frequency found in Fig.~\ref{sFig19}. Notice that as the drive amplitude increases all three lengths increase separating the three balls. We can see that near 1500~mV, the three distances are equally and maximally separated. In the $I-Q$ portraits, we can see that for the largest amplitudes the triangle begins to distort leading to scalene triangles. In addition, the density of points between balls increases signifying that the readout is starting to cause unwanted transitions between the qubit states during the readout.

In order to avoid distortions and unwanted state transitions we chose a moderate drive amplitude of 1150~mV at a cavity drive frequency of 8.71555~GHz, where the separations were reasonable large and the balls nearly formed an equilateral triangle. Under these conditions we can estimate the separation fidelity for the measurement of $|gg\rangle$ by knowing the standard deviation $\sigma$ and separation $S$ of the three balls.\cite{Gard2024} Here, we must sum the errors ($erf(S_i/2\sqrt{2}\sigma_i)$ = 3.0\%, 2.2\%, 2.7\%) coming from the overlap of $|gg\rangle$ with the gaussian balls of $|eg\rangle$, $|ge\rangle$, and $|ee\rangle$. This equates to a separation fidelity of 92.1\%. Had we been able to operate at 1480~mV without adverse effects from increased qubit relaxation, this would have increased to 98.5\%. Measurements of $|gg\rangle$ also provided an estimate of the sample temperature of approximately 60~mK.

In order to fully distinguish the states $|eg\rangle$ and $|ge\rangle$ for monitoring the parametric vacuum Rabi oscillations like those shown in Fig.~4 in main text, we apply a $\pi$-pulse to the $L$ ($R$) qubit before driving the cavity readout tone. This ``shuffles'' the population of $|eg\rangle$ and $|ge\rangle$ to $|gg\rangle$ and $|ee\rangle$ ($|ee\rangle$ and $|gg\rangle$), the two distinguishable outcomes. Shuffling either excited state to $|gg\rangle$ actually increases its fidelity by avoiding state decay during the readout, which was about $1-\exp(-0.5/15)\approx 3\%$ for a 500~ns readout pulse with $T_1=15\,\mu$s. In order to distinguish the $|f\rangle$ states of either qubit when operating both qubits, we had to apply shuffle pulses to separate $|ee\rangle$ from $|fg\rangle$ or $|gf\rangle$. In general, our readout strategy requires measuring statistics on pulse sequences multiple times with and without shuffle pulses. 

\begin{figure}[p!]
\centering
\includegraphics[width=5in]{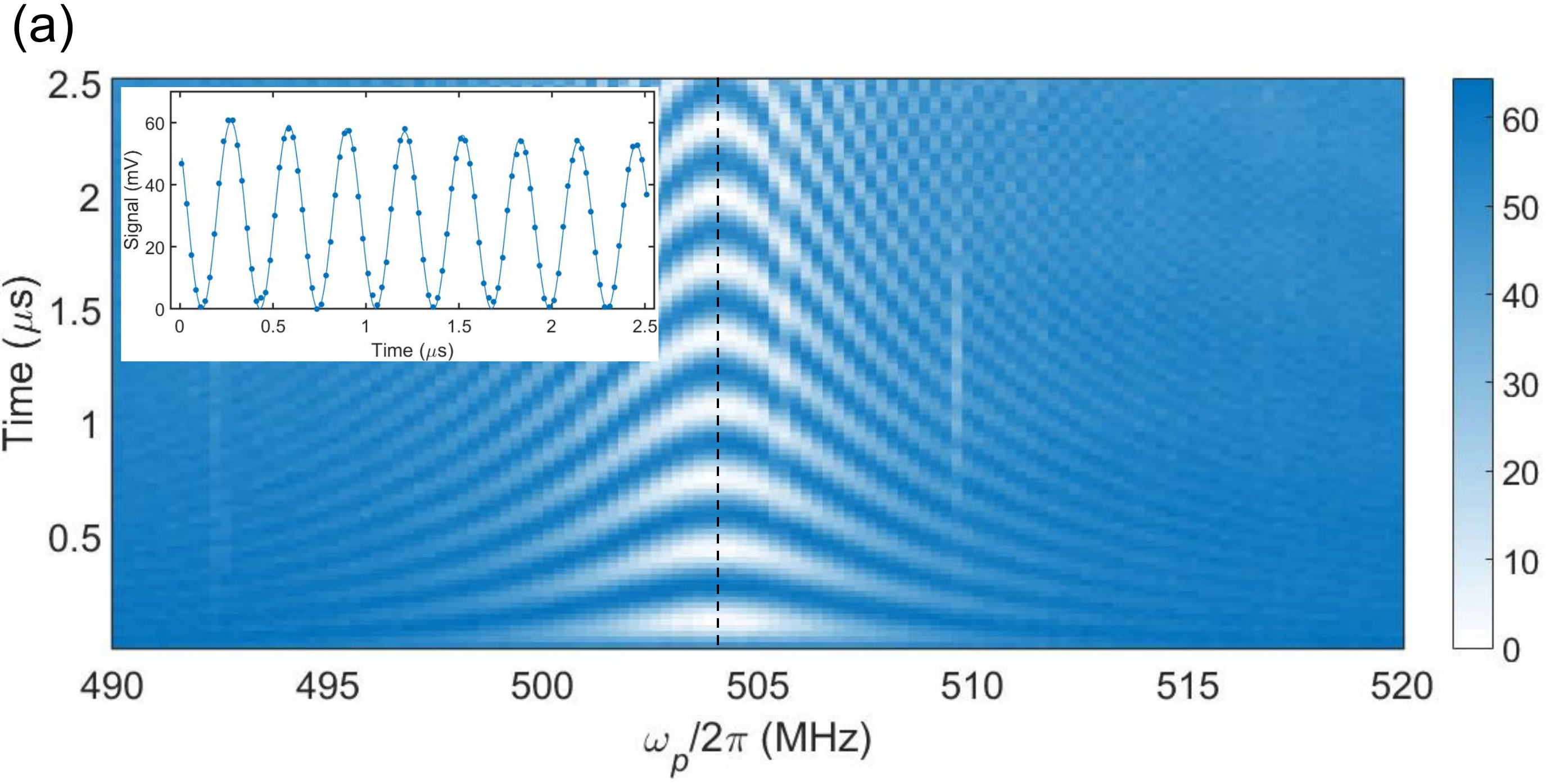}
\includegraphics[width=5in]{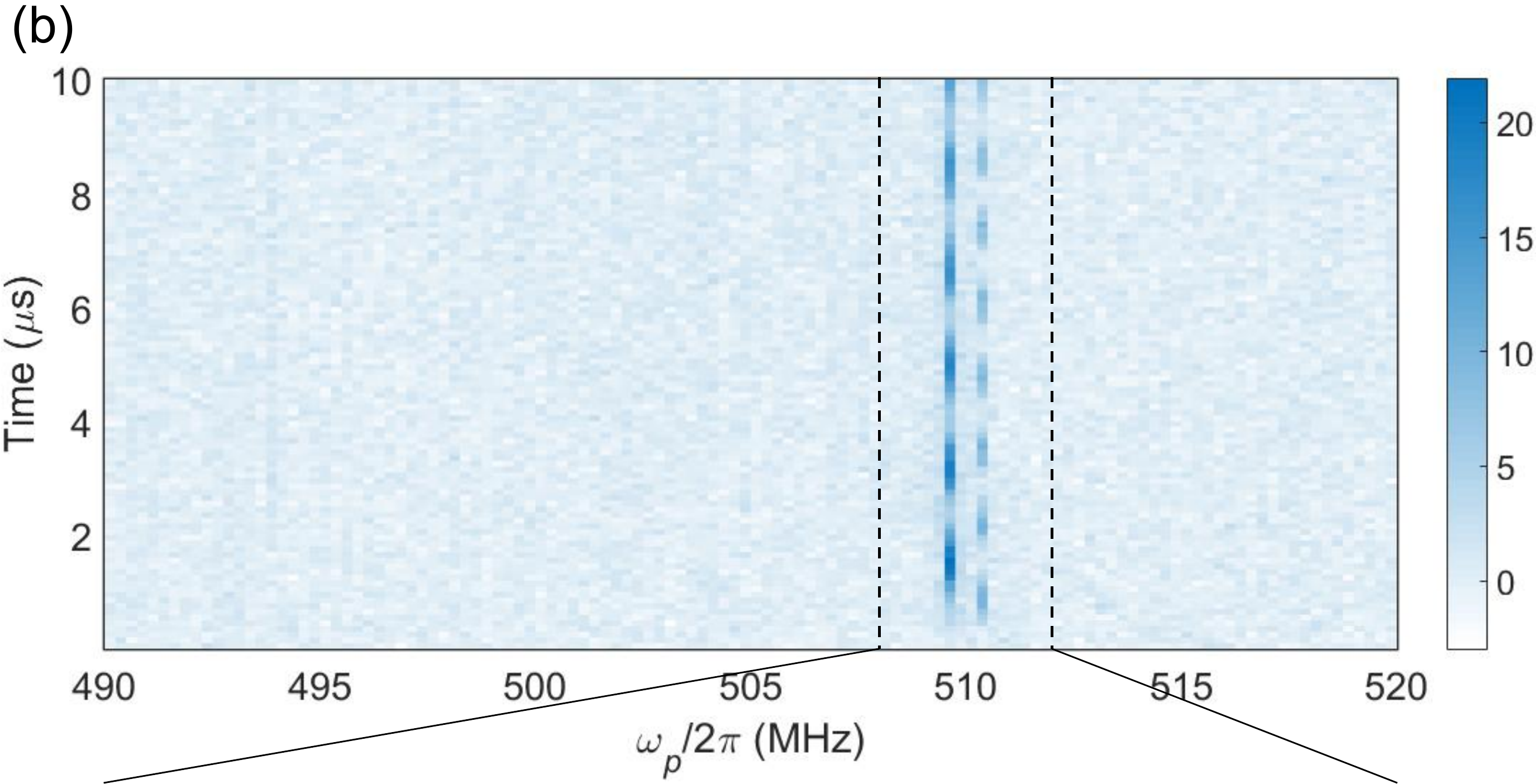}
\includegraphics[width=5in]{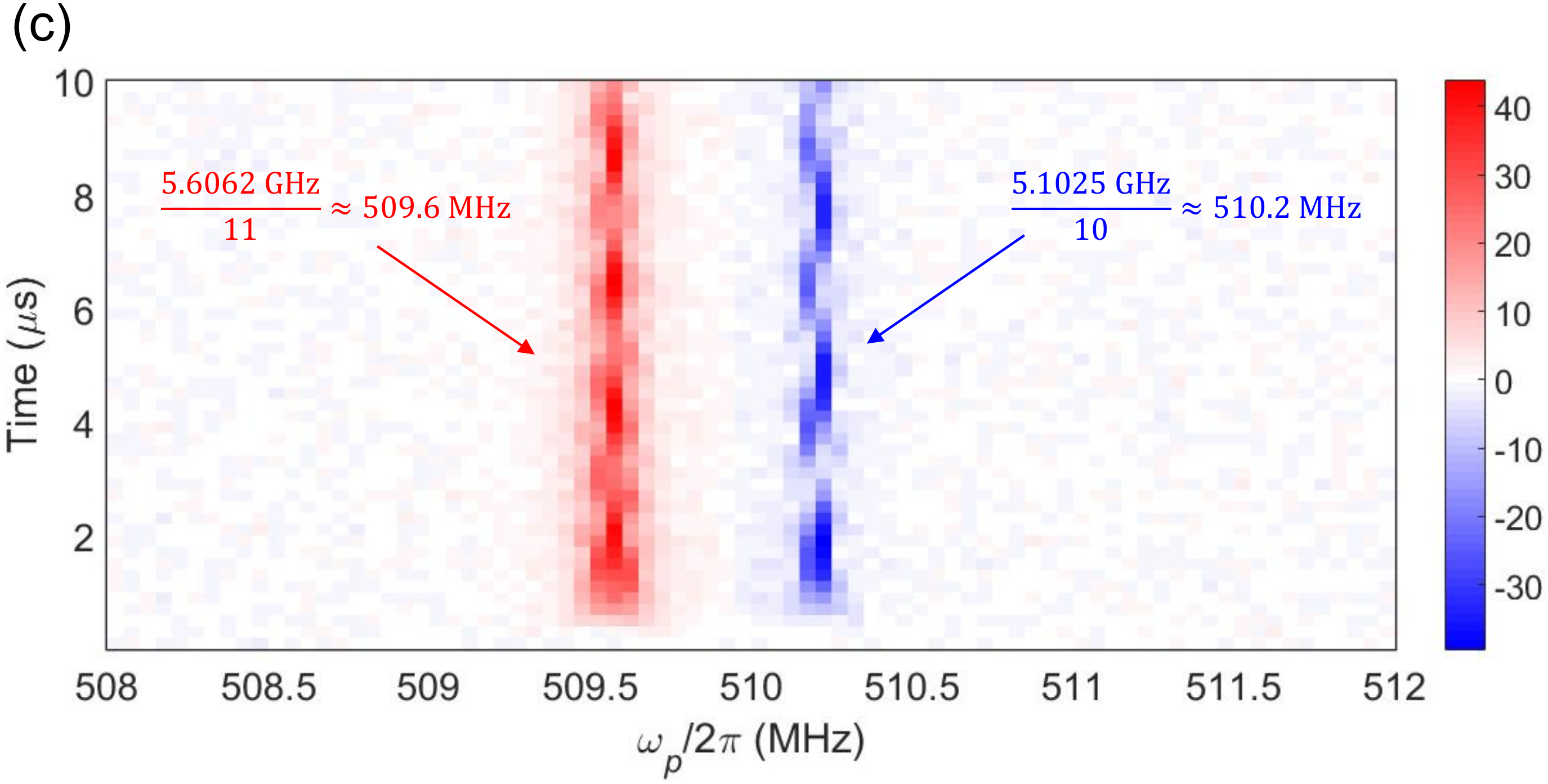}
\caption{\textbf{Detecting multi-photon transitions in the transmons from the parametric pump drive.} (a) Parametric oscillations between $|eg\rangle$ and $|ge\rangle$. The inset shows a line cut along the center of the oscillations. (b) The same measurement, but starting from the state $|gg\rangle$. We clearly can see two regions that show excitations in the measurements. (c) A zoom-in of (b) showing that each qubit ($L$ in blue and $R$ in red) is being Rabi driven by the pump drive at the displayed subharmonic frequencies.}
\label{sFig21}
\end{figure}

\section{Transmon excitations from sub-harmonic pump drives}
\begin{figure}[b!]
\centering
\includegraphics[width=\textwidth]{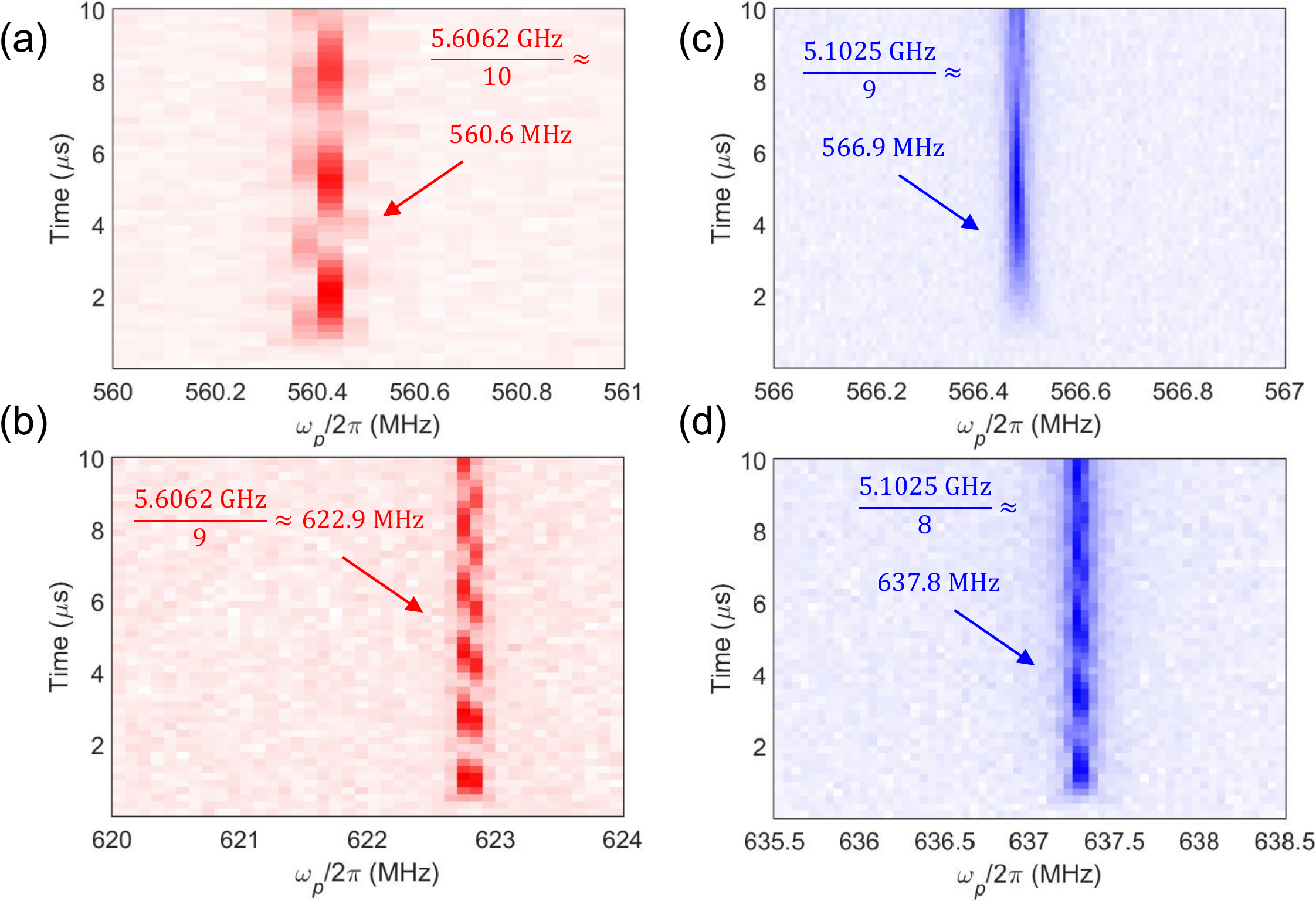}
\caption{\textbf{More zoom-ins of multi-photon transitions.} (a) A 10-photon subharmonic drive on the $R$ qubit. (b) A 9-photon subharmonic drive on the $R$ qubit. (c) A 9-photon subharmonic drive on the $L$ qubit. (d) A 8-photon subharmonic drive on the $L$ qubit.}
\label{sFig22}
\end{figure}

For both kinds of c$Z$ gates described in this work, the operation-induced error rises steeply as the parametric gate pulses get shorter than their optimal value, as seen in Fig.~5(a) of the main text. This divergent error can not be simply explained by the linearly expanding bandwidth in frequency space of shorter shorter microwave pump pulses. To investigate these errors, we performed a diagnostic experiment where we measured the induced vacuum Rabi oscillations over a large bandwidth of microwave frequency for pulsed parametric drive amplitudes close to that used in c$Z$ gates. Upon fine inspection of these data (shown in Fig.~\ref{sFig21}(a)), it is possible to see ``streaks'' at certain pump frequencies. If we retake this data, but start in $|gg\rangle$, then the parametric drives should, under ideal circumstances, \emph{not} change the state whatsoever. However, as we can see in Fig.~\ref{sFig21}(b), we observe Rabi-like oscillations at multiple parametric drive frequencies indicative of two-qubit state transitions. We further interrogate these regions by zooming into their location and by using shuffle pulses (see the previous section~\ref{readout}) in order to determine which qubit is being excited. In Fig.~\ref{sFig21}(c), we can see one set of oscillations shows the $R$ transmon is excited, whereas the other oscillations are coming from the $L$ transmon. By noting the pump frequencies where this occurs, we find that each qubit is being uniquely driven at an integer fraction of their individual excitation frequency. In the case shown here, the $R$ transmon absorbed 11 photons and the $L$ transmon absorbed $10$ photons. Once we understood this problem, we verified these excitations at many more pump locations as highlighted in Fig.~\ref{sFig22}. Remarkably, we were able to find evidence of multi-photon absorption with up to 16 photons! This is conclusive evidence that multi-photon transitions are possible with strong parametric drives and must be considered when performing strong parametric interactions. 

The multi-photon excitation process of transmons described above has a highly nonlinear dependence on drive amplitude. It remains extremely weak at lower to moderate powers, but then begins to grow rapidly at much larger drive powers and has been proposed as a way to maintain control extremely Purcell filtered transmons.\cite{Xia2023}. Thus, for slower gate speeds at lower power, we do not incur any state errors from this process. However, as we reduce the gate duration, the drive power must be increased to larger and larger amplitudes, eventually reducing gate fidelity through these unwanted additional single qubit bit-flip errors. Because this error mechanism is coherent, we could in-principal apply additional single qubit drives to compensate for the multi-photon operations.\cite{Sung2021,Ding2023} It should be emphasized that this issue occurs at well-defined locations in frequency space and could be easier to avoid by further separating the two qubit frequencies. In addition, it is known that our bias coil and coupling design to the SQUID is not optimal for purely driving flux into the SQUID coil. Instead, capacitive coupling from the floating end of the coil to the transmon capacitors can significantly drive there transitions, as verified when we remove any bias line filter. In the future, we plan to significantly decrease this capacitive coupling by galvanically connecting the bias coil and SQUID with a better geometrical design to avoid multi-photon absorption and improve power transfer to flux in the SQUID.\cite{Li2023a}

\section{Automated gate optimization with CMA-ES: Covariance Matrix Adaptation Evolution Strategy} Covariance Matrix Adaptation Evolution Strategy (CMA-ES) is a robust and proficient algorithm tailored for tackling intricate optimization challenges, especially those within the realm of continuous and non-linear optimization. It works especially well when addressing optimization issues characterized by noisy, discontinuous objective functions or those teeming with numerous local optima\cite{wiki:CMA-ES, blog:CMA-ES}. In the field of superconducting qubits, CMA-ES has been used to optimize pulse shapes in a closed-loop fashion to increase the fidelity of single-qubit gates. This process involves treating each output point value of the arbitrary waveform generator as variables.\cite{2021LeakageReduction}

In this work, we've chosen to expand the application of the CMA-ES strategy to fine-tune the parametric two-qubit gates in a timely manner. For this task, the physical parameters of the parametric gates are considered as adjustable variables, including the pump amplitude, frequency, and the width of the pulses, plus the first and second order DRAG (Derivative Removal by Adiabatic Gate) pulse shape control parameters. Although a full detailed description and mathematical discussion of CMA-ES is beyond the scope of this Supplemental Material, here we describe the salient concepts, the general protocol, and our experimental realization.

\subsection{Manual Gate optimization and parameter space}
\label{manopt}
\begin{figure}[b!]
\centering
\includegraphics[width=\textwidth]{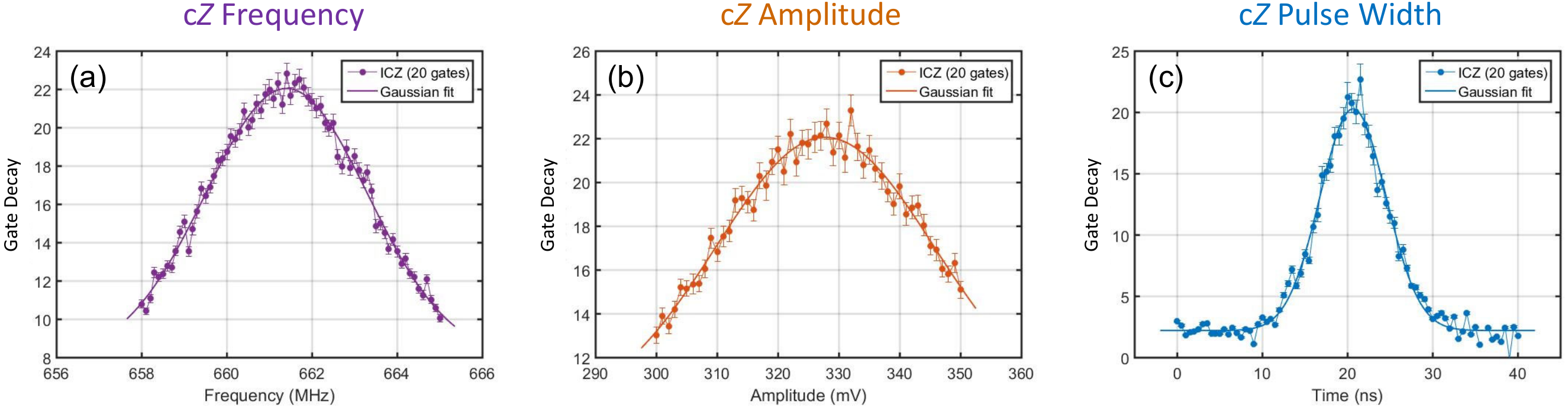}
\caption{\textbf{Optimizing the parametric c$Z$ gate pump parameters by hand.} (a) Gate decay number while adjusting the pump frequency. (b) Gate decay number while adjusting the pump amplitude. (c) Gate decay number while adjusting the pump pulse width.}
\label{sFig23}
\end{figure}

\begin{figure}[t!]
\centering
\includegraphics[width=\textwidth]{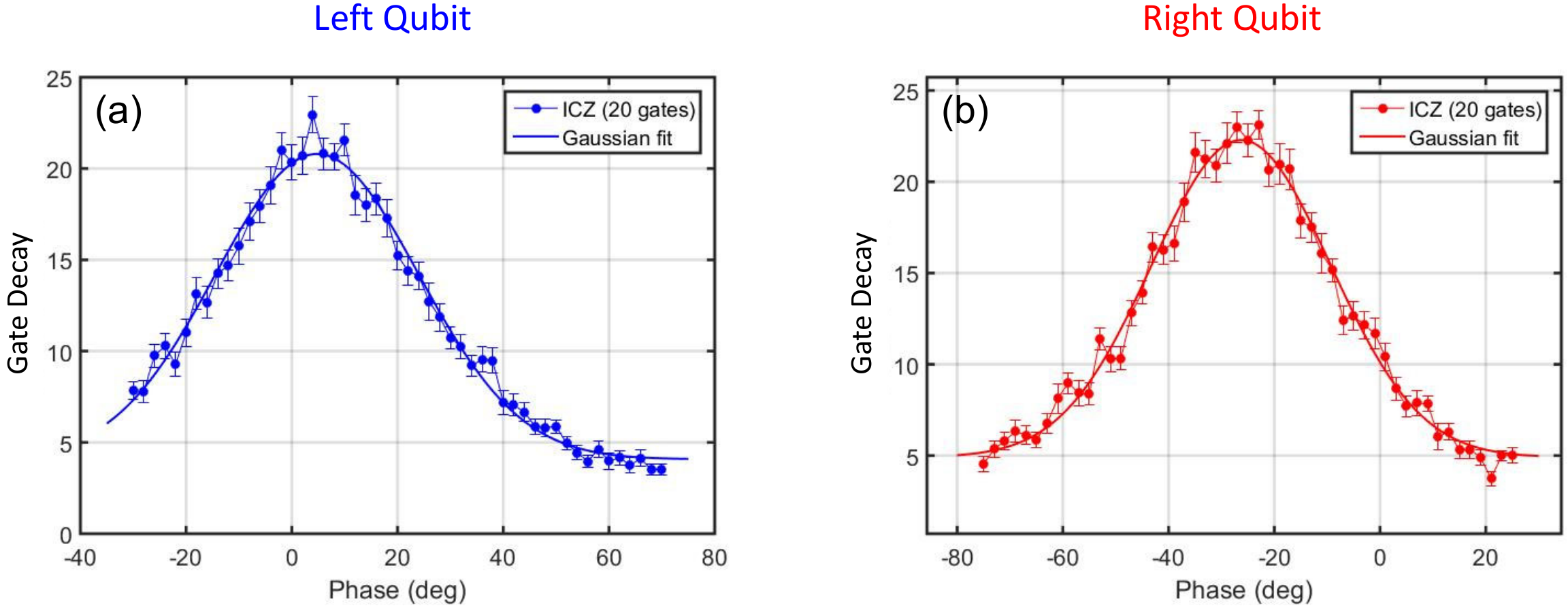}
\caption{\textbf{Optimizing by hand the virtual $Z$ rotations on each qubit during a parametric c$Z$ gate.} (a) Gate decay number while adjusting the $L$ qubit's virtual $Z$-gate. (b) Gate decay number while adjusting the $R$ qubit's virtual $Z$-gate.}
\label{sFig24}
\end{figure}

Before running the CMA-ES protocol, we first optimize the parametric gate parameters ``by hand''. This involves making simple parameter sweeps while evaluating a simple proxy for gate fidelity. The main parameters that characterize the parametric gates are pump amplitude, frequency, and the width of the parametric gate pulse. We chose to use the gate decay number as a proxy for Gate fidelity. This value is calculated as the exponential decay rate (in units of gate number) of the probability of finding the two qubits in state $|gg\rangle$ after a full sequence of gates. These measurements are done with enough points to characterize the probability decay so that we can relatively quickly sweep the parameters of the parametric gate and find optimal values. An example is shown in Fig.~\ref{sFig23}. In addition, because the parametric drive causes single-qubit frequency shifts due to rectification, as discussed above in section~\ref{calib}, we must rotate the basis of each qubit with a virtual $Z$-gate in order to compensate for the real $Z$ rotations imparted on the qubits during the gate operation. We calibrate these virtual $Z$ rotations as we did the other pump parameters. An example is shown in Fig.~\ref{sFig24}.

\subsection{The optimization protocol}
\begin{figure}[t!]
\centering
\includegraphics[width=\textwidth]{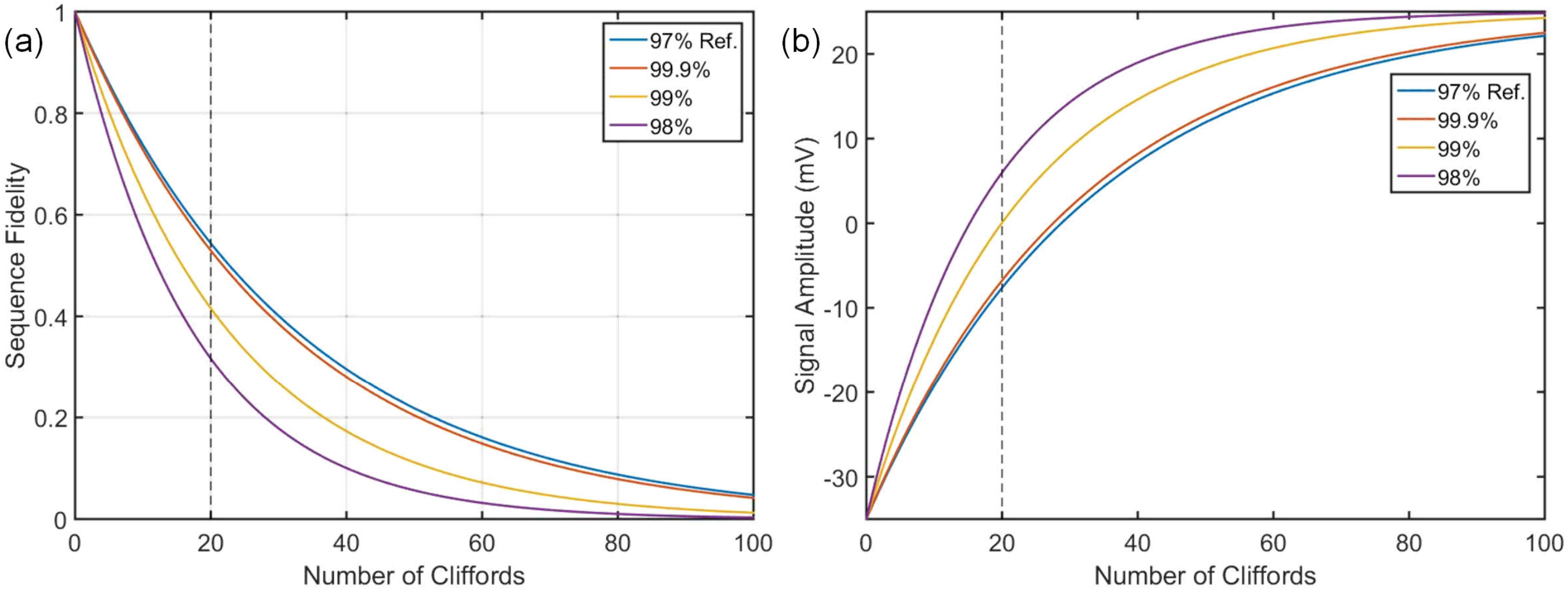}
\caption{\textbf{Visualizing sequence fidelity and raw data.} (a) Calculated sequence fidelity curves $F$ of 99.9\%, 99.0\% and 98.0\% based on a reference decay curve with $P_{\rm ref} = 0.97$. (b) Possible raw data curves corresponding to those shown in (a).}
\label{sFig25}
\end{figure}

In general, the two-qubit gate fidelity can be deduced by the interleaved randomized benchmarking experiments following this equation:
\eq
F = 1 - \left(1 - \frac{P_{\rm int}}{P_{\rm ref}}\right)\left(\frac{2^n-1}{2^n}\right),
\label{fidelity}
\xeq
where $n = 2$ is the number of qubits, and the interleaved decay rate $P_{\rm int}$ and the reference decay rate $P_{\rm ref}$ per gate are determined experimentally. The renormalized data (where the decay starts from 1 and saturates at 0 with no offset) should follow
\eq
S = P^N,
\label{fit1}
\xeq
where $S$ is the readout signal, $N$ is the number of interleaved gates applied, and $0<P<1$ is a fitting parameter that represents the average probability of success. Example curves corresponding to different decay rates and gate fidelities are plotted in Fig.~\ref{sFig25}(a).

Assuming the raw signal drift is negligible, it is possible to compare two sets of gate parameters and determine which one has higher fidelity, without actually having to measure the entire curve or calculating the actual fidelity of the gate. In Fig.~\ref{sFig25}(b) shows possible raw data values, the final voltage readout out of one quadrature of the cavity drive in mV. If we only measure this raw readout signal after $M$ interleaved gates (the vertical dashed line in Fig.~\ref{sFig25} with $M = 20$), then the value at that location will rise and fall with the fidelity. Thus, the gate optimization problem is identical to a maximization problem of a multi-parameter function with a large parameter space. This is where the CMA-ES protocol is ideal; in order to investigate a large parameter space efficiently to find an optimized set of parameters.

During the experimental calibration, the raw data comes in without renormalization, thus the fitting function Eq.~(\ref{fit1}) would become
\eq
S = AP^N + C,
\label{fit2}
\xeq
where $A$ is the amplitude of decay and $C$ the offset constant. Depending on the experimental details, $A$ could be positive or negative, meaning the data could be look like an exponential decay downwards or an exponential growth upwards. For the possible raw signal shown in Fig.~\ref{sFig25}(b), along the dashed line where $M = 20$, the signal amplitude decreases when the fidelity actually increases. Here, the optimization target of CMA-ES will become a minimization problem of the multi-parameter function for this raw data.

In order to make the CMA-ES method sensitive, one should choose a $M$ value that will reflect greater change in signal when fidelity improves. However, the distance between different fidelity curves is highly sensitive to the reference value. Clearly, a reference curve closer to 0.99 will allow more room to approach 0.99. In Fig.~\ref{sFig25}, we choose a reference decay rate of 0.97. Notice that in order to achieve 0.999, one must get very close to the 0.97 curve and the difference between the reference and the other curves seems to be maximum when about 67\% of the curve has decayed, or at the exponential gate decay number discussed above in section~\ref{manopt}. Therefore, knowing the reference curve, it is possible to choose $M$ to be close to the gate decay number. However, more gates also takes more time. Therefore, we practically chose $M = 10-20$ in our calibrations, which saves time but also maintains a decent signal size to enable sufficient optimization.

\subsection{Motivation and Experimental Realization of CMA-ES}
For simplicity, we assume that the two-qubit gate has only 2 parameters to optimize, namely the gate frequency $f$ and its amplitude $a$. Ideally, $f$ and $a$ are supposed to be orthogonal parameters, which means they can be calibrated independent of each other, and changing one parameter doesn't affect the optimization of the other. If this were true, the rough optimization of each individual parameter shown above in section~\ref{manopt} would have been sufficient. However, in real experiments where the loss of the microwave signal itself is a function of $f$ and the sample response is non-linear to the applied gate amplitude $a$, the two parameters are no longer orthogonal. Instead of a one-dimensional scan for $f$ and $a$ respectively, one needs multiple scans to find the optimized values and could end up finding only a local minimum instead of a global optimization point. A full two-dimensional scan could solve the problem, but at the cost of a geometrically increasing search time. Considering more parameters, the optimization process will become an extremely long process with no guarantee of locating the best existing parameter set. The application of CMA-ES could reduce the search time by more than a factor of 10, while maintaining if not exceeding the best optimization results.

Essentially, the CMA-ES optimization we apply is a closed-loop iteration process in which each iteration will attempt to optimize the parameters based on the results of the previous iteration. We start with a set of manually input ``original seed'' parameters, and several iteration control parameters. They include the population of candidate (number of seeds) $m$, the  selection pressure (survival rate) $0<s<1$, the scattering pre-factor $p$, and the initial starting step size $\epsilon$ for the parameters, respectively. Using the (greatly simplified) CMA-ES protocol, we do the following:

\begin{enumerate}
\item[(1)] We start with manually calibrated parameters, $f_0 = 661.5$~MHz and $a_0 = 328.5$~mV, and choose the initial step sizes of $\epsilon_f^0 = 5.0$~MHz and $\epsilon_a^0 = 5.0$~mV, based on the results from section~\ref{manopt}.

\item[(2)] We then generate the first generation of seeds: $f_1^0, f_2^0, ..., f_m^0$, and $a_1^0, a_2^0, ..., a_m^0$, where $m=50$ is the population mentioned above. The seeds are randomly distributed in the space of $f_0\pm\epsilon_f^0$ and $a_0\pm\epsilon_a^0$, respectively. We also decide to use a survival rate of $s = 20\%$

\item[(3)] We pick one set of parameter $P_1^0: \{f_1^0, a_1^0\}$ and build a two-qubit gate. We then generate an interleaved gate sequence with, e.g. $N = 10$ two-qubit gates interleaving $N+1$ randomized Clifford gates. At the end of the sequence we apply the readout pulse and record a signal value. We repeat this experiment for, e.g. $M = 50$ times, and get a mean value of $Q^0_1$.

\item[(4)] For each set of data $P^0_i$, we repeat (3) and get a mean value $Q^0_i$. At the end of the initial iteration, we get a series of $Q^0$ containing $m$ numbers.

\item[(5)] Assuming the optimal condition is to find the minimum. With a survival rate of $s$, only $m\times s = 10$ seeds should survive. We thus pick among $Q^0$ the 10 smallest values (assuming they are sorted) $Q^0_{1,2,...,10}$, and trace down their corresponding parameters $P^0_{1,2,...,10}$. Note that these numberings are different than the order in which experiments are carried out in (3).

\item[(6)] Based on $f^0_{1,2,...,10}$ and $a^0_{1,2,...,10}$, we calculate their mean values $f^0_m$ and $a^0_m$, and their standard deviation $S_f^0$ and $S_a^0$, respectively. Here, we adopt $f^0_m$ and $a^0_m$ to become the new parameters $f_1$ and $a_1$, and use the scattering pre-factor $p$ to calculate the new step sizes: $\epsilon_f^1 = p\times S_f^0$ and $\epsilon_a^1 = p\times S_a^0$.

\item[(7)] Now we can go back to step (2) and start the new iteration, where we generate the new sets of seeds in the new parameter space $f_1\pm\epsilon_f^1$ and $a_1\pm\epsilon_a^1$.
\end{enumerate}


\begin{figure}[t!]
\centering
\includegraphics[width=\textwidth]{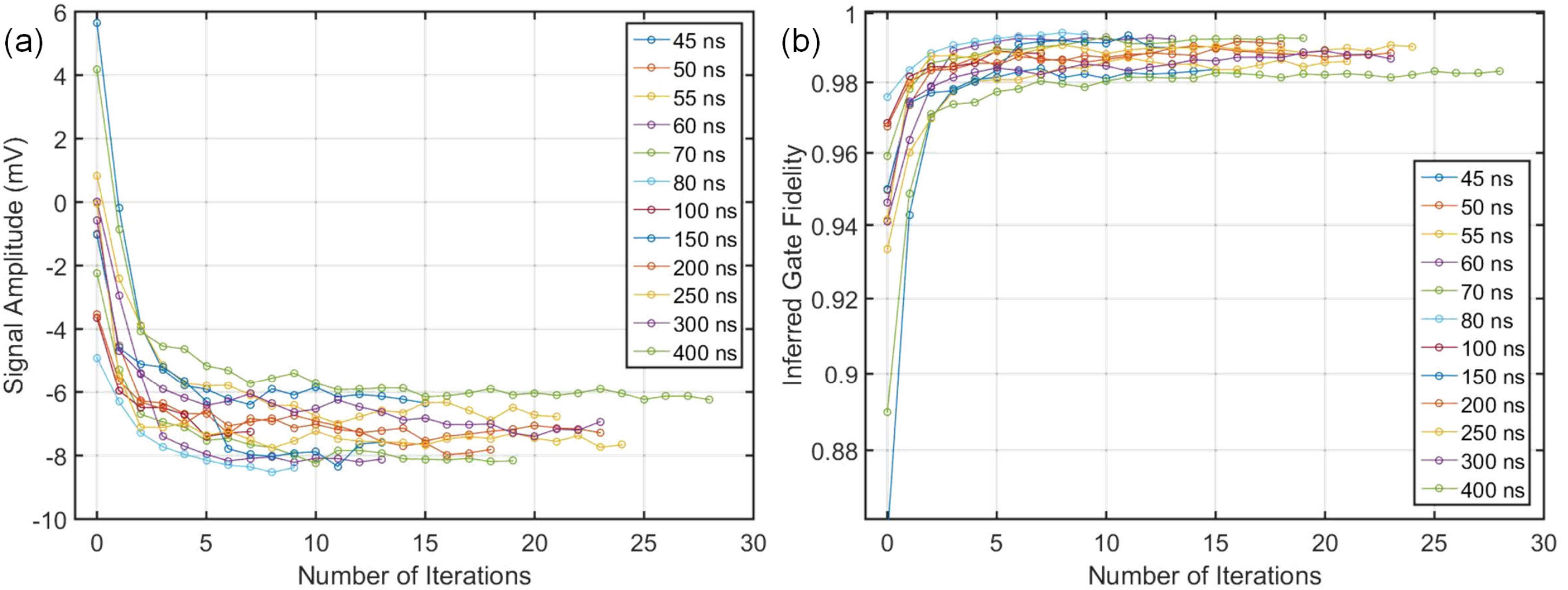}
\caption{\textbf{CMA-ES protocol to optimize gate fidelity.} (a) Raw data of CMA-ES calibration up to 28 periods. (b) CMA-ES calibration up to 28 periods converted to gate fidelities.}
\label{sFig26}
\end{figure}

The change of raw signal versus number iterations is plotted in Fig.~\ref{sFig26}(a) for one gate calibration, where the target is to find the minimum, and $M = 10$ was chosen. The improvement of fidelity is clearly seen (although quantitatively unclear) in the first 15 iterations, after which the improvement saturated into the noise background. This was done for various gate durations as indicated in the figure.

It is possible to convert the CMA-ES calibration results into a corresponding gate fidelity. To do this, we measure and fit the reference curve and obtain a value for $P_{\rm int}$ at the point where the calibration was carried out. This was done (curve not shown) and yielded $P_{\rm ref} = 0.964$. We can then reverse Eq.~(\ref{fit2}) to calculate $P_{\rm int}$:
\eq
P = [(S-C)/A]^{1/N}
\label{fit3}
\xeq
Here $S$ is the CMA-ES raw data, whereas $A$ and $C$ are parameters obtained at the same time as fitting $P_{\rm ref}$. We can then use Eq.~(\ref{fidelity}) to convert the CMA-ES data to fidelity values point by point. The results are shown in Fig.~\ref{sFig26}(b). Notice that for longest gate durations the saturating gate fidelity is consistently lower, agreeing with the decoherence limited predictions shown in Fig.~5 of the main text.

\bibliographystyle{naturemag}
\bibliography{refs_all_Arxiv}

\end{document}